\documentclass[aps,prd,twocolumn,10pt,nofootinbib,showkeys,superscriptaddress]{revtex4-1}
\usepackage{amsmath}
\usepackage{amssymb}
\usepackage{graphicx}
\usepackage{color}
\usepackage{hyperref}

\usepackage[english]{babel}
\usepackage[utf8]{inputenc} 

\newcommand{\nk}{\textbf{k}}
\newcommand{\nq}{\textbf{q}}
\newcommand{\dphi}{\delta \phi}
\newcommand{\x}{\textbf{x}}
\newcommand{\y}{\textbf{y}}
\newcommand{\bra}{\langle}
\newcommand{\ket}{\rangle}
\newcommand{\mH}{\mathcal{H}}
\newcommand{\mP}{\mathcal{P}}
\newcommand{\mR}{\mathcal{R}}

\newcommand{\nn}{\nonumber \\}

\newcommand{\RI}{\text{R,I}}

\graphicspath{ {../} }

\begin{document}

\title{Emergent universe revisited through the CSL theory}

\author{Gabriel R. Bengochea}
\email{gabriel@iafe.uba.ar} \affiliation{Instituto de Astronom\'\i
	a y F\'\i sica del Espacio (IAFE), CONICET - Universidad de Buenos Aires, (1428) Buenos Aires, Argentina}

\author{Mar\'{\i}a P\'{\i}a Piccirilli}
\email{mpp@fcaglp.unlp.edu.ar}
\affiliation{Grupo de Cosmolog\'{\i}a, Facultad
	de Ciencias Astron\'{o}micas y Geof\'{\i}sicas, Universidad Nacional de La
	Plata, Paseo del Bosque S/N 1900 La Plata, Argentina.\\
	CONICET, Godoy Cruz 2290, 1425 Ciudad Aut\'onoma de Buenos Aires, Argentina. }

\author{Gabriel Le\'{o}n}
\email{gleon@fcaglp.unlp.edu.ar }
\affiliation{Grupo de Cosmolog\'{\i}a, Facultad
	de Ciencias Astron\'{o}micas y Geof\'{\i}sicas, Universidad Nacional de La
	Plata, Paseo del Bosque S/N 1900 La Plata, Argentina.\\
	CONICET, Godoy Cruz 2290, 1425 Ciudad Aut\'onoma de Buenos Aires, Argentina. }

\begin{abstract}

In this work we analyze how the spectrum of primordial scalar perturbations is modified, within the emergent universe scenario, when a particular version of the Continuous Spontaneous Localization (CSL) model is incorporated as the generating mechanism of initial perturbations, providing also an explanation to the quantum-to-classical transition of such perturbations.  On the other hand, a phase of super-inflation, prior to slow-roll inflation, is a characteristic feature of the emergent universe hypothesis. In recent works, it was shown that the super-inflation phase could generically induce a suppression of the temperature anisotropies of the CMB at large angular scales. We study here under what conditions the CSL maintains or modifies these characteristics of the emergent universe and their compatibility with the CMB observations.

\end{abstract}

\maketitle

\section{Introduction}
\label{intro}
The success of the standard $\Lambda$CDM cosmological model in explaining the many accurate astronomical observations we have today (e.g. \cite{Planck15,Planck18a,Planck18c,Scolnic18,eBOSS}) includes inflation, a phase of accelerated expansion during the very early epoch of the universe \cite{Starobinsky79,Starobinsky80,Guth81,Mukhanov81,Linde82,Linde83,Albrecht82,Bardeen83,Brandenberger84,Hawking82}.

There are renowned merits attributed to the inflationary paradigm. In addition to solving the horizon problem, predictions from the simplest slow-roll inflationary model, such as a spatially flat geometry characterizing the universe and a quantum origin of the spectrum of primordial perturbations (i.e. a nearly scale-invariant power law), are some of them. Moreover, the predictions are extremely consistent with recent observations from the cosmic microwave background (CMB) radiation \cite{Planck18b}. However, the exploration of some alternatives seems to be interesting, in the light of small features in the CMB power spectra that remain unexplained and also open discussions in some recent works, which we mention below.

Some studies have drawn attention to the lack of large-angle correlations and a weak power in the low-$\ell$ multipole moments of the angular power spectrum, in the observed CMB temperature anisotropies with respect to that predicted within the standard $\Lambda$CDM model. This was first mentioned in the COBE results \cite{cobe} and later confirmed in subsequent generations of satellites \cite{wmap1,wmap2,wmap3,wmap4,planck13,planck1,planck2,planck3}. This feature on the largest angular scales was analyzed, with some controversies included, by several authors (see for instance \cite{copi1,copi2,copi3,copi4,sarkar,efstat}). As is well known, the largest observable angular scales contain direct information on primordial physical processes that occurred during the inflationary era (or prior to it), and could only have undergone modifications by the physics involved in the relatively recent past, e.g. through the late-time integrated Sachs-Wolfe effect. In fact, several authors have shown that some of the observed CMB anomalies could be explained in this way, e.g. \cite{Rakic06,Francis2010}. However, there is still no verdict on whether it is just a statistical fluke or if there really is something new and interesting behind it. Today it constitutes one of the persistent large-angle anomalies in the CMB data that makes up the list of current challenges of the standard $\Lambda$CDM model \cite{copi5,Perivola21}. In search of a convincing explanation, new theoretical ideas have been considered from different approaches. The results for the suppression of the low multipoles in the CMB spectrum can give us clues towards new physics that determined the initial conditions for slow-roll inflation or even tell us something about an earlier phase before slow-roll inflation \cite{Sokolov93,Stevens93,Rocha04,Campa07,Bastero03,Contaldi03,Bridle03,Feng03,Piao04,Boya2006,Cao08,Destri08,Destri10,Handley14,Handley21,Ramirez11,Ramirez12,Biswas13,Liu2013,Liu2014,Ashtekar20,Agullo20,Agullo21}.

As mentioned above, the prediction that our current universe is spatially flat is one of those that is often mentioned in relation to inflationary models, and the data seem to indicate that indeed the observed universe is very close to flat. However, it does not imply that the spatial sections are exactly flat. Cosmological models in the context of non-flat cases have also been explored with interesting results, see e.g. \cite{Tryon73,Vilenkin82,Ellis1987,Ellis1988,Ellis91,Linde95a,Linde95b,White96,Linde98,Linde99,Ellis2002,Linde2003,Lasenby03,Uzan2003,Zhang2004,Ratra17a,Ratra18a}. Recently, in light of the analysis of some observational data, the debate about what is the spatial curvature of the universe has resurfaced, typically quantified via its fractional contribution to the cosmic energy budget today parameterized by $\Omega_K$ \cite{Rasanen2014,Ratra17b,Ratra18b,Ratra18c,Ratra18d,Ratra19a,Ratra19b,Handley2019,Efstathiou19,Riess2019,Silk2020,Valentino2020,Efstathiou20,DiValentino2020a,Benisty20,Vagnozzi20A,Vagnozzi20B,Dhawan21}. This parameter has an important role in determining the evolution of the universe and is closely related with the early universe physics, because if the spatial curvature is positive, then the curvature term will always dominate at early enough times in a slow-rolling inflationary epoch. And indeed, some of these recent analyzes suggest that a small positive spatial curvature could be present favoring, perhaps, the case of a closed universe. It is also interesting to mention that the question of spatial curvature participates in the so-called ``$H_0$ tension'' between sets of early time probes and a number of late time observational data (see e.g. \cite{Verde2019,DiValentino2021,Bolejko2017,Collett2019}). Additionally, spatial curvature also affects the search for the nature of dark energy, an energy component to explain the current stage of accelerated expansion and that would constitute around 70\% of the energy budget of the universe according to the $\Lambda$CDM model \cite{Clarkson07,Clarkson08,Verde10,DiValentino2020b,Valentino2020}.  In particular, it is known that by assuming flatness, when in fact $\Omega_K\neq0$, would induce critically large errors in reconstructing the dark energy equation of state, showing that including curvature as a free parameter is imperative in any future analyses that attempts to determine whether dark energy is a cosmological constant or is something more exotic. And, on the other hand, closed universe models can generally relax the Hubble tension between supernovae observations and the CMB. However, at this stage, there is no conclusive evidence for a positive curvature, but this is at least an attractive possibility supported by the data that warrants further exploration.

An important feature that was highlighted in \cite{Biswas13}, and of particular interest for the present work, is that a phase of super-inflation (i.e. a period where the Hubble parameter increases with time) prior to slow-roll inflation could be related to the suppression of power in the low CMB multipoles. Furthermore, any mechanism that attempts to solve the cosmological singularity problem, within a semiclassical spacetime description, will naturally contain such a phase.

Starting with the very known pioneering works on singularity theorems by Penrose and Hawking \cite{Penrose64,Hawking1965,HawkingPenrose1970}, other authors have made contributions extending their cosmological applications, including the case of inflationary models \cite{HawkingEllis1973,Wald84,Borde1993,Borde1994a,Borde1994b,Borde1996,Borde1997,Guth99, Borde2003}. To evade these singularity theorems, some works have studied and developed alternatives in the context of bouncing models, see e.g. \cite{Molina99,Kanekar2001,Khoury01,Steinh02,Batte04,Bojowald04,Biswas2005,Peter02,Batte15,Lilley15,Branden17,Matsui19,Barrau20}. Another alternative that manages to escape these theorems is the recently developed framework of the \emph{emergent universe} \cite{Ellis1}. The idea of an emergent universe is not new, it can be traced back to seminal works of Einstein and Eddington \cite{Einstein1917,Eddington30}.

The emergent universe (EU) of \cite{Ellis1, Ellis2} is one in which a spatially closed universe (based on General Relativity and dominated by a scalar field minimally coupled to gravity) \emph{emerges} from an initially past-eternal Einstein static state (with a finite initial size), enters a phase of super-inflation and then evolves towards slow-roll inflation to finally give rise to the standard hot-Big Bang. Because of how it is built, there is neither a horizon problem nor singularity. This model has been studied in recent years through different approaches and variants, establishing its stability conditions, analyzing fine-tuning issues, its viability both theoretically and observationally and making it clear which questions are still open; see, for instance \cite{Ellis3,Gibbons1987,Barrow2003,Mukherjee05,Mukherjee06,Banerjee07,Boehmer07,Parisi2007,Campo1,Debnath08,Banerjee08,Beesham09,Campo2,Paul10,Paul2010,Zhang10,Paul2011,Chatto11,Campo3,Campo4,Labrana2012,Cai2012,Rudra12,Ghose12,Liu13,Aguirre13,Cai2014,Ataz14,Zhang14,Bag14,Huang15,Boehmer15,Labrana15,Kho16,Zhang16,Rios16,Khodadi16,Barrau18,Labrana19}. Among the alternative scenarios of the early universe \cite{BrandenRev11,Branden18}, a recent realization of an emergent universe is the so-called string gas cosmology \cite{Branden89,Branden06,Branden08,Branden11}. In this model the universe begins in a long hot and almost static phase, dominated by a thermal gas of closed fundamental strings. Recently, some conjectures such as Swampland \cite{Swamp1} and Trans-Planckian Censorship (TCC) \cite{TCCVafa} have put very strong constraints on possible inflationary models \cite{Swamp2, Swamp3, Brandenb2012, Bedroya20}. However, alternative cosmologies such as bouncing and emergent models are consistent with these conjectures and trivially satisfied. A good understanding of the emerging phase is still missing, but there are some promising approaches \cite{Branden21}.

The next point to consider is the fact that any model that claims to provide a mechanism for the generation of the seeds of cosmic structure, must be able to give a convincing answer to the following issue. In the early stages at the beginning of the universe (and well after the Planck era has ended), the spacetime is assumed to be spatially isotropic and homogeneous. In addition, a standard assumption is that the perturbations of matter fields (e.g. the inflaton field) were in a quantum vacuum state also perfectly symmetric (the symmetry being spatial isotropy and homogeneity), usually described by the so-called Bunch-Davies vacuum. Then, an important puzzle arises, namely to explain the transition from a perfect symmetric state portraying  the early universe to the non-symmetric state that characterizes the current universe, which cannot be attributed to quantum unitary evolution. The standard approach to address that issue is based on the study of the role of \emph{quantum fluctuations} during, for instance, the inflationary epoch. However, since the evolution of any quantum state, according to standard quantum theory, is always dictated by the Schr\"odinger equation (which does not break any initial symmetry of the system or destroy quantum superpositions), the traditional early universe paradigm is incomplete in that sense. In other words, the solely existence of vacuum fluctuations is in no way sufficient to claim that there are actual inhomogeneities of any kind present in the universe. The aforementioned puzzle is usually referred in the literature as the quantum-to-classical transition of the primordial perturbations. In fact, this subject is closely related to what is known as \emph{the measurement problem} in quantum physics \cite{Wigner63,Omnes,Maudlin95,Becker,Norsen,Durr,Albert,okon14}, and is notoriously exposed in the case of the quantum description of the primordial universe \cite{Bell81,PSS06,Sudarsky11, Susana13}. This is because measuring devices and observers who decide when and how to perform some kind of measurements cannot be fundamental notions in a theory which seeks to describe the early universe where neither existed \cite{Hartle93}. Cosmologists have tried to account for the quantum-to-classical transition by different types of arguments. A recent critique of different efforts at explaining this subject can be found in \cite{Berjon21}\footnote{The reader interested in a pedagogical review on this subject, can find it in \cite{Bengochea20}.}.

In order to classify the possible alternatives to address the issue described above, in \cite{Maudlin95} the measurement problem was written in a elegant manner showing that there are three statements that are mutually inconsistent. In short: (A) the physical description provided by the quantum state is complete, (B) quantum states always evolve according to the Schr\"odinger equation, and (C) measurements always have definite results.

If statement (A) is denied, then the quantum state does not contain all the information necessary for the description of a quantum system. In this way, the addition of \emph{hidden variables} and the equations that determine their evolution is required. The best known proposal for this case is the \emph{de Broglie-Bohm model} \cite{bohm}. Some applications to the cosmological case can be found in \cite{Valentini08b,Neto12,goldstein15,Neto18,Valentini19}.

Works based on decoherence \cite{kiefer09,halliwell,kiefer2,polarski} led to a partial understanding of the issue. Nevertheless, this argument by itself (i.e. without any extra assumptions) cannot address the fact that a single (classical) outcome emerges from the quantum theory. In other words, decoherence alone cannot solve the quantum measurement problem \cite{okon16,Adler01, schlosshauer}. Other cosmologists seem to adopt the Everett ``many-worlds" interpretation of quantum mechanics \cite{Everett} plus the decoherence process when confronted with the quantum-to-classical transition in the inflationary universe, e.g. \cite{mukhanov2005}. Regarding this point, we would like to refer the reader to \cite{Sudarsky11,kent,stapp} where arguments against decoherence and the Everett interpretation are also presented. This proposal and the Everettian interpretations are some of the approaches that somehow discard statement (C).

The last remaining choice is to negate statement (B). This path leads to non-standard quantum theories, i.e. theories where the collapse of the wave function is self-induced by some novel mechanism. Known as \emph{objective collapse theories}, from the mid-1970s several authors began to develop modifications to the Schr\"odinger equation, with the aim to alter the evolution of the wave function. In this way, the collapse of the wave function would occur without any reference to external observers or devices present that should perform some sort of measurements \cite{Pearle76,Ghirardi86,Pearle89,Diosi87,Diosi89,Penrose96}. One virtue of collapse models is that they have shown in recent years to have the attractive feature of connecting plausible resolutions of other open problems in a single unified picture \cite{Beneficios}. Reviews on these sort of theories can be found, for instance, in \cite{Bassi1, Bassi2}.

In the present work, we will approach the emergent universe from the perspective of the Continuous Spontaneous Localization (CSL) model \cite{Pearle76,Pearle89}, which will be incorporated into the situation at hand as a mechanism to break the original symmetries of the quantum vacuum state of the field driving the expansion of the early universe, and generating the primordial cosmological perturbations. In this manner, the CSL model naturally provides an explanation of the quantum-to-classical transition of such perturbations. The incorporation of objective collapse schemes and theories in the cosmological context has been studied since 2006 \cite{PSS06}. This has led to numerous investigations in recent years with varied proposals, particularly in the framework of semiclassical gravity (but also with exploratory works in the framework of standard quantization), with very encouraging results \cite{Daniel10,Sudarsky11,Tejedor12,Tejedor12B,Pedro13,Bengochea15,Leon15,Leon16,Leon17,Landau12,Susana13,Benetti16,Bengo17,Pedro18,Benito18,Picci19,Lucila15,Mariani16,Maj17,ModosB,Bouncing16,Josset17,Leon2020}. Other authors have investigated similar ideas and some of these works can be seen, for instance, in \cite{Martin12,Das13,Syksy15,Stephon16,Ellis18}. The debate about the particular details involving the implementation of the CSL theory into the cosmological context is still open. In fact, there is an extensive landscape of possibilities, which constitutes an active line of research at the moment \cite{MartinShadow,Bengo20Letter,Bengo20Long,Martin20R,Bassi21,Martin21,GLGB21}.

In particular, a relevant aspect to examine is the following: when one decides to combine quantum field theory (QFT) with gravitation, one must choose the setting within which such a link is to be made. At the time of writing there is no complete and finished program merging successfully both theories, so a couple of options arise. In Ref. \cite{Bengo20Long} some of these approaches were analyzed, evaluating their pros and cons. There, it was argued that the semiclassical gravity framework appears favored from a theoretical and conceptual point of view when one wants to incorporate collapse models. Therefore, our present analysis will be based on the semiclassical gravity (SCG) framework, in which gravity is treated classically and the matter fields are treated quantum mechanically \cite{PSS06,Tejedor12,Pedro13,Pedro18,Benito18,Benito20}. This approach accepts that gravity is quantum mechanical at the fundamental level, but considers that the characterization of gravity in terms of the metric is only meaningful when the spacetime can be considered classical. Namely, we will be dealing with the description of an epoch well after the full quantum gravity regime has ended (i.e. from when the emergent universe begins to evolve), where the energy scales involved allow one to suppose valid the consideration of the metric as classical and well described by semiclassical gravity equations. Therefore, semiclassical gravity can be treated as an effective description of quantum matter fields inhabiting a classical spacetime. While this approach has received some criticisms \cite{Eppley77,Page81}, those arguments have been refuted \cite{Mattingly05,Mattingly06,Kent18,Tilloy16,Carlip08,Albers08,Ford05Review,Ford05ReviewBook}. In the particular case of CSL theory, its first implementation into the primordial universe, based on the SCG framework, was done in \cite{Pedro13}, and some of us have continued to explore its consequences. For example, in \cite{Lucila15,Maj17,ModosB} it was shown that a strong suppression of primordial B-modes in the CMB is predicted generically. In addition, within that same framework, observational constraints were analyzed in \cite{Picci19}, and in \cite{Leon17} it was found that the condition for eternal inflation can be bypassed. In the next section, some additional motivations for this choice in the present case of analysis will become apparent.

To finish this Introduction, let us mention that in some previous works  \cite{Leon15,Benetti16,Bengo17}, where a spontaneous collapse of the wave function was implemented during a phase of slow-roll inflation, certain features on the low multipoles of the CMB were analyzed as consequence of the collapse. On the other hand, recently in \cite{Labrana15} the emergent universe model, originally put forward in \cite{Ellis1, Ellis2}, was analyzed and it was shown that the super-inflation phase (a characteristic shared by all emergent universe models) could be responsible for part of the anomaly in the low multipoles of the CMB; in particular, for the observed lack of power at large angular scales. Motivated by these results, and under the same assumptions of \cite{Labrana15}, here we calculate the primordial power spectrum of scalar perturbations, but incorporating a particular version of the CSL model for the situation at hand. We will analyze whether the super-inflation phase in the framework of the emergent universe plus CSL continues (or not) to produce the power suppression in the low multipoles. Lastly, we will study under what conditions the CSL maintains or modifies such characteristics and their compatibility with the CMB observations.

Our manuscript is divided as follows. We start in section \ref{secdos} presenting the theoretical framework of the emergent universe plus the CSL proposal; we also obtain the predicted scalar power spectrum. Next, in section \ref{sectres} we present and discuss our results; there, we also perform a further exploration of the corresponding parameter space. Finally, in section \ref{conclusions}, we present our conclusions. Regarding conventions and notation, we use a $(-,+,+,+)$ signature for the spacetime metric and units where $c=1=\hbar$.

\section{Emergent universe in the CSL framework}
\label{secdos}

In this section, we present the implementation of the CSL model into the emergent universe (EU) model.

\subsection{Theoretical background}

As we mentioned in the Introduction, many models of the emergent universe have been studied in recent years. Since we will be closely following the results of Ref. \cite{Labrana15}, here we will do our analysis under the same assumptions considered there, which in turn are based on \cite{Ellis1}. In particular, we assume the action of General Relativity with a scalar field $\phi$, which represents the dominant matter driving the universe early expansion, minimally coupled to gravity, with canonical kinetic term. The scalar potential employed (inspired by $R^2$-inflation) takes the form $V(\phi)=(4\pi G)^{-1}(e^{C\phi}-1)^2$ as the one reconstructed in \cite{Ellis2} (using techniques developed in \cite{Ellis91}). In obtaining that potential, the evolution of the scalar factor $a(t)\simeq a_0+A\:e^{H_0 t}$ was taken into account; where, $a_0 >0$ is the (initial) radius of the Einstein static universe, $C$, $A$ are positive constants, and $H_0$ is the Hubble parameter at the onset of slow-roll inflation\footnote{The type of potential $V$ is as the one shown in Figs. 1 of Refs. \cite{Ellis2,Labrana15}.}.

In the EU model, after leaving its initial static state, the universe enters a slow-rolling regime (at a few e-foldings after leaving its initial static state) where the scale factor grows sufficiently quickly to mitigate neglecting the curvature effects. This period of a de Sitter type of inflation comes naturally to an end (as the scalar field starts oscillating around the minimum of the potential), it is then followed by a re-heating phase, and finally continues to the standard hot Big Bang expansion. In Ref. \cite{Rios16}, it was shown that the temporal evolution, given by Friedmann equation along with the scalar field Klein-Gordon equation, leads the system towards an attractor where $H$ tends to a constant and $\dot{\phi}^2 \to 0$; this is, the system evolves from an Einstein static state to a de Sitter type of expansion.

The dynamics of this model is such that, prior to the traditional slow-roll inflation, there is a phase of super-inflation where the Hubble parameter increases with time, i.e. $\dot{H}>0$. The mechanism which generates this superinflationary period depends on the particular model under consideration, but it is a generic characteristic of the EU scenario. For example, in the models of Refs. \cite{Ellis1,Ellis2}, it is considered a FLRW closed universe where the spatial curvature is responsible for the superinflationary period. However, let us note that we could have chosen another model from those mentioned in the Introduction, provided that the evolution of the background (given by the mentioned scale factor $a(t)$) produces the phases of super-inflation and slow-roll inflation, in which the generation of curvature perturbations is analyzed here.

As discussed in depth in \cite{Ellis2}, even though the traditional emergent universe is with positive spatial curvature, it is quickly negligible in a few e-foldings and furthermore slow-roll inflation can always be made to end for some negative value of $\phi$. On the other hand, it is also possible to find the (finite) $N$ number of e-foldings for the total slow-roll phase within the emergent universe. Analogously to \cite{Labrana15}, we will now make a first approach to the problem at hand and therefore we will neglect the contributions of the space curvature to the primordial perturbation\footnote{See, for instance, Appendix A of Ref. \cite{Labrana15} for details about this point.}.

We follow the standard procedure and separate the scalar field and the metric into a homogeneous background plus small perturbations, i.e. $g_{\mu \nu} = g_{\mu \nu}^{(0)} + \delta g_{\mu \nu}$ and $\phi= \phi_0 + \dphi$.  We will now fix the gauge of the perturbations, and work in the so called longitudinal (or Newtonian) gauge. In this gauge, at first order in the scalar metric perturbations, and assuming no anisotropic stress components, the corresponding line element is
\begin{equation}\label{metricapert}
	ds^2 = a^2 (\eta) [  -(1+2\Psi) d\eta^2 + (1-2 \Psi) \delta_{i,j} dx^i dx^j].
\end{equation}

In these coordinates, the scale factor can be modeled by
\begin{equation}\label{aeta}
	a(\eta) = \frac{a_0}{1-e^{a_0 H_0 \eta}}
\end{equation}
We define $\mH \equiv a'/a$; the prime over variables denotes derivative with respect to conformal time $\eta$.

The metric degrees of freedom will remain classical because of the semiclassical gravity approach, these include the background and the perturbation $\Psi$.  In the matter fields sector, the background scalar field $\phi_0$ will be treated also in a classical fashion; however, the perturbed part $\dphi$ will be subjected to quantization.  Taking into account that the CSL theory modifies the Schr\"odinger  equation, it will be convenient to carry out the quantization in the Schr\"odinger picture. Therefore, we focus on finding the total Hamiltonian of the system.

We introduce the (re-scaled) field variable  $y = a \dphi$. Expanding the action of the system (i.e. a single scalar field minimally coupled to gravity) up to second order in the perturbations, one can find the action associated to $y$.  In this way, the second order action is $S^{(2)} = \int d^4 x \mathcal{L}^{(2)}$, where
  \begin{equation}\label{action2}
	\mathcal{L}^{(2)}_y  = \frac{1}{2} \bigg[  y'^2   - (\nabla y)^2  +  \frac{a''}{a} y^2\bigg].
\end{equation}
We define the canonical momentum $p(\x,\eta) \equiv  \partial \mathcal{L}^{(2)}_y/ \partial y = y'$, in this way, the Hamiltonian density is given by
\begin{equation}\label{hamilty}
	\mathcal{H}^{(2)}_y  = \frac{p^2}{2}     +  \frac{(\nabla y)^2}{2}   - \frac{ y^2}{2} \frac{a''}{a}.
\end{equation}
We now promote the fields $y$ and $p$ to quantum operators satisfying the following equal time commutator relation
\begin{equation}\label{commutator}
	[\hat y (\x,\eta), \hat p (\y,\eta)] = i \delta (\x-\y).
\end{equation}

Our next step is to decompose the field and the conjugated momentum in Fourier modes. This is  justified by the fact that we work with a linear theory and, hence, all the modes evolve independently.  In Fourier space, the total Hamiltonian corresponding to Eq. \eqref{hamilty} takes the form
\begin{equation}\label{hamiltF}
\hat H = \int_{\mathbb{R}^{3+}} d^3 k \quad  \left[  \hat p^*_\nk \hat  p_\nk  + \hat   y^*_\nk  \hat  y_\nk \left( k^2 - \frac{a''}{a}    \right)   \right].
\end{equation}
Furthermore, it will be  convenient to work with real variables. In this way,  we separate the canonical variables into their real and imaginary parts, i.e.
\begin{equation}
	\hat 	y_\nk \equiv \frac{1}{\sqrt{2}} (\hat y_\nk^\text{R} + i \hat  y_\nk^\text{I}  ), \qquad \hat  p_\nk \equiv \frac{1}{\sqrt{2}} (\hat  p_\nk^\text{R} + i \hat  p_\nk^\text{I}  ).
\end{equation}
The quantum commutator in Eq. \eqref{commutator}, implies
\begin{equation}
	[\hat y_\nk^s  , \hat p_\nq^{s'}  ] = i \delta(\nk - \nq) \delta_{s s'}
\end{equation}
where $s=$R,I and $\delta_{s s'}$ is Kronecker's delta. Using this separation the Hamiltonian  becomes $\hat H=  \int_{\mathbb{R}^{3+}} d^3 k (\hat H^\text{R}_\nk  + \hat H^\text{I}_\nk ) $, with the following definitions
\begin{equation}\label{HamiltRI}
	\hat H^{R,I}_\mathbf{k } \equiv \frac{(\hat p_\mathbf{k }^{R,I} )^2  }{2 }+ \frac{(\hat y_\mathbf{k }^{R,I} )^2  }{2 }  \left( k^2 - \frac{a''}{a}    \right) .
\end{equation}

In order to apply the CSL model into the EU scenario,  we will follow the approach first introduced in \cite{Pedro13} for the inflationary regime.  There,  it was found that with an appropriate selection of the field collapse operators and using the corresponding CSL evolution law, it is possible to attain a ``collapse'' in the relevant operators corresponding to the Fourier components of the field. Furthermore,  we will  assume linearity in the collapse generating operator,  therefore,  the reduction mechanism will act on each mode of the field independently, i.e. there will be no mode mixing because of the CSL process.

In view of the above, the evolution of the state vector  characterizing each mode of the  quantum field  as given by the CSL theory is:
\begin{eqnarray}\label{CSLevolution}
|\Phi_{\nk}^{\textrm{R,I}}, \eta \ket &=& \hat T \exp \bigg\{ - \int_{\tau}^{\eta}
d\eta'   \bigg[ i \hat{H}_{\nk}^{\textrm{R,I}} \nn
&+& \frac{1}{4 \lambda_k} (\mathcal{W}_{\nk}^{\RI}(\eta)  - 2 \lambda_k
\hat{y}_{\nk}^{\textrm{R,I}})^2 \bigg] \bigg\} |\Phi_{\nk}^{\textrm{R,I}},
\tau \ket
\end{eqnarray}
where $\hat T$ is the time-ordering operator, and $\tau$ denotes the conformal time at the beginning of the EU regime. Note that the stochastic field $\mathcal{W}_\nk= \mathcal{W}_{\nk}^{\text{R}} + i \mathcal{W}_{\nk}^{\text{I}}   $ depends on $\nk$ and the conformal time.  In other words,  it is reasonable to introduce a stochastic function for each independent degree of freedom given that we are applying the CSL collapse dynamics to each mode of the field.  Consequently, the stochastic field $\mathcal{W}_\nk (\eta)$ might be regarded as a Fourier transform on a stochastic spacetime field $\mathcal{W}(\x,\eta)$. The probability for the stochastic field  is given by the second main CSL equation: the Probability Rule, i.e.
\begin{equation}\label{cslprobabF}
	P(\mathcal{W}_{\nk}^{\RI}) d\mathcal{W}_{\nk}^{\RI} =   \bra \Phi_{\nk}^{\RI} , \eta | \Phi_{\nk}^{\RI}, \eta \ket \prod_{\eta'=\tau}^{\eta-d\eta} \frac{ d \mathcal{W}_{\nk}(\eta')^{\RI}}{\sqrt{2 \pi \lambda_k/d\eta}}.
\end{equation}

As can be seen from the CSL evolution equation \eqref{CSLevolution}, we have chosen the field variable $\hat{y}_\nk^\RI$ as the collapse generating operator.  Technically, this means that the CSL process will drive the initial state vector towards an eigenstate of $\hat{y}_\nk^\RI$. The motivation for this  choice is based on the fact that when one implements the SCG formulation into  cosmological perturbation theory, one obtains, at linear order in Fourier space, the following relation:
 \begin{equation}\label{masterpsi}
	\Psi_{\nk} + \mH^{-1}   \Psi_{\nk}' =    \sqrt{\frac{\mu}{2}}  \frac{\bra \hat y_{\nk} \ket   }{ a    M_P}
\end{equation}
where $\mu \equiv \phi_0^{'2} / (2 \mH^2 M_P^2)$.   We observe in the above expression that the quantum expectation value $\bra \hat y_{\nk} \ket$ acts as a source for the curvature perturbation.  This  might be interpreted  as  indicating that  the collapse  is tied  with some aspect of  the quantum matter that ``gravitates" (i.e. that would  characterize the  interaction between  gravitation and matter degrees of freedom).  Moreover, this view is, in principle, consistent with the proposals by  R. Penrose and L. Diosi suggesting that gravity might play a fundamental role in the so called collapse of the wave function \cite{Diosi87,Penrose96}.

We denote by $\Phi[y,\eta]$ the wave functional characterizing the quantum state of the field. In Fourier space, the wave functional can be factorized into mode components $\Phi[ y_{\nk},\eta] = \Pi_{\nk} \Phi_{\nk}^\textrm{R} [y_{\nk}^\textrm{R},\eta] \times \Phi_{\nk}^\textrm{I} [y_{\nk}^\textrm{I},\eta] $.

It is known that the ground state of the Hamiltonian \eqref{HamiltRI}, characterized by a wave functional  $ \Phi^\RI_0 [y^\RI,\eta]$, is a Gaussian. Also, the Hamiltonian \eqref{HamiltRI} and the CSL evolution equation \eqref{CSLevolution} are quadratic in both $\hat{y}_{\nk}^{\textrm{R,I}}$ and $\hat{p}_{\nk}^{\textrm{R,I}}$; consequently,  the wave functional at any time can be written in the form:
\begin{equation}\label{wf}
	\Phi^\RI [y_{\nk}^\RI,\eta] =
	\exp[-A_k(\eta) (y_{\nk}^\RI )^2 + B_k^\RI(\eta) y_{\nk}^\RI + C_k^\RI (\eta) ].
\end{equation}
The initial state of the field $ |\Phi_{\nk}^{\textrm{R,I}},  \tau \ket$ will be the chosen as the standard Bunch-Davies (BD) vacuum. The corresponding wave functional for the BD vacuum is characterized by the initial conditions
\begin{equation}\label{BDconditions}
	A_k(\tau) = \frac{k}{2}, \qquad B_k^\RI(\tau)=0, \qquad C_k^\RI (\tau)=0.
\end{equation}

\subsection{Power spectrum}

After having introduced the theoretical basis of our model, here we focus on deriving a  prediction for the primordial spectrum, which is the observational quantity of interest.   The standard expression of the primordial spectrum, associated to the curvature perturbation,  is normally expressed in the so-called comoving gauge, while our main equations were obtained in the longitudinal gauge. For a single scalar field, the relation between the curvature perturbation in the comoving gauge $\mR$ and in the longitudinal gauge $\Psi$ is given by \cite{Mukhanov81,mukhanov2005}
\begin{equation}\label{relacionRPsi}
		\mR = \Psi \left(    1 + \frac{1}{\mu}     \right) + \frac{ \mH^{-1} }{\mu}    \Psi'.
\end{equation}

Let us note that, in cosmic time $t$ coordinates, $\mu$ can be expressed as $\mu = \dot{\phi}_0^2/(2 M_P^2 H^2)$. Given that the system has an attractor point such that $H \to$ constant  and $\dot{\phi}_0^2$ tends to an infinitesimal small number, then it follows that $\mu \to 0$. Therefore, in Fourier space, Eq. \eqref{relacionRPsi} implies
 \begin{equation}\label{R}
	\mR_{\nk} \simeq   \frac{1}{\mu}     \left(    \Psi_{\nk}     +  \mH^{-1}  \Psi_{\nk}' \right)  =      \frac{ \bra \hat y_{\nk} \ket}{a M_P \sqrt{2 \mu}}
\end{equation}
where in the last equality we have used our main equation \eqref{masterpsi}.

The scalar power spectrum associated to $\mR_{\nk}$ is defined as
\begin{equation}\label{PSdef}
	\overline{\mR_{\nk}\mR^*_{\nq}} \equiv \frac{2 \pi^2}{k^3} \mP_{s} (k) \delta(\nk-\nq)
\end{equation}
where $\mP_{s} (k)$ is the dimensionless power spectrum. The bar appearing in \eqref{PSdef} denotes an ensemble average over possible realizations of the stochastic field $\mR_{\nk}$. In the CSL model, each realization will be associated to a particular realization of the stochastic process characterizing the collapse.   We can use approximation \eqref{R} to compute the  spectrum associated to $\mR_{\nk}$, i.e.
\begin{equation}\label{PSR}
	\overline{\mR_\nk \mR_\nq^*} =  \frac{\mH^2}{ a^2 \phi_0'^2}   \overline{\bra \hat y_\nk \ket \bra \hat y_\nq \ket^*}.
\end{equation}

From definition \eqref{PSdef} and Eq. \eqref{PSR},  we can identify an equivalent scalar power spectrum as:
\begin{equation}\label{masterPS}
	\mP_{s} (k) \delta(\nk-\nq) = \frac{k^3 \mH^2}{2 \pi^2 a^2 \phi_0'^2}   \overline{\bra \hat y_\nk \ket \bra \hat y_\nq \ket^*}.
\end{equation}
In terms of the Real and Imaginary parts of $\hat y_\nk$, the ensemble average in \eqref{masterPS} is
\begin{equation}\label{igualdad00}
	\overline{ \bra \hat y_{\nk} \ket \bra \hat y_{\nq} \ket^* } =  (\overline{\bra \hat y_{\nk}^\textrm{R} \ket^2} + \overline{\bra \hat y_{\nk}^\textrm{I} \ket^2}) \delta (\nk-\nq).
\end{equation}
Furthermore, $\overline{ \bra \hat y_{\nk}^\textrm{R} \ket^2 }= \overline{\bra \hat y_{\nk}^\textrm{I} \ket^2}$, thus we will omit the indexes R,I from now on.

Using the main equations of the CSL model, Eqs. \eqref{CSLevolution} and \eqref{cslprobabF} one obtains:
\begin{equation}\label{igualdad0}
	\overline{\bra \hat y_{\nk} \ket^2} = \overline{\bra \hat y_{\nk}^2 \ket} - \frac{1}{\textrm{Re} [A_k(\eta)]}.
\end{equation}
Substituting Eqs. \eqref{igualdad00} and \eqref{igualdad0} into Eq. \eqref{masterPS}, we find that the power spectrum can be expressed as:
\begin{equation}\label{masterPS2}
		\mP_{s} (k) = \frac{k^3 \mH^2}{2 \pi^2 a^2 \phi_0'^2}  \left(  \overline{\bra \hat y_{\nk}^2 \ket} - \frac{1}{4 \textrm{Re} [A_k(\eta)]}  \right).
\end{equation}
We observe that  the prediction for the power spectrum depends on the terms $ \overline{\bra \hat y_{\nk}^2\ket }$ and $(\textrm{Re} [A_k(\eta)])^{-1} $, which can be obtained from the CSL equations.

The quantity $(\textrm{Re} [A_k(\eta)])^{-1} $ represents the variance of the field variable, which in turn is related to  the width of the wave functional \eqref{wf}. The evolution equation for this quantity can be found by  taking  the time derivative of \eqref{CSLevolution}, and applying the resulting operator to the wave functional \eqref{wf}. Then, regrouping terms of order $y^2$, $y^1$ and $y^0$, the evolution equations corresponding to these terms become decoupled. In particular,  the evolution equation associated to $y^2$ only contains  $A_k(\eta)$, so  it decouples from the other variables $B_k(\eta)$ and $C_k(\eta)$.  The evolution equation is then
\begin{equation}\label{evolAk}
	A_k'  =   - 2i A_k^2 + \frac{i}{2}  \left(  k^2 - \frac{a''}{a}   \right) + \lambda_k.
\end{equation}
Performing the change of variable $A_k \equiv f'/(2i f)$, Eq. \eqref{evolAk} can be expressed as
\begin{equation}\label{evolfk}
f'' + \left(q^2 - \frac{a''}{a} \right) f = 0
\end{equation}
where:
\begin{equation}\label{defq}
	q^2 \equiv k^2 \left(1 - 2i \frac{ \lambda_k}{k^2}\right).
\end{equation}
The solution to Eq. \eqref{evolfk}, which satisfies the BD initial condition corresponding to $A_k(\tau) = k/2$, is
\begin{equation}\label{solf}
	f = \frac{e^{-i q \eta}}{\sqrt{2k} (1- e^{a_0 H_0 \eta})  } \: \:_2F_1  (q_-, q_+,b; e^{a_0 H_0 \eta})
\end{equation}
where
$\:_2F_1$ is the hypergeometric function, and
\begin{equation}\label{defamasamenos}
	q_{\pm} \equiv - 1 - \frac{i q}{a_0 H_0} \pm \left[ 1 - \left(  \frac{q}{a_0 H_0}  \right)^2 \right]^{1/2}
\end{equation}
\begin{equation}\label{defb}
	b \equiv 1 - \frac{2iq}{a_0 H_0}.
\end{equation}
With solution $f$, one can return to the original variable $A_k$ and obtain the sought quantity $(\textrm{Re} [A_k(\eta)])^{-1} $.  As a matter of fact, one has
\begin{equation}\label{ReAk}
	\textrm{Re} [A_k(\eta)] = \frac{W}{|f|^2 4 i}
\end{equation}
where $W \equiv f' f^* - f'^{*} f$ is the corresponding Wronskian.  Note that if $\lambda_k = 0$, then $W = i$ for all $\eta$, and $q = k$.

The other important term in the power spectrum is $ \overline{\bra \hat y_{\nk}^2\ket }$.  In order to find this quantity, it will be useful to define the following objects: $Q
\equiv \overline{\bra \hat y_{\nk}^2 \ket}$, $R\equiv \overline{\bra \hat p_{\nk}^2 \ket}$ and $S\equiv \overline{\bra \hat p_{\nk} \hat y_{\nk} + \hat y_{\nk}
	\hat p_{\nk} \ket}$. The evolution equations for $Q,R$ and $S$ obtained from the CSL equations are:
\begin{equation}\label{QRS}
Q' = S, \enskip R'=-w_k(\eta) S + \lambda_k, \enskip
S'=2R - 2 Q w_k (\eta)
\end{equation}
with $w_k (\eta) \equiv k^2 - a''/a$. Therefore, we have a linear system of coupled differential equations, whose general solution is a particular solution to the system plus a solution to the homogeneous equation (with $\lambda_k=0$).  In this way, the solution can be written as:
\begin{equation}\label{Q}
	Q (\eta) = C_1 y_1^2 + C_2 y_2^2 + C_3 y_1 y_2 + Q_p
\end{equation}
where the constants $C_1, C_2$ and $C_3$ are found by imposing the initial conditions corresponding to the Bunch-Davies vacuum state: $Q(\tau)=1/2k,
R(\tau) = k/2$, $S(\tau)=0$. The functions $y_1$ and $y_2$ are two linearly independent solutions of $y'' =- w_k  y $, and the function $Q_p$ is a particular solution of
\begin{equation}\label{eqQp}
	Q_p''' + 4w_k Q_p' + 2 w_k' Q_p = 2 \lambda_k.
\end{equation}

The exact solutions $y_1$ and $y_2$ are
\begin{equation}\label{y}
	y_1 (\eta) =  \frac{e^{-i k \eta}}{\sqrt{2k} (1- e^{a_0 H_0 \eta})  } \: \:_2F_1  (k_-, k_+,b; e^{a_0 H_0 \eta})
\end{equation}
and $y_2 =y_1^*$, also $k_\pm$ and $b$ are defined in the same manner as in \eqref{defamasamenos} \eqref{defb} but replacing $q \to k$.

On the other hand, the exact solution of Eq. \eqref{eqQp} is difficult to find, but we can find approximate solutions in the regimes of interest. In particular, we are interested in the static regime, which corresponds also to the regime where the BD initial conditions are imposed.  The other regime involved is the de Sitter phase, where the power spectrum  is evaluated for the purpose of comparing it with the standard prediction.  Therefore, in the static regime $w_k \simeq k^2$ while in the de Sitter phase $w_k \simeq k^2 - 2/\eta^2$.  It is remarkable that, in these two regimes,  $Q_p$ can be approximated by the same solution, i.e.
\begin{equation}\label{solQp}
	Q_p(\eta) \simeq \frac{\lambda_k \eta}{2 k^2}.
\end{equation}
Thus, the constants obtained from imposing the initial conditions are,
\begin{equation}\label{Cs}
	C_1 = \frac{-i \lambda_k}{4 k^2} e^{2 i k \tau}, \qquad C_2 = C_1^*, \qquad C_3 =  1 - \frac{\lambda_k \tau}{k}.
\end{equation}

Now that we have all the elements needed for obtaining the power spectrum \eqref{masterPS2}, it is straightforward to check first that if $\lambda_k = 0$ then $\mP_{s} =0$, because $Q(\eta) = (4 \textrm{Re} [A_k(\eta)])^{-1}$ exactly in that case. This result is also consistent with our view in which, if there is no collapse, then the metric perturbations are zero, i.e. there are no  inhomogeneities in the spacetime.

On the other hand, considering the modes in the super-Hubble limit ($-k\eta \to 0$), the power spectrum \eqref{masterPS2} can be approximated by
\begin{equation}\label{PSfinal}
		\mP_s (k) = A_s \chi^2 |F(\chi)|^2 C(k)
\end{equation}
where
\begin{equation}\label{defAs}
	A_s \equiv \frac{H_0^4}{4 \pi^2 \dot{\phi}_0^2}
\end{equation}
\begin{equation}\label{defchi}
	\chi \equiv \frac{k}{a_0 H_0}
\end{equation}
\begin{equation}\label{defF}
	F(\chi) \equiv  \frac{2 \Gamma(1 - 2 i \chi) }{ \Gamma(2 -  i \chi - \sqrt{1-\chi^2}) \Gamma(2 -  i \chi + \sqrt{1-\chi^2}) }
\end{equation}
\begin{equation}\label{Ckappgeneric}
	C(k) \simeq 1 + \frac{\lambda_k |\tau|}{k}+ \frac{\lambda_k}{k^2} \sin 2\delta
\end{equation}
\begin{equation}\label{defdelta}
	\delta \equiv \arctan \left( \frac{\text{Im} F}{\text{Re} F}   \right) - \chi a_0|\tau|.
\end{equation}

Note that in the definition of the amplitude $A_s$, we have used that the quantity $\mH^2/\phi_0'^2 = H^2/\dot{\phi}_0^2$ tends to a constant, given the existence of the aforementioned attractor point for the dynamical background variables in the limit $-k \eta \to 0$.

Thus, we have found the main prediction of this section, namely the primordial scalar power spectrum originated by the CSL mechanism within the EU model.

\section{Results}
\label{sectres}

In this section, we shall proceed to examine the observational effects of implementing the CSL model in the EU scenario. It will be useful to set as a reference model the one described in \cite{Labrana15}, we will refer to it as the {\em original model}.

Furthermore, as also argued in \cite{Labrana15}, we can generalize the power spectrum obtained in order to include the small scale dependence normally associated with the scalar spectral index $n_s$.  Thus, for the present section we will use the following expression for the primordial power spectrum (PPS):
\begin{equation}\label{PPSexacta}
	\mP_s (k) = A_s   \chi^2 |F(\chi)|^2 C(k)  \left( \frac{k}{k_P} \right)^{n_s-1}
\end{equation}
where  $k_P$ is a pivot scale, which is traditionally set as $k_P = 0.05$ Mpc$^{-1}$

The next step in our analysis is to introduce a parameterization of $\lambda_k$, in order to explore the possible observational features of our model. We propose the linear parameterization in $k$ given by
\begin{equation}
  \label{eq:param_lambdak}
  \lambda_k = \lambda_0 \, ( k + B)
\end{equation}
with $\lambda_0$ acting as a proportionality constant plus a parameter $B\geq 0$. In fact, if $B=0$, we recover a very similar expression for the PPS as the one obtained in the \textit{original model}. In this way, $B$ quantifies small deviations from the \textit{original model} reflecting the inclusion of the CSL model.  A very similar parameterization was also implemented in a recent work involving the CSL proposal during inflation \cite{Leon2020}. Also, we will fix the value of the proportionality constant as $\lambda_0 = 10^{-14}$ s$^{-1}$. This choice is motivated by the fact that such a value is within the range allowed by laboratory experiments testing non-relativistic versions of the CSL model \cite{sandro2017}, where $\lambda_0$ corresponds to the CSL parameter for these kind of models. In the units used in the present paper, the former choice is equivalent to  $\lambda_0= 1.029\, {\rm Mpc^{-1}}$. Given that $k$ has also units of ${\rm Mpc^{-1}}$, the constant $B$ has units of ${\rm Mpc^{-1}}$ too.  For ease of notation, from now on we will neglect the units of these quantities with the understanding that the corresponding units have been well established.

At this point, we would like to discuss the following issue. Numerical calculations set a restriction for implementing the exact formula \eqref{PPSexacta}. That is, as long as the value of $k$ increases, the Gamma functions become exponentially small beyond the capability of machine representation. This demands a cutoff  value $k_{max}$, which
in principle we set it to $0.015$, to be consistent with the value chosen in the \textit{original model} \cite{Labrana15}.
On the other hand, we can also approximate expression \eqref{PPSexacta}, obtaining
\begin{equation}\label{PSapp}
  \mP_{s}(k)  \simeq A_s \frac{\chi^2}{(1+\chi)^2 }  \frac{\lambda_k \tau}{k}\left( \frac{k}{k_P} \right)^{n_s-1}.
\end{equation}
The advantage of this approximation is that no numerical restrictions are imposed, hence it can be used in the numerical calculations up to the end of the observable window $k=1$.

\begin{figure}[h]
  \centering
  \includegraphics[angle=-90,width=0.45\textwidth]{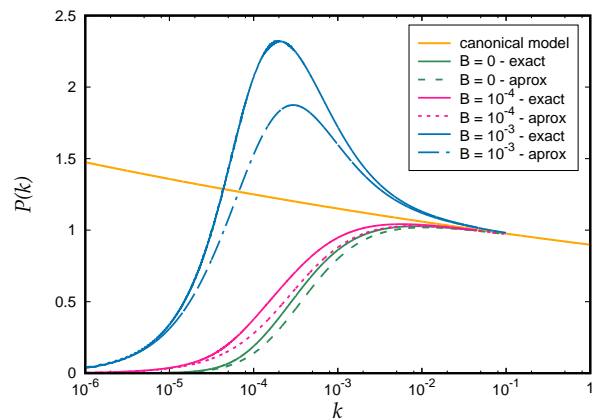}
  \caption{Primordial power spectrum for the EU model including the CSL
   proposal.  Here we set $a_0H_0= 2 \times 10^{-4}$ from \cite{Labrana15} and consider
   different values of the $B$ parameter  (units in $\rm Mpc^{-1}$).  Solid
    line represents the exact formula, while dashed line corresponds to the
    approximate expression. The {\em canonical model} is also plotted as a reference.
    The PPS which results from the CSL model
    tends to the canonical model for some scale between
    $0.01 < k <0.1$. The green lines correspond to $B=0$, which is essentially the {\em original model}.  }
  \label{fig:P_prim}
\end{figure}

Consequently,  we have two options for performing the analysis. The first one is to use the exact expression for the PPS \eqref{PPSexacta}  up to $k_{max}$,
and then perform an analytic continuation such that it approaches smoothly to the standard expression corresponding to the canonical model, i.e. to the PPS of the standard cosmological model. The second option we can consider is to use the approximate formula \eqref{PSapp} for the whole $k$ range. Both alternatives can be seen in Fig. \ref{fig:P_prim}, which shows the PPS for different values of the $B$ parameter comparing both the exact and approximate expressions. After an exhaustive exploration, we find that the values $10^{-3}$ and $10^{-4}$ are representative to show the behavior of the power spectrum curve. For each case, both calculations, exact (solid line) and approximate (dashed line), are shown.

Moreover, we will include in each Figure a plot called {\em canonical model}, which is used as a second reference (in addition to the {\em original model}). As its name suggests, the canonical model corresponds to the standard $\Lambda$CDM cosmological model, with the cosmological parameters determined by latest data from \textit{Planck} collaboration \cite{Planck18c}. In particular, we focus on the angular spectrum corresponding to the temperature and E--mode polarization auto-correlation and cross correlation functions. These data leads to the following set of parameters at the $68\%$ confidence level: $\Omega_bh^2 = 0.02236$, $\Omega_{c}h^2 = 0.1202$, $H_{\rm today} = 67.27 \,{\rm km \, s^{-1} \, Mpc^{-1}}$, $A_s = 2.101 \,\times\, 10^{-9}$, $n_s = 0.9649$, $\tau_d = 0.0544$ (called the optical depth parameter), and considering no running of the scalar spectral index.

The canonical model plot will be useful not only as a guide to quickly spot the novel features introduced by our model, but also to confirm the fact that, for given a value of $a_0H_0$, a certain scale $k_{max}$ can be found where the PPS meets naturally the standard $\Lambda$CDM model. This was the spirit in which the value of $k_ {max}=0.015$  was adopted, as mentioned above.

\begin{figure}[h]
  \centering
  \includegraphics[angle=-90,width=0.45\textwidth]{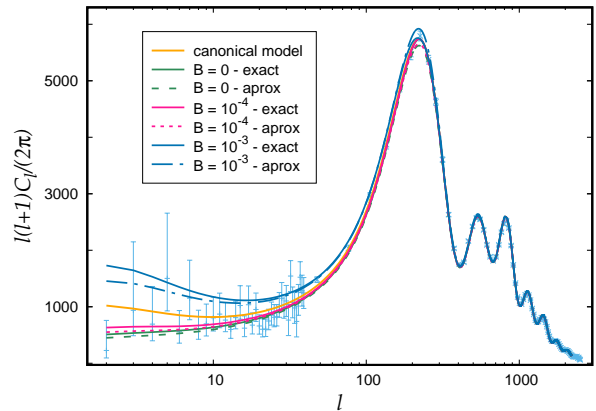}
  \caption{Angular power spectrum of the  temperature anisotropy
    auto-correlation function.  The canonical
    model is represented in orange. The solid and dashed lines correspond to the exact and approximated expressions for the PPS obtained from the CSL model, respectively. Here we considered $a_0H_0 = 2 \times 10^{-4}$, and three values for $B= 0,10^{-4},10^{-3}$. The plot depicts the
    behaviour of our model's predictions, progressively separating from
    the {\em original model} ($B=0$) and approaching to the canonical model either from above or below. The effect is
    mainly observed at the lowest multipoles, while larger multipoles are
    unaffected. \textit{Planck's} data and error bars are shown in order to
    provide some intuition of the effects the model might bring in when fitting
    observational data.}

  \label{fig:Cls-datos}
\end{figure}

The next step is to analyze the effects of varying  the model parameters  in the angular power spectrum. To accomplish this, we  perform the corresponding modifications in
the Code for Anisotropies in the Microwave Background (CAMB) software \cite{Lewis:1999bs}. Figure \ref{fig:Cls-datos} depicts the resulting angular spectrum for  different values of $B$.

In order to have a better visualization on the impact our model may have on observational signatures, we show  the best--fit $\Lambda$CDM prediction for the angular spectrum and the corresponding \textit{Planck} data\footnote{These data points correspond to R3.01 baseline Planck TT, TE, EE+lowE+lensing for multipoles between $2 < l <
  2508$. Considering lensing effects in our calculations did not show any
  difference in the result. Therefore, we consider that these data are adequate for making comparisons with our model.} together with their error bars (shown in blue).  The effect of varying $B$ is mainly seen in the low multipoles ($l < 50$).  If $B$ tends to zero,  the  CSL spectrum approaches to the one of the {\em original model},  and at the same time, it separates from the canonical model at the bottom of the graph. On the contrary, if $B$ is increased,  the CSL spectrum splits from the canonical model at the top part of the plot corresponding to the lowest multipoles.

The previous analysis indicates that our model  has the potential to exhibit different features in the low--$l$ range, approaching the $\Lambda$CDM angular spectrum from upwards or below, and having the \textit{original model} as the lower limit.  In other words, including the CSL mechanism in the EU model could result not only  in a suppression of the temperature anisotropies of the CMB at large angular scales, but also in an excess.

\subsection{Further parameter space exploration}\label{variables}

From the previous discussion, it is clear that the implementation of the CSL model into the EU scenario could be in good agreement with observational data. However, it is also important to check whether one can, in principle, truly distinguish the predictions between the canonical model and the one proposed in this work.  Recall that we have introduced a new parameter $B$, while the combination $a_0H_0$ comes from the {\em original model}.  Therefore, we are interested in testing the robustness of $B$  and $a_0H_0$ in the predicted angular spectrum. In other words, we will vary $B$ and $a_0 H_0$ enough to see how much our predicted spectrum deviates from the standard one.

\begin{figure}[h]
  \centering
 \includegraphics[angle=-90,width=0.45\textwidth]{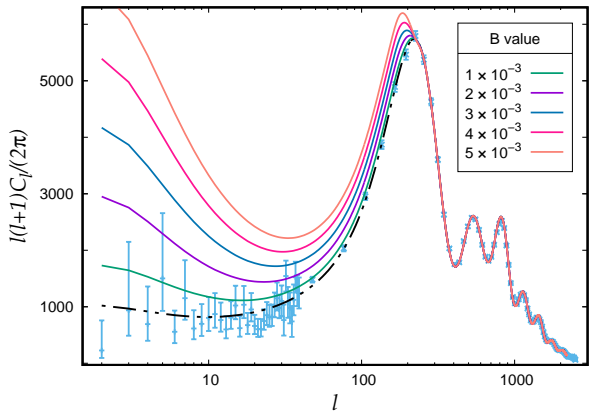}
 \caption{Exploring a wide range of $B$ values clearly shows that not any
   value would fit observational data (even when our model mostly affects the
   lowest multipoles, which are masked by the cosmic variance).  Different $B$ values of
   order $10^{-3}$ are tested, fixing $a_0H_0= 2 \times 10^{-4}$.
The solid line represents the prediction from  the CSL model (exact formula). The
   dashed black line depicts the prediction from the  {\em canonical
     model}. Increasing the value of  $B$ shows a progressive departure from
   the standard prediction and the data. This implies  an excess in the CMB
   angular power spectrum anisotropies.}
  \label{fig:Cls-Bvar_a0H0std}
\end{figure}

At first glance, one could argue that any well fitted value of $B$ would make the model consistent with observations, because any new features introduced in the spectrum could remain masked under the so called cosmic variance. However, exploring a wide range of the $B$ parameter space, shows that, even though highest multipoles are unaffected, the first acoustic peak is half missed for increasing values of $B$, see Fig. \ref{fig:Cls-Bvar_a0H0std}. This suggests that $B$ has an upper limit and its value can be constrained with observational data; therefore, the model has predictability.

\begin{figure}[h]
  \centering
  \includegraphics[angle=-90,width=0.45\textwidth]{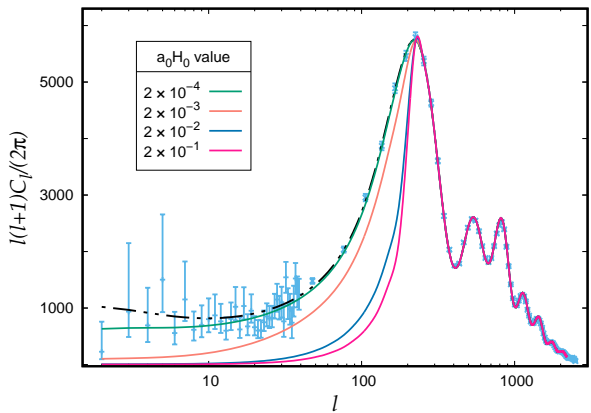}
  \caption{
     Four different orders of magnitude corresponding to $a_0H_0$ are explored, with fixed
    $B=10^{-4}$ (which has shown good compatibility with data). The
    highest values show a strong suppression in the angular power
    spectrum not compatible with observational data. This indicates that
    $a_0H_0$ plays the role of a free parameter and could be estimated
    with statistical analysis.  Dashed black line represents the {\em
      canonical model}. \textit{Planck's} data are shown in blue points along with their error bars.}
  \label{fig:Cls-a0H0var}
\end{figure}

Up to this point, we have worked with the same fixed value of  $a_0H_0$ as the one considered in the \textit{original model} \cite{Labrana15}. Henceforth,  it is interesting to explore the possibility to regard  the combination $a_0H_0$ as a free parameter. In Fig. \ref{fig:Cls-a0H0var}, we vary $a_0H_0$ along four orders of magnitude. Increasing  $a_0H_0$, implies a strong suppression in the angular power spectrum. This fact indicates that not any value of $a_0H_0$ would make the model consistent with the data.  Furthermore, it also indicates that the suppression observed in the {\em original model} could be in part explained by a particular combination of the cosmological parameters that has been chosen there.

\begin{figure}[h]
  \centering
  \includegraphics[angle=-90,width=0.45\textwidth]{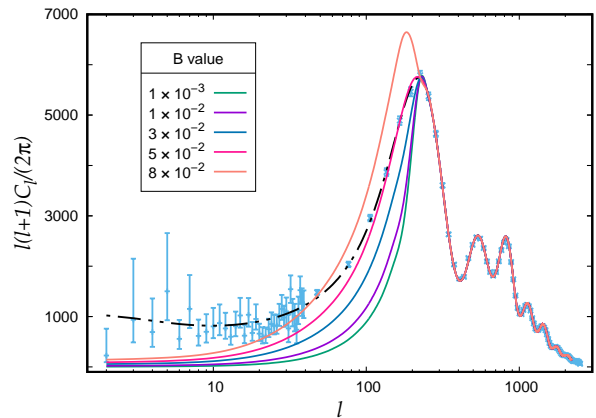}
  \caption{Different values for the $B$ parameter are explored
    considering $a_0H_0= 2 \times 10^{-2}$ which has proven no
    good fit with data. The plot shows that $B$ is a free
    parameter which can be constrained and is well
    restricted by the data, but does not compensate the suppression given by
    the choice of  $a_0H_0$. For higher $B$ values, not only the lowest
    multipoles are missed, but also the first acoustic peak fails to be
    matched. In fact, this constitutes a promising result for accurate parameter
    estimation with observational data under statistical analysis.  The
    {\em canonical model} is shown in point--dashed black line for
    reference.  }
  \label{fig:Cls-Bvar}
\end{figure}

The final issue we want to address is: given a particular value of $a_0H_0$ that is not compatible with observational data, could it be compensated in some way by varying the $B$ parameter introduced by the CSL model? Figure \ref{fig:Cls-Bvar} encompasses this question by plotting different values of $B$ for a fixed $a_0H_0= 2 \times 10^{-2}$ (which has been previously seen not compatible with observational data). Even though the variation of $B$ introduces new features and the model predictions might approach to \textit{Planck's} data at one end, other sectors of the data are highly missed. Consequently, we can safely state that the $B$ parameter cannot compensate the suppression induced by a non-favoured value of $a_0H_0$.

\section{Conclusions}
\label{conclusions}

In recent years, the \emph{emergent universe} (EU) has been studied from different perspectives as a viable cosmological model, which could not only avoid the horizon problem and the initial singularity of the standard approach, but could also account for some anomalies in the observations of the CMB at large angular scales.

Motivated by the recent results of \cite{Labrana15}, where it was shown that a phase of super-inflation prior to that of standard slow-roll inflation (originated in the context of the emergent universe presented in \cite{Ellis1, Ellis2}) could explain the power suppression observed in the low multipoles of the CMB, we decided to explore in this work the same emergent universe but within the framework of a particular objective collapse theory, known as Continuous Spontaneous Localization (CSL). The CSL theory has been studied with encouraging results in cosmological applications for several years, as a mechanism to break the original symmetries of the quantum vacuum state of the fields, generating the primordial cosmological perturbations and giving an explanation to the quantum-to-classical transition of such perturbations.

We obtained a prediction for the primordial power spectrum. The computed spectrum can be consistent with CMB observational data, if a particular parameterization for the collapse rate $\lambda_k$ is assumed. The parametrization we considered was also proposed and analyzed in previous works but in the framework of standard inflation. Such a parameterization introduces an extra free parameter $B$, in addition to the parameter $a_0H_0$ of the emergent model studied in \cite{Labrana15}. We also found that the predictions and results of such a work can be recovered, when $B=0$ is chosen.

From the analysis in Sec. \ref{sectres}, we have found that implementing the CSL collapse proposal to the emergent universe scenario introduces extra modifications at the lowest multipoles. Specifically, through the CSL-parameter $B$, the angular spectrum in the low multipoles sector, exhibits a suppression or an increment. This is a different feature from what is generically produced in  models with a super-inflation phase which only decrease the curve spectrum at large angular scales. On the other hand, in the model proposed in this work, whether there is an excess or suppression, will be determined by a combination of the parameters derived from the {\em original model} and the novel one introduced by the collapse rate parameterization. This fact enables our model to introduce some new features in the angular spectrum, particularly in the sector of interest where the so-called low--$l$ anomaly is located.

On the other hand, increasing the value of the combination $a_0H_0$ produces a suppression in the lowest multipoles. We have also seen that the variation of the $B$ parameter produces opposite (and similiar) effects to $a_0H_0$. Nevertheless, a suppression of the angular spectrum given by $a_0H_0$ cannot be compensated with an increase of $B$ .

One important aspect of our model is that the primordial power spectrum obtained smoothly approaches to the one from the standard $\Lambda\rm CDM$ model within the observational range of interest. This characteristic constitutes a natural modification of the (temperature) angular spectrum, affecting mainly the lowest multipoles without altering the highest ones (which are already constrained by observations to a high degree of accuracy).

Exploration of the free parameters shows that not any value will make the proposed model consistent with observational data. This suggests that there are good opportunities of effectively constraining the parameter space with the full machinery of statistical analysis based on Monte Carlo--Markov chains for cosmology. Such a scenario gives good predictability to the CSL collapse proposal in the emergent universe model. In fact, we expect that the extra free parameter $B$ should take a small value (between $10^{-3}$ and $10^{-4}$) but not centered at $B=0$, which clearly distinguishes our proposal from the one explored in \cite{Labrana15}.


\begin{acknowledgements}
G.R.B. is supported by CONICET (Argentina) and he acknowledges support from grant PIP 112-2017-0100220CO of CONICET (Argentina). G.L. and M.P.P. are supported by CONICET (Argentina), they also acknowledge support from the following project grants: Universidad Nacional de La Plata I+D G175 and PIP 112-2020-0100729CO of CONICET (Argentina). M.P.P. thanks Susana Landau for useful discussions regarding the impact of non--standard cosmological models on CMB angular power spectrum. The authors acknowledge the use of the supercluster MIZTLI of UNAM through project LANCAD-UNAM-DGTIC-132 and thank the people of DGTIC-UNAM for technical and computational support.

\end{acknowledgements}

\bibliography{bibliografia}

\begin{thebibliography}{269}
\expandafter\ifx\csname natexlab\endcsname\relax\def\natexlab#1{#1}\fi
\expandafter\ifx\csname bibnamefont\endcsname\relax
  \def\bibnamefont#1{#1}\fi
\expandafter\ifx\csname bibfnamefont\endcsname\relax
  \def\bibfnamefont#1{#1}\fi
\expandafter\ifx\csname citenamefont\endcsname\relax
  \def\citenamefont#1{#1}\fi
\expandafter\ifx\csname url\endcsname\relax
  \def\url#1{\texttt{#1}}\fi
\expandafter\ifx\csname urlprefix\endcsname\relax\def\urlprefix{URL }\fi
\providecommand{\bibinfo}[2]{#2}
\providecommand{\eprint}[2][]{\url{#2}}

\bibitem[{\citenamefont{Ade et~al.}(2016{\natexlab{a}})}]{Planck15}
\bibinfo{author}{\bibfnamefont{P.~A.~R.} \bibnamefont{Ade}}
  \bibnamefont{et~al.} (\bibinfo{collaboration}{Planck}),
  \bibinfo{journal}{Astron. Astrophys.} \textbf{\bibinfo{volume}{594}},
  \bibinfo{pages}{A13} (\bibinfo{year}{2016}{\natexlab{a}}),
  \eprint{1502.01589}.

\bibitem[{\citenamefont{Aghanim et~al.}(2020{\natexlab{a}})}]{Planck18a}
\bibinfo{author}{\bibfnamefont{N.}~\bibnamefont{Aghanim}} \bibnamefont{et~al.}
  (\bibinfo{collaboration}{Planck}), \bibinfo{journal}{Astron. Astrophys.}
  \textbf{\bibinfo{volume}{641}}, \bibinfo{pages}{A1}
  (\bibinfo{year}{2020}{\natexlab{a}}), \eprint{1807.06205}.

\bibitem[{\citenamefont{Aghanim et~al.}(2020{\natexlab{b}})}]{Planck18c}
\bibinfo{author}{\bibfnamefont{N.}~\bibnamefont{Aghanim}} \bibnamefont{et~al.}
  (\bibinfo{collaboration}{Planck}), \bibinfo{journal}{Astron. Astrophys.}
  \textbf{\bibinfo{volume}{641}}, \bibinfo{pages}{A6}
  (\bibinfo{year}{2020}{\natexlab{b}}), \eprint{1807.06209}.

\bibitem[{\citenamefont{Scolnic et~al.}(2018)}]{Scolnic18}
\bibinfo{author}{\bibfnamefont{D.~M.} \bibnamefont{Scolnic}}
  \bibnamefont{et~al.}, \bibinfo{journal}{Astrophys. J.}
  \textbf{\bibinfo{volume}{859}}, \bibinfo{pages}{101} (\bibinfo{year}{2018}),
  \eprint{1710.00845}.

\bibitem[{\citenamefont{Alam et~al.}(2021)}]{eBOSS}
\bibinfo{author}{\bibfnamefont{S.}~\bibnamefont{Alam}} \bibnamefont{et~al.}
  (\bibinfo{collaboration}{eBOSS}), \bibinfo{journal}{Phys. Rev. D}
  \textbf{\bibinfo{volume}{103}}, \bibinfo{pages}{083533}
  (\bibinfo{year}{2021}), \eprint{2007.08991}.

\bibitem[{\citenamefont{Starobinsky}(1979)}]{Starobinsky79}
\bibinfo{author}{\bibfnamefont{A.~A.} \bibnamefont{Starobinsky}},
  \bibinfo{journal}{JETP Lett.} \textbf{\bibinfo{volume}{30}},
  \bibinfo{pages}{682} (\bibinfo{year}{1979}).

\bibitem[{\citenamefont{Starobinsky}(1980)}]{Starobinsky80}
\bibinfo{author}{\bibfnamefont{A.~A.} \bibnamefont{Starobinsky}},
  \bibinfo{journal}{Phys. Lett.} \textbf{\bibinfo{volume}{B91}},
  \bibinfo{pages}{99} (\bibinfo{year}{1980}).

\bibitem[{\citenamefont{Guth}(1981)}]{Guth81}
\bibinfo{author}{\bibfnamefont{A.~H.} \bibnamefont{Guth}},
  \bibinfo{journal}{Phys. Rev.} \textbf{\bibinfo{volume}{D23}},
  \bibinfo{pages}{347} (\bibinfo{year}{1981}).

\bibitem[{\citenamefont{Mukhanov and Chibisov}(1981)}]{Mukhanov81}
\bibinfo{author}{\bibfnamefont{V.~F.} \bibnamefont{Mukhanov}} \bibnamefont{and}
  \bibinfo{author}{\bibfnamefont{G.~V.} \bibnamefont{Chibisov}},
  \bibinfo{journal}{JETP Lett.} \textbf{\bibinfo{volume}{33}},
  \bibinfo{pages}{532} (\bibinfo{year}{1981}), \bibinfo{note}{[Pisma Zh. Eksp.
  Teor. Fiz.33,549(1981)]}.

\bibitem[{\citenamefont{Linde}(1982)}]{Linde82}
\bibinfo{author}{\bibfnamefont{A.~D.} \bibnamefont{Linde}},
  \bibinfo{journal}{Phys. Lett.} \textbf{\bibinfo{volume}{B108}},
  \bibinfo{pages}{389} (\bibinfo{year}{1982}).

\bibitem[{\citenamefont{Linde}(1983)}]{Linde83}
\bibinfo{author}{\bibfnamefont{A.~D.} \bibnamefont{Linde}},
  \bibinfo{journal}{Phys. Lett. B} \textbf{\bibinfo{volume}{129}},
  \bibinfo{pages}{177} (\bibinfo{year}{1983}).

\bibitem[{\citenamefont{Albrecht and Steinhardt}(1982)}]{Albrecht82}
\bibinfo{author}{\bibfnamefont{A.}~\bibnamefont{Albrecht}} \bibnamefont{and}
  \bibinfo{author}{\bibfnamefont{P.~J.} \bibnamefont{Steinhardt}},
  \bibinfo{journal}{Phys. Rev. Lett.} \textbf{\bibinfo{volume}{48}},
  \bibinfo{pages}{1220} (\bibinfo{year}{1982}).

\bibitem[{\citenamefont{Bardeen et~al.}(1983)\citenamefont{Bardeen, Steinhardt,
  and Turner}}]{Bardeen83}
\bibinfo{author}{\bibfnamefont{J.~M.} \bibnamefont{Bardeen}},
  \bibinfo{author}{\bibfnamefont{P.~J.} \bibnamefont{Steinhardt}},
  \bibnamefont{and} \bibinfo{author}{\bibfnamefont{M.~S.}
  \bibnamefont{Turner}}, \bibinfo{journal}{Phys. Rev.}
  \textbf{\bibinfo{volume}{D28}}, \bibinfo{pages}{679} (\bibinfo{year}{1983}).

\bibitem[{\citenamefont{{Brandenberger}}(1984)}]{Brandenberger84}
\bibinfo{author}{\bibfnamefont{R.~H.} \bibnamefont{{Brandenberger}}},
  \bibinfo{journal}{Nuclear Physics B} \textbf{\bibinfo{volume}{245}},
  \bibinfo{pages}{328} (\bibinfo{year}{1984}).

\bibitem[{\citenamefont{Hawking}(1982)}]{Hawking82}
\bibinfo{author}{\bibfnamefont{S.~W.} \bibnamefont{Hawking}},
  \bibinfo{journal}{Phys. Lett.} \textbf{\bibinfo{volume}{115B}},
  \bibinfo{pages}{295} (\bibinfo{year}{1982}).

\bibitem[{\citenamefont{Akrami et~al.}(2020{\natexlab{a}})}]{Planck18b}
\bibinfo{author}{\bibfnamefont{Y.}~\bibnamefont{Akrami}} \bibnamefont{et~al.}
  (\bibinfo{collaboration}{Planck}), \bibinfo{journal}{Astron. Astrophys.}
  \textbf{\bibinfo{volume}{641}}, \bibinfo{pages}{A10}
  (\bibinfo{year}{2020}{\natexlab{a}}), \eprint{1807.06211}.

\bibitem[{\citenamefont{Hinshaw et~al.}(1996)\citenamefont{Hinshaw, Banday,
  Bennett, Gorski, Kogut, Lineweaver, Smoot, and Wright}}]{cobe}
\bibinfo{author}{\bibfnamefont{G.}~\bibnamefont{Hinshaw}},
  \bibinfo{author}{\bibfnamefont{A.~J.} \bibnamefont{Banday}},
  \bibinfo{author}{\bibfnamefont{C.~L.} \bibnamefont{Bennett}},
  \bibinfo{author}{\bibfnamefont{K.~M.} \bibnamefont{Gorski}},
  \bibinfo{author}{\bibfnamefont{A.}~\bibnamefont{Kogut}},
  \bibinfo{author}{\bibfnamefont{C.~H.} \bibnamefont{Lineweaver}},
  \bibinfo{author}{\bibfnamefont{G.~F.} \bibnamefont{Smoot}}, \bibnamefont{and}
  \bibinfo{author}{\bibfnamefont{E.~L.} \bibnamefont{Wright}},
  \bibinfo{journal}{Astrophys. J.} \textbf{\bibinfo{volume}{464}},
  \bibinfo{pages}{L25} (\bibinfo{year}{1996}), \eprint{astro-ph/9601061}.

\bibitem[{\citenamefont{Spergel et~al.}(2003)}]{wmap1}
\bibinfo{author}{\bibfnamefont{D.~N.} \bibnamefont{Spergel}}
  \bibnamefont{et~al.} (\bibinfo{collaboration}{WMAP}),
  \bibinfo{journal}{Astrophys. J. Suppl.} \textbf{\bibinfo{volume}{148}},
  \bibinfo{pages}{175} (\bibinfo{year}{2003}), \eprint{astro-ph/0302209}.

\bibitem[{\citenamefont{Spergel et~al.}(2007)}]{wmap2}
\bibinfo{author}{\bibfnamefont{D.~N.} \bibnamefont{Spergel}}
  \bibnamefont{et~al.} (\bibinfo{collaboration}{WMAP}),
  \bibinfo{journal}{Astrophys. J. Suppl.} \textbf{\bibinfo{volume}{170}},
  \bibinfo{pages}{377} (\bibinfo{year}{2007}), \eprint{astro-ph/0603449}.

\bibitem[{\citenamefont{{Larson} et~al.}(2011)\citenamefont{{Larson},
  {Dunkley}, {Hinshaw}, {Komatsu}, {Nolta}, {Bennett}, {Gold}, {Halpern},
  {Hill}, {Jarosik} et~al.}}]{wmap3}
\bibinfo{author}{\bibfnamefont{D.}~\bibnamefont{{Larson}}},
  \bibinfo{author}{\bibfnamefont{J.}~\bibnamefont{{Dunkley}}},
  \bibinfo{author}{\bibfnamefont{G.}~\bibnamefont{{Hinshaw}}},
  \bibinfo{author}{\bibfnamefont{E.}~\bibnamefont{{Komatsu}}},
  \bibinfo{author}{\bibfnamefont{M.~R.} \bibnamefont{{Nolta}}},
  \bibinfo{author}{\bibfnamefont{C.~L.} \bibnamefont{{Bennett}}},
  \bibinfo{author}{\bibfnamefont{B.}~\bibnamefont{{Gold}}},
  \bibinfo{author}{\bibfnamefont{M.}~\bibnamefont{{Halpern}}},
  \bibinfo{author}{\bibfnamefont{R.~S.} \bibnamefont{{Hill}}},
  \bibinfo{author}{\bibfnamefont{N.}~\bibnamefont{{Jarosik}}},
  \bibnamefont{et~al.}, \bibinfo{journal}{Astrophys. J. Suppl.}
  \textbf{\bibinfo{volume}{192}}, \bibinfo{eid}{16} (\bibinfo{year}{2011}),
  \eprint{1001.4635}.

\bibitem[{\citenamefont{{Komatsu} et~al.}(2011)\citenamefont{{Komatsu},
  {Smith}, {Dunkley}, {Bennett}, {Gold}, {Hinshaw}, {Jarosik}, {Larson},
  {Nolta}, {Page} et~al.}}]{wmap4}
\bibinfo{author}{\bibfnamefont{E.}~\bibnamefont{{Komatsu}}},
  \bibinfo{author}{\bibfnamefont{K.~M.} \bibnamefont{{Smith}}},
  \bibinfo{author}{\bibfnamefont{J.}~\bibnamefont{{Dunkley}}},
  \bibinfo{author}{\bibfnamefont{C.~L.} \bibnamefont{{Bennett}}},
  \bibinfo{author}{\bibfnamefont{B.}~\bibnamefont{{Gold}}},
  \bibinfo{author}{\bibfnamefont{G.}~\bibnamefont{{Hinshaw}}},
  \bibinfo{author}{\bibfnamefont{N.}~\bibnamefont{{Jarosik}}},
  \bibinfo{author}{\bibfnamefont{D.}~\bibnamefont{{Larson}}},
  \bibinfo{author}{\bibfnamefont{M.~R.} \bibnamefont{{Nolta}}},
  \bibinfo{author}{\bibfnamefont{L.}~\bibnamefont{{Page}}},
  \bibnamefont{et~al.}, \bibinfo{journal}{Astrophys. J. Suppl.}
  \textbf{\bibinfo{volume}{192}}, \bibinfo{eid}{18} (\bibinfo{year}{2011}),
  \eprint{1001.4538}.

\bibitem[{\citenamefont{Ade et~al.}(2014{\natexlab{a}})}]{planck13}
\bibinfo{author}{\bibfnamefont{P.~A.~R.} \bibnamefont{Ade}}
  \bibnamefont{et~al.} (\bibinfo{collaboration}{Planck}),
  \bibinfo{journal}{Astron. Astrophys.} \textbf{\bibinfo{volume}{571}},
  \bibinfo{pages}{A1} (\bibinfo{year}{2014}{\natexlab{a}}), \eprint{1303.5062}.

\bibitem[{\citenamefont{Ade et~al.}(2014{\natexlab{b}})}]{planck1}
\bibinfo{author}{\bibfnamefont{P.~A.~R.} \bibnamefont{Ade}}
  \bibnamefont{et~al.} (\bibinfo{collaboration}{Planck}),
  \bibinfo{journal}{Astron. Astrophys.} \textbf{\bibinfo{volume}{571}},
  \bibinfo{pages}{A15} (\bibinfo{year}{2014}{\natexlab{b}}),
  \eprint{1303.5075}.

\bibitem[{\citenamefont{Ade et~al.}(2016{\natexlab{b}})}]{planck2}
\bibinfo{author}{\bibfnamefont{P.~A.~R.} \bibnamefont{Ade}}
  \bibnamefont{et~al.} (\bibinfo{collaboration}{Planck}),
  \bibinfo{journal}{Astron. Astrophys.} \textbf{\bibinfo{volume}{594}},
  \bibinfo{pages}{A16} (\bibinfo{year}{2016}{\natexlab{b}}),
  \eprint{1506.07135}.

\bibitem[{\citenamefont{Akrami et~al.}(2020{\natexlab{b}})}]{planck3}
\bibinfo{author}{\bibfnamefont{Y.}~\bibnamefont{Akrami}} \bibnamefont{et~al.}
  (\bibinfo{collaboration}{Planck}), \bibinfo{journal}{Astron. Astrophys.}
  \textbf{\bibinfo{volume}{641}}, \bibinfo{pages}{A7}
  (\bibinfo{year}{2020}{\natexlab{b}}), \eprint{1906.02552}.

\bibitem[{\citenamefont{Copi et~al.}(2007)\citenamefont{Copi, Huterer, Schwarz,
  and Starkman}}]{copi1}
\bibinfo{author}{\bibfnamefont{C.}~\bibnamefont{Copi}},
  \bibinfo{author}{\bibfnamefont{D.}~\bibnamefont{Huterer}},
  \bibinfo{author}{\bibfnamefont{D.}~\bibnamefont{Schwarz}}, \bibnamefont{and}
  \bibinfo{author}{\bibfnamefont{G.}~\bibnamefont{Starkman}},
  \bibinfo{journal}{Phys. Rev.} \textbf{\bibinfo{volume}{D75}},
  \bibinfo{pages}{023507} (\bibinfo{year}{2007}), \eprint{astro-ph/0605135}.

\bibitem[{\citenamefont{Copi et~al.}(2009)\citenamefont{Copi, Huterer, Schwarz,
  and Starkman}}]{copi2}
\bibinfo{author}{\bibfnamefont{C.~J.} \bibnamefont{Copi}},
  \bibinfo{author}{\bibfnamefont{D.}~\bibnamefont{Huterer}},
  \bibinfo{author}{\bibfnamefont{D.~J.} \bibnamefont{Schwarz}},
  \bibnamefont{and} \bibinfo{author}{\bibfnamefont{G.~D.}
  \bibnamefont{Starkman}}, \bibinfo{journal}{Mon. Not. Roy. Astron. Soc.}
  \textbf{\bibinfo{volume}{399}}, \bibinfo{pages}{295} (\bibinfo{year}{2009}),
  \eprint{0808.3767}.

\bibitem[{\citenamefont{Copi et~al.}(2010)\citenamefont{Copi, Huterer, Schwarz,
  and Starkman}}]{copi3}
\bibinfo{author}{\bibfnamefont{C.~J.} \bibnamefont{Copi}},
  \bibinfo{author}{\bibfnamefont{D.}~\bibnamefont{Huterer}},
  \bibinfo{author}{\bibfnamefont{D.~J.} \bibnamefont{Schwarz}},
  \bibnamefont{and} \bibinfo{author}{\bibfnamefont{G.~D.}
  \bibnamefont{Starkman}}, \bibinfo{journal}{Adv. Astron.}
  \textbf{\bibinfo{volume}{2010}}, \bibinfo{pages}{847541}
  (\bibinfo{year}{2010}), \eprint{1004.5602}.

\bibitem[{\citenamefont{Copi et~al.}(2015)\citenamefont{Copi, Huterer, Schwarz,
  and Starkman}}]{copi4}
\bibinfo{author}{\bibfnamefont{C.~J.} \bibnamefont{Copi}},
  \bibinfo{author}{\bibfnamefont{D.}~\bibnamefont{Huterer}},
  \bibinfo{author}{\bibfnamefont{D.~J.} \bibnamefont{Schwarz}},
  \bibnamefont{and} \bibinfo{author}{\bibfnamefont{G.~D.}
  \bibnamefont{Starkman}}, \bibinfo{journal}{Mon. Not. Roy. Astron. Soc.}
  \textbf{\bibinfo{volume}{451}}, \bibinfo{pages}{2978} (\bibinfo{year}{2015}),
  \eprint{1310.3831}.

\bibitem[{\citenamefont{Sarkar et~al.}(2011)\citenamefont{Sarkar, Huterer,
  Copi, Starkman, and Schwarz}}]{sarkar}
\bibinfo{author}{\bibfnamefont{D.}~\bibnamefont{Sarkar}},
  \bibinfo{author}{\bibfnamefont{D.}~\bibnamefont{Huterer}},
  \bibinfo{author}{\bibfnamefont{C.~J.} \bibnamefont{Copi}},
  \bibinfo{author}{\bibfnamefont{G.~D.} \bibnamefont{Starkman}},
  \bibnamefont{and} \bibinfo{author}{\bibfnamefont{D.~J.}
  \bibnamefont{Schwarz}}, \bibinfo{journal}{Astropart. Phys.}
  \textbf{\bibinfo{volume}{34}}, \bibinfo{pages}{591} (\bibinfo{year}{2011}),
  \eprint{1004.3784}.

\bibitem[{\citenamefont{Efstathiou et~al.}(2010)\citenamefont{Efstathiou, Ma,
  and Hanson}}]{efstat}
\bibinfo{author}{\bibfnamefont{G.}~\bibnamefont{Efstathiou}},
  \bibinfo{author}{\bibfnamefont{Y.-Z.} \bibnamefont{Ma}}, \bibnamefont{and}
  \bibinfo{author}{\bibfnamefont{D.}~\bibnamefont{Hanson}},
  \bibinfo{journal}{Mon. Not. Roy. Astron. Soc.}
  \textbf{\bibinfo{volume}{407}}, \bibinfo{pages}{2530} (\bibinfo{year}{2010}),
  \eprint{0911.5399}.

\bibitem[{\citenamefont{Rakic et~al.}(2006)\citenamefont{Rakic, Rasanen, and
  Schwarz}}]{Rakic06}
\bibinfo{author}{\bibfnamefont{A.}~\bibnamefont{Rakic}},
  \bibinfo{author}{\bibfnamefont{S.}~\bibnamefont{Rasanen}}, \bibnamefont{and}
  \bibinfo{author}{\bibfnamefont{D.~J.} \bibnamefont{Schwarz}},
  \bibinfo{journal}{Mon. Not. Roy. Astron. Soc.}
  \textbf{\bibinfo{volume}{369}}, \bibinfo{pages}{L27} (\bibinfo{year}{2006}),
  \eprint{astro-ph/0601445}.

\bibitem[{\citenamefont{{Francis} and {Peacock}}(2010)}]{Francis2010}
\bibinfo{author}{\bibfnamefont{C.~L.} \bibnamefont{{Francis}}}
  \bibnamefont{and} \bibinfo{author}{\bibfnamefont{J.~A.}
  \bibnamefont{{Peacock}}}, \bibinfo{journal}{Mon. Not. Roy. Astron. Soc.}
  \textbf{\bibinfo{volume}{406}}, \bibinfo{pages}{14} (\bibinfo{year}{2010}),
  \eprint{0909.2495}.

\bibitem[{\citenamefont{Schwarz et~al.}(2016)\citenamefont{Schwarz, Copi,
  Huterer, and Starkman}}]{copi5}
\bibinfo{author}{\bibfnamefont{D.~J.} \bibnamefont{Schwarz}},
  \bibinfo{author}{\bibfnamefont{C.~J.} \bibnamefont{Copi}},
  \bibinfo{author}{\bibfnamefont{D.}~\bibnamefont{Huterer}}, \bibnamefont{and}
  \bibinfo{author}{\bibfnamefont{G.~D.} \bibnamefont{Starkman}},
  \bibinfo{journal}{Class. Quant. Grav.} \textbf{\bibinfo{volume}{33}},
  \bibinfo{pages}{184001} (\bibinfo{year}{2016}), \eprint{1510.07929}.

\bibitem[{\citenamefont{Perivolaropoulos and Skara}(2021)}]{Perivola21}
\bibinfo{author}{\bibfnamefont{L.}~\bibnamefont{Perivolaropoulos}}
  \bibnamefont{and} \bibinfo{author}{\bibfnamefont{F.}~\bibnamefont{Skara}}
  (\bibinfo{year}{2021}), \eprint{2105.05208}.

\bibitem[{\citenamefont{{Sokolov}}(1993)}]{Sokolov93}
\bibinfo{author}{\bibfnamefont{I.~Y.} \bibnamefont{{Sokolov}}},
  \bibinfo{journal}{Soviet Journal of Experimental and Theoretical Physics
  Letters} \textbf{\bibinfo{volume}{57}}, \bibinfo{pages}{617}
  (\bibinfo{year}{1993}).

\bibitem[{\citenamefont{Stevens et~al.}(1993)\citenamefont{Stevens, Scott, and
  Silk}}]{Stevens93}
\bibinfo{author}{\bibfnamefont{D.}~\bibnamefont{Stevens}},
  \bibinfo{author}{\bibfnamefont{D.}~\bibnamefont{Scott}}, \bibnamefont{and}
  \bibinfo{author}{\bibfnamefont{J.}~\bibnamefont{Silk}},
  \bibinfo{journal}{Phys. Rev. Lett.} \textbf{\bibinfo{volume}{71}},
  \bibinfo{pages}{20} (\bibinfo{year}{1993}).

\bibitem[{\citenamefont{Rocha et~al.}(2004)\citenamefont{Rocha, Cayón, Bowen,
  Canavezes, Silk, Banday, and Górski}}]{Rocha04}
\bibinfo{author}{\bibfnamefont{G.}~\bibnamefont{Rocha}},
  \bibinfo{author}{\bibfnamefont{L.}~\bibnamefont{Cayón}},
  \bibinfo{author}{\bibfnamefont{R.}~\bibnamefont{Bowen}},
  \bibinfo{author}{\bibfnamefont{A.}~\bibnamefont{Canavezes}},
  \bibinfo{author}{\bibfnamefont{J.}~\bibnamefont{Silk}},
  \bibinfo{author}{\bibfnamefont{A.~J.} \bibnamefont{Banday}},
  \bibnamefont{and} \bibinfo{author}{\bibfnamefont{K.~M.}
  \bibnamefont{Górski}}, \bibinfo{journal}{Monthly Notices of the Royal
  Astronomical Society} \textbf{\bibinfo{volume}{351}}, \bibinfo{pages}{769}
  (\bibinfo{year}{2004}).

\bibitem[{\citenamefont{Campanelli et~al.}(2007)\citenamefont{Campanelli, Cea,
  and Tedesco}}]{Campa07}
\bibinfo{author}{\bibfnamefont{L.}~\bibnamefont{Campanelli}},
  \bibinfo{author}{\bibfnamefont{P.}~\bibnamefont{Cea}}, \bibnamefont{and}
  \bibinfo{author}{\bibfnamefont{L.}~\bibnamefont{Tedesco}},
  \bibinfo{journal}{Phys. Rev. D} \textbf{\bibinfo{volume}{76}},
  \bibinfo{pages}{063007} (\bibinfo{year}{2007}).

\bibitem[{\citenamefont{Bastero-Gil et~al.}(2003)\citenamefont{Bastero-Gil,
  Freese, and Mersini-Houghton}}]{Bastero03}
\bibinfo{author}{\bibfnamefont{M.}~\bibnamefont{Bastero-Gil}},
  \bibinfo{author}{\bibfnamefont{K.}~\bibnamefont{Freese}}, \bibnamefont{and}
  \bibinfo{author}{\bibfnamefont{L.}~\bibnamefont{Mersini-Houghton}},
  \bibinfo{journal}{Phys. Rev. D} \textbf{\bibinfo{volume}{68}},
  \bibinfo{pages}{123514} (\bibinfo{year}{2003}), \eprint{hep-ph/0306289}.

\bibitem[{\citenamefont{Contaldi et~al.}(2003)\citenamefont{Contaldi, Peloso,
  Kofman, and Linde}}]{Contaldi03}
\bibinfo{author}{\bibfnamefont{C.~R.} \bibnamefont{Contaldi}},
  \bibinfo{author}{\bibfnamefont{M.}~\bibnamefont{Peloso}},
  \bibinfo{author}{\bibfnamefont{L.}~\bibnamefont{Kofman}}, \bibnamefont{and}
  \bibinfo{author}{\bibfnamefont{A.~D.} \bibnamefont{Linde}},
  \bibinfo{journal}{JCAP} \textbf{\bibinfo{volume}{07}}, \bibinfo{pages}{002}
  (\bibinfo{year}{2003}), \eprint{astro-ph/0303636}.

\bibitem[{\citenamefont{Bridle et~al.}(2003)\citenamefont{Bridle, Lewis,
  Weller, and Efstathiou}}]{Bridle03}
\bibinfo{author}{\bibfnamefont{S.~L.} \bibnamefont{Bridle}},
  \bibinfo{author}{\bibfnamefont{A.~M.} \bibnamefont{Lewis}},
  \bibinfo{author}{\bibfnamefont{J.}~\bibnamefont{Weller}}, \bibnamefont{and}
  \bibinfo{author}{\bibfnamefont{G.}~\bibnamefont{Efstathiou}},
  \bibinfo{journal}{Monthly Notices of the Royal Astronomical Society}
  \textbf{\bibinfo{volume}{342}}, \bibinfo{pages}{L72} (\bibinfo{year}{2003}).

\bibitem[{\citenamefont{Feng and Zhang}(2003)}]{Feng03}
\bibinfo{author}{\bibfnamefont{B.}~\bibnamefont{Feng}} \bibnamefont{and}
  \bibinfo{author}{\bibfnamefont{X.}~\bibnamefont{Zhang}},
  \bibinfo{journal}{Phys. Lett. B} \textbf{\bibinfo{volume}{570}},
  \bibinfo{pages}{145} (\bibinfo{year}{2003}), \eprint{astro-ph/0305020}.

\bibitem[{\citenamefont{Piao et~al.}(2004)\citenamefont{Piao, Feng, and
  Zhang}}]{Piao04}
\bibinfo{author}{\bibfnamefont{Y.-S.} \bibnamefont{Piao}},
  \bibinfo{author}{\bibfnamefont{B.}~\bibnamefont{Feng}}, \bibnamefont{and}
  \bibinfo{author}{\bibfnamefont{X.}~\bibnamefont{Zhang}},
  \bibinfo{journal}{Phys. Rev. D} \textbf{\bibinfo{volume}{69}},
  \bibinfo{pages}{103520} (\bibinfo{year}{2004}).

\bibitem[{\citenamefont{Boyanovsky et~al.}(2006)\citenamefont{Boyanovsky,
  de~Vega, and Sanchez}}]{Boya2006}
\bibinfo{author}{\bibfnamefont{D.}~\bibnamefont{Boyanovsky}},
  \bibinfo{author}{\bibfnamefont{H.~J.} \bibnamefont{de~Vega}},
  \bibnamefont{and} \bibinfo{author}{\bibfnamefont{N.~G.}
  \bibnamefont{Sanchez}}, \bibinfo{journal}{Phys. Rev. D}
  \textbf{\bibinfo{volume}{74}}, \bibinfo{pages}{123007}
  (\bibinfo{year}{2006}), \eprint{astro-ph/0607487}.

\bibitem[{\citenamefont{Cao et~al.}(2008)\citenamefont{Cao, de~Vega, and
  Sanchez}}]{Cao08}
\bibinfo{author}{\bibfnamefont{F.~J.} \bibnamefont{Cao}},
  \bibinfo{author}{\bibfnamefont{H.~J.} \bibnamefont{de~Vega}},
  \bibnamefont{and} \bibinfo{author}{\bibfnamefont{N.~G.}
  \bibnamefont{Sanchez}}, \bibinfo{journal}{Phys. Rev. D}
  \textbf{\bibinfo{volume}{78}}, \bibinfo{pages}{083508}
  (\bibinfo{year}{2008}), \eprint{0809.0623}.

\bibitem[{\citenamefont{Destri et~al.}(2008)\citenamefont{Destri, de~Vega, and
  Sanchez}}]{Destri08}
\bibinfo{author}{\bibfnamefont{C.}~\bibnamefont{Destri}},
  \bibinfo{author}{\bibfnamefont{H.~J.} \bibnamefont{de~Vega}},
  \bibnamefont{and} \bibinfo{author}{\bibfnamefont{N.~G.}
  \bibnamefont{Sanchez}}, \bibinfo{journal}{Phys. Rev. D}
  \textbf{\bibinfo{volume}{78}}, \bibinfo{pages}{023013}
  (\bibinfo{year}{2008}), \eprint{0804.2387}.

\bibitem[{\citenamefont{Destri et~al.}(2010)\citenamefont{Destri, de~Vega, and
  Sanchez}}]{Destri10}
\bibinfo{author}{\bibfnamefont{C.}~\bibnamefont{Destri}},
  \bibinfo{author}{\bibfnamefont{H.~J.} \bibnamefont{de~Vega}},
  \bibnamefont{and} \bibinfo{author}{\bibfnamefont{N.~G.}
  \bibnamefont{Sanchez}}, \bibinfo{journal}{Phys. Rev. D}
  \textbf{\bibinfo{volume}{81}}, \bibinfo{pages}{063520}
  (\bibinfo{year}{2010}), \eprint{0912.2994}.

\bibitem[{\citenamefont{Handley et~al.}(2014)\citenamefont{Handley, Brechet,
  Lasenby, and Hobson}}]{Handley14}
\bibinfo{author}{\bibfnamefont{W.~J.} \bibnamefont{Handley}},
  \bibinfo{author}{\bibfnamefont{S.~D.} \bibnamefont{Brechet}},
  \bibinfo{author}{\bibfnamefont{A.~N.} \bibnamefont{Lasenby}},
  \bibnamefont{and} \bibinfo{author}{\bibfnamefont{M.~P.}
  \bibnamefont{Hobson}}, \bibinfo{journal}{Phys. Rev. D}
  \textbf{\bibinfo{volume}{89}}, \bibinfo{pages}{063505}
  (\bibinfo{year}{2014}), \eprint{1401.2253}.

\bibitem[{\citenamefont{Gessey-Jones and Handley}(2021)}]{Handley21}
\bibinfo{author}{\bibfnamefont{T.}~\bibnamefont{Gessey-Jones}}
  \bibnamefont{and} \bibinfo{author}{\bibfnamefont{W.~J.}
  \bibnamefont{Handley}}, \bibinfo{journal}{Phys. Rev. D}
  \textbf{\bibinfo{volume}{104}}, \bibinfo{pages}{063532}
  (\bibinfo{year}{2021}), \eprint{2104.03016}.

\bibitem[{\citenamefont{Ramirez and Schwarz}(2012)}]{Ramirez11}
\bibinfo{author}{\bibfnamefont{E.}~\bibnamefont{Ramirez}} \bibnamefont{and}
  \bibinfo{author}{\bibfnamefont{D.~J.} \bibnamefont{Schwarz}},
  \bibinfo{journal}{Phys. Rev. D} \textbf{\bibinfo{volume}{85}},
  \bibinfo{pages}{103516} (\bibinfo{year}{2012}), \eprint{1111.7131}.

\bibitem[{\citenamefont{Ramirez}(2012)}]{Ramirez12}
\bibinfo{author}{\bibfnamefont{E.}~\bibnamefont{Ramirez}},
  \bibinfo{journal}{Phys. Rev. D} \textbf{\bibinfo{volume}{85}},
  \bibinfo{pages}{103517} (\bibinfo{year}{2012}).

\bibitem[{\citenamefont{Biswas and Mazumdar}(2014)}]{Biswas13}
\bibinfo{author}{\bibfnamefont{T.}~\bibnamefont{Biswas}} \bibnamefont{and}
  \bibinfo{author}{\bibfnamefont{A.}~\bibnamefont{Mazumdar}},
  \bibinfo{journal}{Class. Quant. Grav.} \textbf{\bibinfo{volume}{31}},
  \bibinfo{pages}{025019} (\bibinfo{year}{2014}), \eprint{1304.3648}.

\bibitem[{\citenamefont{Liu et~al.}(2013)\citenamefont{Liu, Guo, and
  Piao}}]{Liu2013}
\bibinfo{author}{\bibfnamefont{Z.-G.} \bibnamefont{Liu}},
  \bibinfo{author}{\bibfnamefont{Z.-K.} \bibnamefont{Guo}}, \bibnamefont{and}
  \bibinfo{author}{\bibfnamefont{Y.-S.} \bibnamefont{Piao}},
  \bibinfo{journal}{Phys. Rev. D} \textbf{\bibinfo{volume}{88}},
  \bibinfo{pages}{063539} (\bibinfo{year}{2013}).

\bibitem[{\citenamefont{Liu et~al.}(2014)\citenamefont{Liu, Guo, and
  Piao}}]{Liu2014}
\bibinfo{author}{\bibfnamefont{Z.-G.} \bibnamefont{Liu}},
  \bibinfo{author}{\bibfnamefont{Z.-K.} \bibnamefont{Guo}}, \bibnamefont{and}
  \bibinfo{author}{\bibfnamefont{Y.-S.} \bibnamefont{Piao}},
  \bibinfo{journal}{Eur. Phys. J. C} \textbf{\bibinfo{volume}{74}},
  \bibinfo{pages}{3006} (\bibinfo{year}{2014}), \eprint{1311.1599}.

\bibitem[{\citenamefont{Ashtekar et~al.}(2020)\citenamefont{Ashtekar, Gupt,
  Jeong, and Sreenath}}]{Ashtekar20}
\bibinfo{author}{\bibfnamefont{A.}~\bibnamefont{Ashtekar}},
  \bibinfo{author}{\bibfnamefont{B.}~\bibnamefont{Gupt}},
  \bibinfo{author}{\bibfnamefont{D.}~\bibnamefont{Jeong}}, \bibnamefont{and}
  \bibinfo{author}{\bibfnamefont{V.}~\bibnamefont{Sreenath}},
  \bibinfo{journal}{Phys. Rev. Lett.} \textbf{\bibinfo{volume}{125}},
  \bibinfo{pages}{051302} (\bibinfo{year}{2020}), \eprint{2001.11689}.

\bibitem[{\citenamefont{Agullo et~al.}(2021{\natexlab{a}})\citenamefont{Agullo,
  Kranas, and Sreenath}}]{Agullo20}
\bibinfo{author}{\bibfnamefont{I.}~\bibnamefont{Agullo}},
  \bibinfo{author}{\bibfnamefont{D.}~\bibnamefont{Kranas}}, \bibnamefont{and}
  \bibinfo{author}{\bibfnamefont{V.}~\bibnamefont{Sreenath}},
  \bibinfo{journal}{Gen. Rel. Grav.} \textbf{\bibinfo{volume}{53}},
  \bibinfo{pages}{17} (\bibinfo{year}{2021}{\natexlab{a}}),
  \eprint{2005.01796}.

\bibitem[{\citenamefont{Agullo et~al.}(2021{\natexlab{b}})\citenamefont{Agullo,
  Kranas, and Sreenath}}]{Agullo21}
\bibinfo{author}{\bibfnamefont{I.}~\bibnamefont{Agullo}},
  \bibinfo{author}{\bibfnamefont{D.}~\bibnamefont{Kranas}}, \bibnamefont{and}
  \bibinfo{author}{\bibfnamefont{V.}~\bibnamefont{Sreenath}},
  \bibinfo{journal}{Class. Quant. Grav.} \textbf{\bibinfo{volume}{38}},
  \bibinfo{pages}{065010} (\bibinfo{year}{2021}{\natexlab{b}}),
  \eprint{2006.09605}.

\bibitem[{\citenamefont{{Tryon}}(1973)}]{Tryon73}
\bibinfo{author}{\bibfnamefont{E.~P.} \bibnamefont{{Tryon}}},
  \bibinfo{journal}{\nat} \textbf{\bibinfo{volume}{246}}, \bibinfo{pages}{396}
  (\bibinfo{year}{1973}).

\bibitem[{\citenamefont{{Vilenkin}}(1982)}]{Vilenkin82}
\bibinfo{author}{\bibfnamefont{A.}~\bibnamefont{{Vilenkin}}},
  \bibinfo{journal}{Physics Letters B} \textbf{\bibinfo{volume}{117}},
  \bibinfo{pages}{25} (\bibinfo{year}{1982}).

\bibitem[{\citenamefont{{Ellis}}(1987)}]{Ellis1987}
\bibinfo{author}{\bibfnamefont{G.~F.~R.} \bibnamefont{{Ellis}}},
  \bibinfo{journal}{\apj} \textbf{\bibinfo{volume}{314}}, \bibinfo{pages}{1}
  (\bibinfo{year}{1987}).

\bibitem[{\citenamefont{{Ellis}}(1988)}]{Ellis1988}
\bibinfo{author}{\bibfnamefont{G.~F.~R.} \bibnamefont{{Ellis}}},
  \bibinfo{journal}{Classical and Quantum Gravity}
  \textbf{\bibinfo{volume}{5}}, \bibinfo{pages}{891} (\bibinfo{year}{1988}).

\bibitem[{\citenamefont{{Ellis} et~al.}(1991)\citenamefont{{Ellis}, {Lyth}, and
  {Miji{\'c}}}}]{Ellis91}
\bibinfo{author}{\bibfnamefont{G.~F.~R.} \bibnamefont{{Ellis}}},
  \bibinfo{author}{\bibfnamefont{D.~H.} \bibnamefont{{Lyth}}},
  \bibnamefont{and} \bibinfo{author}{\bibfnamefont{M.~B.}
  \bibnamefont{{Miji{\'c}}}}, \bibinfo{journal}{Physics Letters B}
  \textbf{\bibinfo{volume}{271}}, \bibinfo{pages}{52} (\bibinfo{year}{1991}).

\bibitem[{\citenamefont{Linde}(1995)}]{Linde95a}
\bibinfo{author}{\bibfnamefont{A.~D.} \bibnamefont{Linde}},
  \bibinfo{journal}{Phys. Lett. B} \textbf{\bibinfo{volume}{351}},
  \bibinfo{pages}{99} (\bibinfo{year}{1995}), \eprint{hep-th/9503097}.

\bibitem[{\citenamefont{Linde and Mezhlumian}(1995)}]{Linde95b}
\bibinfo{author}{\bibfnamefont{A.~D.} \bibnamefont{Linde}} \bibnamefont{and}
  \bibinfo{author}{\bibfnamefont{A.}~\bibnamefont{Mezhlumian}},
  \bibinfo{journal}{Phys. Rev. D} \textbf{\bibinfo{volume}{52}},
  \bibinfo{pages}{6789} (\bibinfo{year}{1995}), \eprint{astro-ph/9506017}.

\bibitem[{\citenamefont{White and Scott}(1996)}]{White96}
\bibinfo{author}{\bibfnamefont{M.~J.} \bibnamefont{White}} \bibnamefont{and}
  \bibinfo{author}{\bibfnamefont{D.}~\bibnamefont{Scott}},
  \bibinfo{journal}{Astrophys. J.} \textbf{\bibinfo{volume}{459}},
  \bibinfo{pages}{415} (\bibinfo{year}{1996}), \eprint{astro-ph/9508157}.

\bibitem[{\citenamefont{Linde}(1998)}]{Linde98}
\bibinfo{author}{\bibfnamefont{A.~D.} \bibnamefont{Linde}},
  \bibinfo{journal}{Phys. Rev. D} \textbf{\bibinfo{volume}{58}},
  \bibinfo{pages}{083514} (\bibinfo{year}{1998}), \eprint{gr-qc/9802038}.

\bibitem[{\citenamefont{Linde}(1999)}]{Linde99}
\bibinfo{author}{\bibfnamefont{A.~D.} \bibnamefont{Linde}},
  \bibinfo{journal}{Phys. Rev. D} \textbf{\bibinfo{volume}{59}},
  \bibinfo{pages}{023503} (\bibinfo{year}{1999}), \eprint{hep-ph/9807493}.

\bibitem[{\citenamefont{Ellis et~al.}(2002)\citenamefont{Ellis, Stoeger,
  McEwan, and Dunsby}}]{Ellis2002}
\bibinfo{author}{\bibfnamefont{G.~F.~R.} \bibnamefont{Ellis}},
  \bibinfo{author}{\bibfnamefont{W.~R.} \bibnamefont{Stoeger},
  \bibfnamefont{S.~J.}},
  \bibinfo{author}{\bibfnamefont{P.}~\bibnamefont{McEwan}}, \bibnamefont{and}
  \bibinfo{author}{\bibfnamefont{P.}~\bibnamefont{Dunsby}},
  \bibinfo{journal}{Gen. Rel. Grav.} \textbf{\bibinfo{volume}{34}},
  \bibinfo{pages}{1445} (\bibinfo{year}{2002}), \eprint{gr-qc/0109023}.

\bibitem[{\citenamefont{Linde}(2003)}]{Linde2003}
\bibinfo{author}{\bibfnamefont{A.~D.} \bibnamefont{Linde}},
  \bibinfo{journal}{JCAP} \textbf{\bibinfo{volume}{05}}, \bibinfo{pages}{002}
  (\bibinfo{year}{2003}), \eprint{astro-ph/0303245}.

\bibitem[{\citenamefont{Lasenby and Doran}(2005)}]{Lasenby03}
\bibinfo{author}{\bibfnamefont{A.}~\bibnamefont{Lasenby}} \bibnamefont{and}
  \bibinfo{author}{\bibfnamefont{C.}~\bibnamefont{Doran}},
  \bibinfo{journal}{Phys. Rev. D} \textbf{\bibinfo{volume}{71}},
  \bibinfo{pages}{063502} (\bibinfo{year}{2005}), \eprint{astro-ph/0307311}.

\bibitem[{\citenamefont{Uzan et~al.}(2003)\citenamefont{Uzan, Kirchner, and
  Ellis}}]{Uzan2003}
\bibinfo{author}{\bibfnamefont{J.-P.} \bibnamefont{Uzan}},
  \bibinfo{author}{\bibfnamefont{U.}~\bibnamefont{Kirchner}}, \bibnamefont{and}
  \bibinfo{author}{\bibfnamefont{G.~F.~R.} \bibnamefont{Ellis}},
  \bibinfo{journal}{Mon. Not. Roy. Astron. Soc.}
  \textbf{\bibinfo{volume}{344}}, \bibinfo{pages}{L65} (\bibinfo{year}{2003}),
  \eprint{astro-ph/0302597}.

\bibitem[{\citenamefont{{Zhang} and {Sun}}(2004)}]{Zhang2004}
\bibinfo{author}{\bibfnamefont{D.-H.} \bibnamefont{{Zhang}}} \bibnamefont{and}
  \bibinfo{author}{\bibfnamefont{C.-Y.} \bibnamefont{{Sun}}},
  \bibinfo{journal}{Chinese Physics Letters} \textbf{\bibinfo{volume}{21}},
  \bibinfo{pages}{1865} (\bibinfo{year}{2004}).

\bibitem[{\citenamefont{Ratra}(2017)}]{Ratra17a}
\bibinfo{author}{\bibfnamefont{B.}~\bibnamefont{Ratra}},
  \bibinfo{journal}{Phys. Rev. D} \textbf{\bibinfo{volume}{96}},
  \bibinfo{pages}{103534} (\bibinfo{year}{2017}), \eprint{1707.03439}.

\bibitem[{\citenamefont{Ooba et~al.}(2018{\natexlab{a}})\citenamefont{Ooba,
  Ratra, and Sugiyama}}]{Ratra18a}
\bibinfo{author}{\bibfnamefont{J.}~\bibnamefont{Ooba}},
  \bibinfo{author}{\bibfnamefont{B.}~\bibnamefont{Ratra}}, \bibnamefont{and}
  \bibinfo{author}{\bibfnamefont{N.}~\bibnamefont{Sugiyama}},
  \bibinfo{journal}{Astrophys. J.} \textbf{\bibinfo{volume}{864}},
  \bibinfo{pages}{80} (\bibinfo{year}{2018}{\natexlab{a}}),
  \eprint{1707.03452}.

\bibitem[{\citenamefont{R\"as\"anen et~al.}(2015)\citenamefont{R\"as\"anen,
  Bolejko, and Finoguenov}}]{Rasanen2014}
\bibinfo{author}{\bibfnamefont{S.}~\bibnamefont{R\"as\"anen}},
  \bibinfo{author}{\bibfnamefont{K.}~\bibnamefont{Bolejko}}, \bibnamefont{and}
  \bibinfo{author}{\bibfnamefont{A.}~\bibnamefont{Finoguenov}},
  \bibinfo{journal}{Phys. Rev. Lett.} \textbf{\bibinfo{volume}{115}},
  \bibinfo{pages}{101301} (\bibinfo{year}{2015}), \eprint{1412.4976}.

\bibitem[{\citenamefont{Park and Ratra}(2019{\natexlab{a}})}]{Ratra17b}
\bibinfo{author}{\bibfnamefont{C.-G.} \bibnamefont{Park}} \bibnamefont{and}
  \bibinfo{author}{\bibfnamefont{B.}~\bibnamefont{Ratra}},
  \bibinfo{journal}{Astrophys. J.} \textbf{\bibinfo{volume}{882}},
  \bibinfo{pages}{158} (\bibinfo{year}{2019}{\natexlab{a}}),
  \eprint{1801.00213}.

\bibitem[{\citenamefont{Ryan et~al.}(2018)\citenamefont{Ryan, Doshi, and
  Ratra}}]{Ratra18b}
\bibinfo{author}{\bibfnamefont{J.}~\bibnamefont{Ryan}},
  \bibinfo{author}{\bibfnamefont{S.}~\bibnamefont{Doshi}}, \bibnamefont{and}
  \bibinfo{author}{\bibfnamefont{B.}~\bibnamefont{Ratra}},
  \bibinfo{journal}{Mon. Not. Roy. Astron. Soc.}
  \textbf{\bibinfo{volume}{480}}, \bibinfo{pages}{759} (\bibinfo{year}{2018}),
  \eprint{1805.06408}.

\bibitem[{\citenamefont{Yu et~al.}(2018)\citenamefont{Yu, Ratra, and
  Wang}}]{Ratra18c}
\bibinfo{author}{\bibfnamefont{H.}~\bibnamefont{Yu}},
  \bibinfo{author}{\bibfnamefont{B.}~\bibnamefont{Ratra}}, \bibnamefont{and}
  \bibinfo{author}{\bibfnamefont{F.-Y.} \bibnamefont{Wang}},
  \bibinfo{journal}{Astrophys. J.} \textbf{\bibinfo{volume}{856}},
  \bibinfo{pages}{3} (\bibinfo{year}{2018}), \eprint{1711.03437}.

\bibitem[{\citenamefont{Ooba et~al.}(2018{\natexlab{b}})\citenamefont{Ooba,
  Ratra, and Sugiyama}}]{Ratra18d}
\bibinfo{author}{\bibfnamefont{J.}~\bibnamefont{Ooba}},
  \bibinfo{author}{\bibfnamefont{B.}~\bibnamefont{Ratra}}, \bibnamefont{and}
  \bibinfo{author}{\bibfnamefont{N.}~\bibnamefont{Sugiyama}},
  \bibinfo{journal}{Astrophys. J.} \textbf{\bibinfo{volume}{864}},
  \bibinfo{pages}{80} (\bibinfo{year}{2018}{\natexlab{b}}),
  \eprint{1707.03452}.

\bibitem[{\citenamefont{Park and Ratra}(2019{\natexlab{b}})}]{Ratra19a}
\bibinfo{author}{\bibfnamefont{C.-G.} \bibnamefont{Park}} \bibnamefont{and}
  \bibinfo{author}{\bibfnamefont{B.}~\bibnamefont{Ratra}},
  \bibinfo{journal}{Astrophys. Space Sci.} \textbf{\bibinfo{volume}{364}},
  \bibinfo{pages}{134} (\bibinfo{year}{2019}{\natexlab{b}}),
  \eprint{1809.03598}.

\bibitem[{\citenamefont{Ryan et~al.}(2019)\citenamefont{Ryan, Chen, and
  Ratra}}]{Ratra19b}
\bibinfo{author}{\bibfnamefont{J.}~\bibnamefont{Ryan}},
  \bibinfo{author}{\bibfnamefont{Y.}~\bibnamefont{Chen}}, \bibnamefont{and}
  \bibinfo{author}{\bibfnamefont{B.}~\bibnamefont{Ratra}},
  \bibinfo{journal}{Mon. Not. Roy. Astron. Soc.}
  \textbf{\bibinfo{volume}{488}}, \bibinfo{pages}{3844} (\bibinfo{year}{2019}),
  \eprint{1902.03196}.

\bibitem[{\citenamefont{Handley}(2021)}]{Handley2019}
\bibinfo{author}{\bibfnamefont{W.}~\bibnamefont{Handley}},
  \bibinfo{journal}{Phys. Rev. D} \textbf{\bibinfo{volume}{103}},
  \bibinfo{pages}{L041301} (\bibinfo{year}{2021}), \eprint{1908.09139}.

\bibitem[{\citenamefont{Efstathiou and Gratton}(2019)}]{Efstathiou19}
\bibinfo{author}{\bibfnamefont{G.}~\bibnamefont{Efstathiou}} \bibnamefont{and}
  \bibinfo{author}{\bibfnamefont{S.}~\bibnamefont{Gratton}}
  (\bibinfo{year}{2019}), \eprint{1910.00483}.

\bibitem[{\citenamefont{Riess et~al.}(2019)\citenamefont{Riess, Casertano,
  Yuan, Macri, and Scolnic}}]{Riess2019}
\bibinfo{author}{\bibfnamefont{A.~G.} \bibnamefont{Riess}},
  \bibinfo{author}{\bibfnamefont{S.}~\bibnamefont{Casertano}},
  \bibinfo{author}{\bibfnamefont{W.}~\bibnamefont{Yuan}},
  \bibinfo{author}{\bibfnamefont{L.~M.} \bibnamefont{Macri}}, \bibnamefont{and}
  \bibinfo{author}{\bibfnamefont{D.}~\bibnamefont{Scolnic}},
  \bibinfo{journal}{Astrophys. J.} \textbf{\bibinfo{volume}{876}},
  \bibinfo{pages}{85} (\bibinfo{year}{2019}), \eprint{1903.07603}.

\bibitem[{\citenamefont{Di~Valentino et~al.}(2019)\citenamefont{Di~Valentino,
  Melchiorri, and Silk}}]{Silk2020}
\bibinfo{author}{\bibfnamefont{E.}~\bibnamefont{Di~Valentino}},
  \bibinfo{author}{\bibfnamefont{A.}~\bibnamefont{Melchiorri}},
  \bibnamefont{and} \bibinfo{author}{\bibfnamefont{J.}~\bibnamefont{Silk}},
  \bibinfo{journal}{Nature Astron.} \textbf{\bibinfo{volume}{4}},
  \bibinfo{pages}{196} (\bibinfo{year}{2019}), \eprint{1911.02087}.

\bibitem[{\citenamefont{Di~Valentino
  et~al.}(2021{\natexlab{a}})\citenamefont{Di~Valentino, Melchiorri, and
  Silk}}]{Valentino2020}
\bibinfo{author}{\bibfnamefont{E.}~\bibnamefont{Di~Valentino}},
  \bibinfo{author}{\bibfnamefont{A.}~\bibnamefont{Melchiorri}},
  \bibnamefont{and} \bibinfo{author}{\bibfnamefont{J.}~\bibnamefont{Silk}},
  \bibinfo{journal}{Astrophys. J. Lett.} \textbf{\bibinfo{volume}{908}},
  \bibinfo{pages}{L9} (\bibinfo{year}{2021}{\natexlab{a}}),
  \eprint{2003.04935}.

\bibitem[{\citenamefont{Efstathiou and Gratton}(2020)}]{Efstathiou20}
\bibinfo{author}{\bibfnamefont{G.}~\bibnamefont{Efstathiou}} \bibnamefont{and}
  \bibinfo{author}{\bibfnamefont{S.}~\bibnamefont{Gratton}},
  \bibinfo{journal}{Mon. Not. Roy. Astron. Soc.}
  \textbf{\bibinfo{volume}{496}}, \bibinfo{pages}{L91} (\bibinfo{year}{2020}),
  \eprint{2002.06892}.

\bibitem[{\citenamefont{Di~Valentino
  et~al.}(2021{\natexlab{b}})}]{DiValentino2020a}
\bibinfo{author}{\bibfnamefont{E.}~\bibnamefont{Di~Valentino}}
  \bibnamefont{et~al.}, \bibinfo{journal}{Astropart. Phys.}
  \textbf{\bibinfo{volume}{131}}, \bibinfo{pages}{102607}
  (\bibinfo{year}{2021}{\natexlab{b}}), \eprint{2008.11286}.

\bibitem[{\citenamefont{Benisty and Staicova}(2021)}]{Benisty20}
\bibinfo{author}{\bibfnamefont{D.}~\bibnamefont{Benisty}} \bibnamefont{and}
  \bibinfo{author}{\bibfnamefont{D.}~\bibnamefont{Staicova}},
  \bibinfo{journal}{Astron. Astrophys.} \textbf{\bibinfo{volume}{647}},
  \bibinfo{pages}{A38} (\bibinfo{year}{2021}), \eprint{2009.10701}.

\bibitem[{\citenamefont{Vagnozzi
  et~al.}(2021{\natexlab{a}})\citenamefont{Vagnozzi, Di~Valentino, Gariazzo,
  Melchiorri, Mena, and Silk}}]{Vagnozzi20A}
\bibinfo{author}{\bibfnamefont{S.}~\bibnamefont{Vagnozzi}},
  \bibinfo{author}{\bibfnamefont{E.}~\bibnamefont{Di~Valentino}},
  \bibinfo{author}{\bibfnamefont{S.}~\bibnamefont{Gariazzo}},
  \bibinfo{author}{\bibfnamefont{A.}~\bibnamefont{Melchiorri}},
  \bibinfo{author}{\bibfnamefont{O.}~\bibnamefont{Mena}}, \bibnamefont{and}
  \bibinfo{author}{\bibfnamefont{J.}~\bibnamefont{Silk}},
  \bibinfo{journal}{Phys. Dark Univ.} \textbf{\bibinfo{volume}{33}},
  \bibinfo{pages}{100851} (\bibinfo{year}{2021}{\natexlab{a}}),
  \eprint{2010.02230}.

\bibitem[{\citenamefont{Vagnozzi
  et~al.}(2021{\natexlab{b}})\citenamefont{Vagnozzi, Loeb, and
  Moresco}}]{Vagnozzi20B}
\bibinfo{author}{\bibfnamefont{S.}~\bibnamefont{Vagnozzi}},
  \bibinfo{author}{\bibfnamefont{A.}~\bibnamefont{Loeb}}, \bibnamefont{and}
  \bibinfo{author}{\bibfnamefont{M.}~\bibnamefont{Moresco}},
  \bibinfo{journal}{Astrophys. J.} \textbf{\bibinfo{volume}{908}},
  \bibinfo{pages}{84} (\bibinfo{year}{2021}{\natexlab{b}}),
  \eprint{2011.11645}.

\bibitem[{\citenamefont{Dhawan et~al.}(2021)\citenamefont{Dhawan, Alsing, and
  Vagnozzi}}]{Dhawan21}
\bibinfo{author}{\bibfnamefont{S.}~\bibnamefont{Dhawan}},
  \bibinfo{author}{\bibfnamefont{J.}~\bibnamefont{Alsing}}, \bibnamefont{and}
  \bibinfo{author}{\bibfnamefont{S.}~\bibnamefont{Vagnozzi}},
  \bibinfo{journal}{Mon. Not. Roy. Astron. Soc.}
  \textbf{\bibinfo{volume}{506}}, \bibinfo{pages}{L1} (\bibinfo{year}{2021}),
  \eprint{2104.02485}.

\bibitem[{\citenamefont{Verde et~al.}(2019)\citenamefont{Verde, Treu, and
  Riess}}]{Verde2019}
\bibinfo{author}{\bibfnamefont{L.}~\bibnamefont{Verde}},
  \bibinfo{author}{\bibfnamefont{T.}~\bibnamefont{Treu}}, \bibnamefont{and}
  \bibinfo{author}{\bibfnamefont{A.~G.} \bibnamefont{Riess}},
  \bibinfo{journal}{Nature Astron.} \textbf{\bibinfo{volume}{3}},
  \bibinfo{pages}{891} (\bibinfo{year}{2019}), \eprint{1907.10625}.

\bibitem[{\citenamefont{Di~Valentino
  et~al.}(2021{\natexlab{c}})\citenamefont{Di~Valentino, Mena, Pan, Visinelli,
  Yang, Melchiorri, Mota, Riess, and Silk}}]{DiValentino2021}
\bibinfo{author}{\bibfnamefont{E.}~\bibnamefont{Di~Valentino}},
  \bibinfo{author}{\bibfnamefont{O.}~\bibnamefont{Mena}},
  \bibinfo{author}{\bibfnamefont{S.}~\bibnamefont{Pan}},
  \bibinfo{author}{\bibfnamefont{L.}~\bibnamefont{Visinelli}},
  \bibinfo{author}{\bibfnamefont{W.}~\bibnamefont{Yang}},
  \bibinfo{author}{\bibfnamefont{A.}~\bibnamefont{Melchiorri}},
  \bibinfo{author}{\bibfnamefont{D.~F.} \bibnamefont{Mota}},
  \bibinfo{author}{\bibfnamefont{A.~G.} \bibnamefont{Riess}}, \bibnamefont{and}
  \bibinfo{author}{\bibfnamefont{J.}~\bibnamefont{Silk}},
  \bibinfo{journal}{Class. Quant. Grav.} \textbf{\bibinfo{volume}{38}},
  \bibinfo{pages}{153001} (\bibinfo{year}{2021}{\natexlab{c}}),
  \eprint{2103.01183}.

\bibitem[{\citenamefont{Bolejko}(2018)}]{Bolejko2017}
\bibinfo{author}{\bibfnamefont{K.}~\bibnamefont{Bolejko}},
  \bibinfo{journal}{Phys. Rev. D} \textbf{\bibinfo{volume}{97}},
  \bibinfo{pages}{103529} (\bibinfo{year}{2018}), \eprint{1712.02967}.

\bibitem[{\citenamefont{Collett et~al.}(2019)\citenamefont{Collett, Montanari,
  and Rasanen}}]{Collett2019}
\bibinfo{author}{\bibfnamefont{T.}~\bibnamefont{Collett}},
  \bibinfo{author}{\bibfnamefont{F.}~\bibnamefont{Montanari}},
  \bibnamefont{and} \bibinfo{author}{\bibfnamefont{S.}~\bibnamefont{Rasanen}},
  \bibinfo{journal}{Phys. Rev. Lett.} \textbf{\bibinfo{volume}{123}},
  \bibinfo{pages}{231101} (\bibinfo{year}{2019}), \eprint{1905.09781}.

\bibitem[{\citenamefont{Clarkson et~al.}(2007)\citenamefont{Clarkson, Cortes,
  and Bassett}}]{Clarkson07}
\bibinfo{author}{\bibfnamefont{C.}~\bibnamefont{Clarkson}},
  \bibinfo{author}{\bibfnamefont{M.}~\bibnamefont{Cortes}}, \bibnamefont{and}
  \bibinfo{author}{\bibfnamefont{B.~A.} \bibnamefont{Bassett}},
  \bibinfo{journal}{JCAP} \textbf{\bibinfo{volume}{08}}, \bibinfo{pages}{011}
  (\bibinfo{year}{2007}), \eprint{astro-ph/0702670}.

\bibitem[{\citenamefont{Hlozek et~al.}(2008)\citenamefont{Hlozek, Cortes,
  Clarkson, and Bassett}}]{Clarkson08}
\bibinfo{author}{\bibfnamefont{R.}~\bibnamefont{Hlozek}},
  \bibinfo{author}{\bibfnamefont{M.}~\bibnamefont{Cortes}},
  \bibinfo{author}{\bibfnamefont{C.}~\bibnamefont{Clarkson}}, \bibnamefont{and}
  \bibinfo{author}{\bibfnamefont{B.}~\bibnamefont{Bassett}},
  \bibinfo{journal}{Gen. Rel. Grav.} \textbf{\bibinfo{volume}{40}},
  \bibinfo{pages}{285} (\bibinfo{year}{2008}), \eprint{0801.3847}.

\bibitem[{\citenamefont{Barenboim et~al.}(2010)\citenamefont{Barenboim,
  Mart\'\i{}nez, Mena, and Verde}}]{Verde10}
\bibinfo{author}{\bibfnamefont{G.}~\bibnamefont{Barenboim}},
  \bibinfo{author}{\bibfnamefont{E.~F.} \bibnamefont{Mart\'\i{}nez}},
  \bibinfo{author}{\bibfnamefont{O.}~\bibnamefont{Mena}}, \bibnamefont{and}
  \bibinfo{author}{\bibfnamefont{L.}~\bibnamefont{Verde}},
  \bibinfo{journal}{JCAP} \textbf{\bibinfo{volume}{03}}, \bibinfo{pages}{008}
  (\bibinfo{year}{2010}), \eprint{0910.0252}.

\bibitem[{\citenamefont{Di~Valentino
  et~al.}(2021{\natexlab{d}})\citenamefont{Di~Valentino, Melchiorri, Mena, Pan,
  and Yang}}]{DiValentino2020b}
\bibinfo{author}{\bibfnamefont{E.}~\bibnamefont{Di~Valentino}},
  \bibinfo{author}{\bibfnamefont{A.}~\bibnamefont{Melchiorri}},
  \bibinfo{author}{\bibfnamefont{O.}~\bibnamefont{Mena}},
  \bibinfo{author}{\bibfnamefont{S.}~\bibnamefont{Pan}}, \bibnamefont{and}
  \bibinfo{author}{\bibfnamefont{W.}~\bibnamefont{Yang}},
  \bibinfo{journal}{Mon. Not. Roy. Astron. Soc.}
  \textbf{\bibinfo{volume}{502}}, \bibinfo{pages}{L23}
  (\bibinfo{year}{2021}{\natexlab{d}}), \eprint{2011.00283}.

\bibitem[{\citenamefont{Penrose}(1965)}]{Penrose64}
\bibinfo{author}{\bibfnamefont{R.}~\bibnamefont{Penrose}},
  \bibinfo{journal}{Phys. Rev. Lett.} \textbf{\bibinfo{volume}{14}},
  \bibinfo{pages}{57} (\bibinfo{year}{1965}).

\bibitem[{\citenamefont{{Hawking}}(1965)}]{Hawking1965}
\bibinfo{author}{\bibfnamefont{S.~W.} \bibnamefont{{Hawking}}},
  \bibinfo{journal}{\prl} \textbf{\bibinfo{volume}{15}}, \bibinfo{pages}{689}
  (\bibinfo{year}{1965}).

\bibitem[{\citenamefont{{Hawking} and {Penrose}}(1970)}]{HawkingPenrose1970}
\bibinfo{author}{\bibfnamefont{S.~W.} \bibnamefont{{Hawking}}}
  \bibnamefont{and}
  \bibinfo{author}{\bibfnamefont{R.}~\bibnamefont{{Penrose}}},
  \bibinfo{journal}{Proceedings of the Royal Society of London Series A}
  \textbf{\bibinfo{volume}{314}}, \bibinfo{pages}{529} (\bibinfo{year}{1970}).

\bibitem[{\citenamefont{Hawking and Ellis}(2011)}]{HawkingEllis1973}
\bibinfo{author}{\bibfnamefont{S.~W.} \bibnamefont{Hawking}} \bibnamefont{and}
  \bibinfo{author}{\bibfnamefont{G.~F.~R.} \bibnamefont{Ellis}},
  \emph{\bibinfo{title}{{The Large Scale Structure of Space-Time}}}, Cambridge
  Monographs on Mathematical Physics (\bibinfo{publisher}{Cambridge University
  Press}, \bibinfo{year}{2011}).

\bibitem[{\citenamefont{Wald}(1984)}]{Wald84}
\bibinfo{author}{\bibfnamefont{R.~M.} \bibnamefont{Wald}},
  \emph{\bibinfo{title}{{General Relativity}}} (\bibinfo{publisher}{Chicago
  Univ. Pr.}, \bibinfo{address}{Chicago, USA}, \bibinfo{year}{1984}).

\bibitem[{\citenamefont{Borde and Vilenkin}(1994{\natexlab{a}})}]{Borde1993}
\bibinfo{author}{\bibfnamefont{A.}~\bibnamefont{Borde}} \bibnamefont{and}
  \bibinfo{author}{\bibfnamefont{A.}~\bibnamefont{Vilenkin}},
  \bibinfo{journal}{Phys. Rev. Lett.} \textbf{\bibinfo{volume}{72}},
  \bibinfo{pages}{3305} (\bibinfo{year}{1994}{\natexlab{a}}),
  \eprint{gr-qc/9312022}.

\bibitem[{\citenamefont{Borde and Vilenkin}(1994{\natexlab{b}})}]{Borde1994a}
\bibinfo{author}{\bibfnamefont{A.}~\bibnamefont{Borde}} \bibnamefont{and}
  \bibinfo{author}{\bibfnamefont{A.}~\bibnamefont{Vilenkin}}, in
  \emph{\bibinfo{booktitle}{{8th Nishinomiya-Yukawa Memorial Symposium:
  Relativistic Cosmology}}} (\bibinfo{year}{1994}{\natexlab{b}}),
  \eprint{gr-qc/9403004}.

\bibitem[{\citenamefont{Borde}(1994)}]{Borde1994b}
\bibinfo{author}{\bibfnamefont{A.}~\bibnamefont{Borde}},
  \bibinfo{journal}{Phys. Rev. D} \textbf{\bibinfo{volume}{50}},
  \bibinfo{pages}{3692} (\bibinfo{year}{1994}), \eprint{gr-qc/9403049}.

\bibitem[{\citenamefont{Borde and Vilenkin}(1996)}]{Borde1996}
\bibinfo{author}{\bibfnamefont{A.}~\bibnamefont{Borde}} \bibnamefont{and}
  \bibinfo{author}{\bibfnamefont{A.}~\bibnamefont{Vilenkin}},
  \bibinfo{journal}{Int. J. Mod. Phys. D} \textbf{\bibinfo{volume}{5}},
  \bibinfo{pages}{813} (\bibinfo{year}{1996}), \eprint{gr-qc/9612036}.

\bibitem[{\citenamefont{Borde and Vilenkin}(1997)}]{Borde1997}
\bibinfo{author}{\bibfnamefont{A.}~\bibnamefont{Borde}} \bibnamefont{and}
  \bibinfo{author}{\bibfnamefont{A.}~\bibnamefont{Vilenkin}},
  \bibinfo{journal}{Phys. Rev. D} \textbf{\bibinfo{volume}{56}},
  \bibinfo{pages}{717} (\bibinfo{year}{1997}), \eprint{gr-qc/9702019}.

\bibitem[{\citenamefont{Guth}(2001)}]{Guth99}
\bibinfo{author}{\bibfnamefont{A.~H.} \bibnamefont{Guth}},
  \bibinfo{journal}{Annals N. Y. Acad. Sci.} \textbf{\bibinfo{volume}{950}},
  \bibinfo{pages}{66} (\bibinfo{year}{2001}), \eprint{astro-ph/0101507}.

\bibitem[{\citenamefont{Borde et~al.}(2003)\citenamefont{Borde, Guth, and
  Vilenkin}}]{Borde2003}
\bibinfo{author}{\bibfnamefont{A.}~\bibnamefont{Borde}},
  \bibinfo{author}{\bibfnamefont{A.~H.} \bibnamefont{Guth}}, \bibnamefont{and}
  \bibinfo{author}{\bibfnamefont{A.}~\bibnamefont{Vilenkin}},
  \bibinfo{journal}{Phys. Rev. Lett.} \textbf{\bibinfo{volume}{90}},
  \bibinfo{pages}{151301} (\bibinfo{year}{2003}), \eprint{gr-qc/0110012}.

\bibitem[{\citenamefont{Molina-Paris and Visser}(1999)}]{Molina99}
\bibinfo{author}{\bibfnamefont{C.}~\bibnamefont{Molina-Paris}}
  \bibnamefont{and} \bibinfo{author}{\bibfnamefont{M.}~\bibnamefont{Visser}},
  \bibinfo{journal}{Phys. Lett. B} \textbf{\bibinfo{volume}{455}},
  \bibinfo{pages}{90} (\bibinfo{year}{1999}), \eprint{gr-qc/9810023}.

\bibitem[{\citenamefont{Kanekar et~al.}(2001)\citenamefont{Kanekar, Sahni, and
  Shtanov}}]{Kanekar2001}
\bibinfo{author}{\bibfnamefont{N.}~\bibnamefont{Kanekar}},
  \bibinfo{author}{\bibfnamefont{V.}~\bibnamefont{Sahni}}, \bibnamefont{and}
  \bibinfo{author}{\bibfnamefont{Y.}~\bibnamefont{Shtanov}},
  \bibinfo{journal}{Phys. Rev. D} \textbf{\bibinfo{volume}{63}},
  \bibinfo{pages}{083520} (\bibinfo{year}{2001}), \eprint{astro-ph/0101448}.

\bibitem[{\citenamefont{Khoury et~al.}(2001)\citenamefont{Khoury, Ovrut,
  Steinhardt, and Turok}}]{Khoury01}
\bibinfo{author}{\bibfnamefont{J.}~\bibnamefont{Khoury}},
  \bibinfo{author}{\bibfnamefont{B.~A.} \bibnamefont{Ovrut}},
  \bibinfo{author}{\bibfnamefont{P.~J.} \bibnamefont{Steinhardt}},
  \bibnamefont{and} \bibinfo{author}{\bibfnamefont{N.}~\bibnamefont{Turok}},
  \bibinfo{journal}{Phys. Rev. D} \textbf{\bibinfo{volume}{64}},
  \bibinfo{pages}{123522} (\bibinfo{year}{2001}), \eprint{hep-th/0103239}.

\bibitem[{\citenamefont{Steinhardt and Turok}(2002)}]{Steinh02}
\bibinfo{author}{\bibfnamefont{P.~J.} \bibnamefont{Steinhardt}}
  \bibnamefont{and} \bibinfo{author}{\bibfnamefont{N.}~\bibnamefont{Turok}},
  \bibinfo{journal}{Phys. Rev. D} \textbf{\bibinfo{volume}{65}},
  \bibinfo{pages}{126003} (\bibinfo{year}{2002}), \eprint{hep-th/0111098}.

\bibitem[{\citenamefont{Battefeld et~al.}(2004)\citenamefont{Battefeld, Patil,
  and Brandenberger}}]{Batte04}
\bibinfo{author}{\bibfnamefont{T.~J.} \bibnamefont{Battefeld}},
  \bibinfo{author}{\bibfnamefont{S.~P.} \bibnamefont{Patil}}, \bibnamefont{and}
  \bibinfo{author}{\bibfnamefont{R.}~\bibnamefont{Brandenberger}},
  \bibinfo{journal}{Phys. Rev. D} \textbf{\bibinfo{volume}{70}},
  \bibinfo{pages}{066006} (\bibinfo{year}{2004}), \eprint{hep-th/0401010}.

\bibitem[{\citenamefont{Bojowald et~al.}(2004)\citenamefont{Bojowald, Maartens,
  and Singh}}]{Bojowald04}
\bibinfo{author}{\bibfnamefont{M.}~\bibnamefont{Bojowald}},
  \bibinfo{author}{\bibfnamefont{R.}~\bibnamefont{Maartens}}, \bibnamefont{and}
  \bibinfo{author}{\bibfnamefont{P.}~\bibnamefont{Singh}},
  \bibinfo{journal}{Phys. Rev. D} \textbf{\bibinfo{volume}{70}},
  \bibinfo{pages}{083517} (\bibinfo{year}{2004}), \eprint{hep-th/0407115}.

\bibitem[{\citenamefont{Biswas et~al.}(2006)\citenamefont{Biswas, Mazumdar, and
  Siegel}}]{Biswas2005}
\bibinfo{author}{\bibfnamefont{T.}~\bibnamefont{Biswas}},
  \bibinfo{author}{\bibfnamefont{A.}~\bibnamefont{Mazumdar}}, \bibnamefont{and}
  \bibinfo{author}{\bibfnamefont{W.}~\bibnamefont{Siegel}},
  \bibinfo{journal}{JCAP} \textbf{\bibinfo{volume}{03}}, \bibinfo{pages}{009}
  (\bibinfo{year}{2006}), \eprint{hep-th/0508194}.

\bibitem[{\citenamefont{Peter and Pinto-Neto}(2002)}]{Peter02}
\bibinfo{author}{\bibfnamefont{P.}~\bibnamefont{Peter}} \bibnamefont{and}
  \bibinfo{author}{\bibfnamefont{N.}~\bibnamefont{Pinto-Neto}},
  \bibinfo{journal}{Phys. Rev. D} \textbf{\bibinfo{volume}{66}},
  \bibinfo{pages}{063509} (\bibinfo{year}{2002}), \eprint{hep-th/0203013}.

\bibitem[{\citenamefont{Battefeld and Peter}(2015)}]{Batte15}
\bibinfo{author}{\bibfnamefont{D.}~\bibnamefont{Battefeld}} \bibnamefont{and}
  \bibinfo{author}{\bibfnamefont{P.}~\bibnamefont{Peter}},
  \bibinfo{journal}{Phys. Rept.} \textbf{\bibinfo{volume}{571}},
  \bibinfo{pages}{1} (\bibinfo{year}{2015}), \eprint{1406.2790}.

\bibitem[{\citenamefont{Lilley and Peter}(2015)}]{Lilley15}
\bibinfo{author}{\bibfnamefont{M.}~\bibnamefont{Lilley}} \bibnamefont{and}
  \bibinfo{author}{\bibfnamefont{P.}~\bibnamefont{Peter}},
  \bibinfo{journal}{Comptes Rendus Physique} \textbf{\bibinfo{volume}{16}},
  \bibinfo{pages}{1038} (\bibinfo{year}{2015}), \eprint{1503.06578}.

\bibitem[{\citenamefont{Brandenberger and Peter}(2017)}]{Branden17}
\bibinfo{author}{\bibfnamefont{R.}~\bibnamefont{Brandenberger}}
  \bibnamefont{and} \bibinfo{author}{\bibfnamefont{P.}~\bibnamefont{Peter}},
  \bibinfo{journal}{Found. Phys.} \textbf{\bibinfo{volume}{47}},
  \bibinfo{pages}{797} (\bibinfo{year}{2017}), \eprint{1603.05834}.

\bibitem[{\citenamefont{Matsui et~al.}(2019)\citenamefont{Matsui, Takahashi,
  and Terada}}]{Matsui19}
\bibinfo{author}{\bibfnamefont{H.}~\bibnamefont{Matsui}},
  \bibinfo{author}{\bibfnamefont{F.}~\bibnamefont{Takahashi}},
  \bibnamefont{and} \bibinfo{author}{\bibfnamefont{T.}~\bibnamefont{Terada}},
  \bibinfo{journal}{Phys. Lett. B} \textbf{\bibinfo{volume}{795}},
  \bibinfo{pages}{152} (\bibinfo{year}{2019}), \eprint{1904.12312}.

\bibitem[{\citenamefont{Barrau}(2020)}]{Barrau20}
\bibinfo{author}{\bibfnamefont{A.}~\bibnamefont{Barrau}},
  \bibinfo{journal}{Eur. Phys. J. C} \textbf{\bibinfo{volume}{80}},
  \bibinfo{pages}{579} (\bibinfo{year}{2020}), \eprint{2005.04693}.

\bibitem[{\citenamefont{Ellis and Maartens}(2004)}]{Ellis1}
\bibinfo{author}{\bibfnamefont{G.~F.~R.} \bibnamefont{Ellis}} \bibnamefont{and}
  \bibinfo{author}{\bibfnamefont{R.}~\bibnamefont{Maartens}},
  \bibinfo{journal}{Class. Quant. Grav.} \textbf{\bibinfo{volume}{21}},
  \bibinfo{pages}{223} (\bibinfo{year}{2004}), \eprint{gr-qc/0211082}.

\bibitem[{\citenamefont{{Einstein}}(1917)}]{Einstein1917}
\bibinfo{author}{\bibfnamefont{A.}~\bibnamefont{{Einstein}}},
  \bibinfo{journal}{Sitzungsberichte der K{\"o}niglich Preu{\ss}ischen Akademie
  der Wissenschaften (Berlin} pp. \bibinfo{pages}{142--152}
  (\bibinfo{year}{1917}).

\bibitem[{\citenamefont{Eddington}(1930)}]{Eddington30}
\bibinfo{author}{\bibfnamefont{A.~S.} \bibnamefont{Eddington}},
  \bibinfo{journal}{Mon. Not. Roy. Astron. Soc.} \textbf{\bibinfo{volume}{90}},
  \bibinfo{pages}{668} (\bibinfo{year}{1930}).

\bibitem[{\citenamefont{Ellis et~al.}(2004)\citenamefont{Ellis, Murugan, and
  Tsagas}}]{Ellis2}
\bibinfo{author}{\bibfnamefont{G.~F.~R.} \bibnamefont{Ellis}},
  \bibinfo{author}{\bibfnamefont{J.}~\bibnamefont{Murugan}}, \bibnamefont{and}
  \bibinfo{author}{\bibfnamefont{C.~G.} \bibnamefont{Tsagas}},
  \bibinfo{journal}{Class. Quant. Grav.} \textbf{\bibinfo{volume}{21}},
  \bibinfo{pages}{233} (\bibinfo{year}{2004}), \eprint{gr-qc/0307112}.

\bibitem[{\citenamefont{Mulryne et~al.}(2005)\citenamefont{Mulryne, Tavakol,
  Lidsey, and Ellis}}]{Ellis3}
\bibinfo{author}{\bibfnamefont{D.~J.} \bibnamefont{Mulryne}},
  \bibinfo{author}{\bibfnamefont{R.}~\bibnamefont{Tavakol}},
  \bibinfo{author}{\bibfnamefont{J.~E.} \bibnamefont{Lidsey}},
  \bibnamefont{and} \bibinfo{author}{\bibfnamefont{G.~F.~R.}
  \bibnamefont{Ellis}}, \bibinfo{journal}{Phys. Rev. D}
  \textbf{\bibinfo{volume}{71}}, \bibinfo{pages}{123512}
  (\bibinfo{year}{2005}), \eprint{astro-ph/0502589}.

\bibitem[{\citenamefont{{Gibbons}}(1987)}]{Gibbons1987}
\bibinfo{author}{\bibfnamefont{G.~W.} \bibnamefont{{Gibbons}}},
  \bibinfo{journal}{Nuclear Physics B} \textbf{\bibinfo{volume}{292}},
  \bibinfo{pages}{784} (\bibinfo{year}{1987}).

\bibitem[{\citenamefont{Barrow et~al.}(2003)\citenamefont{Barrow, Ellis,
  Maartens, and Tsagas}}]{Barrow2003}
\bibinfo{author}{\bibfnamefont{J.~D.} \bibnamefont{Barrow}},
  \bibinfo{author}{\bibfnamefont{G.~F.~R.} \bibnamefont{Ellis}},
  \bibinfo{author}{\bibfnamefont{R.}~\bibnamefont{Maartens}}, \bibnamefont{and}
  \bibinfo{author}{\bibfnamefont{C.~G.} \bibnamefont{Tsagas}},
  \bibinfo{journal}{Class. Quant. Grav.} \textbf{\bibinfo{volume}{20}},
  \bibinfo{pages}{L155} (\bibinfo{year}{2003}), \eprint{gr-qc/0302094}.

\bibitem[{\citenamefont{Mukherjee et~al.}(2005)\citenamefont{Mukherjee, Paul,
  Maharaj, and Beesham}}]{Mukherjee05}
\bibinfo{author}{\bibfnamefont{S.}~\bibnamefont{Mukherjee}},
  \bibinfo{author}{\bibfnamefont{B.~C.} \bibnamefont{Paul}},
  \bibinfo{author}{\bibfnamefont{S.~D.} \bibnamefont{Maharaj}},
  \bibnamefont{and} \bibinfo{author}{\bibfnamefont{A.}~\bibnamefont{Beesham}}
  (\bibinfo{year}{2005}), \eprint{gr-qc/0505103}.

\bibitem[{\citenamefont{Mukherjee et~al.}(2006)\citenamefont{Mukherjee, Paul,
  Dadhich, Maharaj, and Beesham}}]{Mukherjee06}
\bibinfo{author}{\bibfnamefont{S.}~\bibnamefont{Mukherjee}},
  \bibinfo{author}{\bibfnamefont{B.~C.} \bibnamefont{Paul}},
  \bibinfo{author}{\bibfnamefont{N.~K.} \bibnamefont{Dadhich}},
  \bibinfo{author}{\bibfnamefont{S.~D.} \bibnamefont{Maharaj}},
  \bibnamefont{and} \bibinfo{author}{\bibfnamefont{A.}~\bibnamefont{Beesham}},
  \bibinfo{journal}{Class. Quant. Grav.} \textbf{\bibinfo{volume}{23}},
  \bibinfo{pages}{6927} (\bibinfo{year}{2006}), \eprint{gr-qc/0605134}.

\bibitem[{\citenamefont{Banerjee et~al.}(2007)\citenamefont{Banerjee,
  Bandyopadhyay, and Chakraborty}}]{Banerjee07}
\bibinfo{author}{\bibfnamefont{A.}~\bibnamefont{Banerjee}},
  \bibinfo{author}{\bibfnamefont{T.}~\bibnamefont{Bandyopadhyay}},
  \bibnamefont{and}
  \bibinfo{author}{\bibfnamefont{S.}~\bibnamefont{Chakraborty}},
  \bibinfo{journal}{Grav. Cosmol.} \textbf{\bibinfo{volume}{13}},
  \bibinfo{pages}{290} (\bibinfo{year}{2007}), \eprint{0705.3933}.

\bibitem[{\citenamefont{Boehmer et~al.}(2007)\citenamefont{Boehmer,
  Hollenstein, and Lobo}}]{Boehmer07}
\bibinfo{author}{\bibfnamefont{C.~G.} \bibnamefont{Boehmer}},
  \bibinfo{author}{\bibfnamefont{L.}~\bibnamefont{Hollenstein}},
  \bibnamefont{and} \bibinfo{author}{\bibfnamefont{F.~S.~N.}
  \bibnamefont{Lobo}}, \bibinfo{journal}{Phys. Rev. D}
  \textbf{\bibinfo{volume}{76}}, \bibinfo{pages}{084005}
  (\bibinfo{year}{2007}), \eprint{0706.1663}.

\bibitem[{\citenamefont{Parisi et~al.}(2007)\citenamefont{Parisi, Bruni,
  Maartens, and Vandersloot}}]{Parisi2007}
\bibinfo{author}{\bibfnamefont{L.}~\bibnamefont{Parisi}},
  \bibinfo{author}{\bibfnamefont{M.}~\bibnamefont{Bruni}},
  \bibinfo{author}{\bibfnamefont{R.}~\bibnamefont{Maartens}}, \bibnamefont{and}
  \bibinfo{author}{\bibfnamefont{K.}~\bibnamefont{Vandersloot}},
  \bibinfo{journal}{Class. Quant. Grav.} \textbf{\bibinfo{volume}{24}},
  \bibinfo{pages}{6243} (\bibinfo{year}{2007}), \eprint{0706.4431}.

\bibitem[{\citenamefont{del Campo et~al.}(2007)\citenamefont{del Campo,
  Herrera, and Labrana}}]{Campo1}
\bibinfo{author}{\bibfnamefont{S.}~\bibnamefont{del Campo}},
  \bibinfo{author}{\bibfnamefont{R.}~\bibnamefont{Herrera}}, \bibnamefont{and}
  \bibinfo{author}{\bibfnamefont{P.}~\bibnamefont{Labrana}},
  \bibinfo{journal}{JCAP} \textbf{\bibinfo{volume}{11}}, \bibinfo{pages}{030}
  (\bibinfo{year}{2007}), \eprint{0711.1559}.

\bibitem[{\citenamefont{Debnath}(2008)}]{Debnath08}
\bibinfo{author}{\bibfnamefont{U.}~\bibnamefont{Debnath}},
  \bibinfo{journal}{Class. Quant. Grav.} \textbf{\bibinfo{volume}{25}},
  \bibinfo{pages}{205019} (\bibinfo{year}{2008}), \eprint{0808.2379}.

\bibitem[{\citenamefont{Banerjee et~al.}(2008)\citenamefont{Banerjee,
  Bandyopadhyay, and Chakraborty}}]{Banerjee08}
\bibinfo{author}{\bibfnamefont{A.}~\bibnamefont{Banerjee}},
  \bibinfo{author}{\bibfnamefont{T.}~\bibnamefont{Bandyopadhyay}},
  \bibnamefont{and}
  \bibinfo{author}{\bibfnamefont{S.}~\bibnamefont{Chakraborty}},
  \bibinfo{journal}{Gen. Rel. Grav.} \textbf{\bibinfo{volume}{40}},
  \bibinfo{pages}{1603} (\bibinfo{year}{2008}), \eprint{0711.4188}.

\bibitem[{\citenamefont{Beesham et~al.}(2009)\citenamefont{Beesham, Chervon,
  and Maharaj}}]{Beesham09}
\bibinfo{author}{\bibfnamefont{A.}~\bibnamefont{Beesham}},
  \bibinfo{author}{\bibfnamefont{S.~V.} \bibnamefont{Chervon}},
  \bibnamefont{and} \bibinfo{author}{\bibfnamefont{S.~D.}
  \bibnamefont{Maharaj}}, \bibinfo{journal}{Class. Quant. Grav.}
  \textbf{\bibinfo{volume}{26}}, \bibinfo{pages}{075017}
  (\bibinfo{year}{2009}), \eprint{0904.0773}.

\bibitem[{\citenamefont{del Campo et~al.}(2009)\citenamefont{del Campo,
  Herrera, and Labrana}}]{Campo2}
\bibinfo{author}{\bibfnamefont{S.}~\bibnamefont{del Campo}},
  \bibinfo{author}{\bibfnamefont{R.}~\bibnamefont{Herrera}}, \bibnamefont{and}
  \bibinfo{author}{\bibfnamefont{P.}~\bibnamefont{Labrana}},
  \bibinfo{journal}{JCAP} \textbf{\bibinfo{volume}{07}}, \bibinfo{pages}{006}
  (\bibinfo{year}{2009}), \eprint{0905.0614}.

\bibitem[{\citenamefont{Paul and Ghose}(2010)}]{Paul10}
\bibinfo{author}{\bibfnamefont{B.~C.} \bibnamefont{Paul}} \bibnamefont{and}
  \bibinfo{author}{\bibfnamefont{S.}~\bibnamefont{Ghose}},
  \bibinfo{journal}{Gen. Rel. Grav.} \textbf{\bibinfo{volume}{42}},
  \bibinfo{pages}{795} (\bibinfo{year}{2010}), \eprint{0809.4131}.

\bibitem[{\citenamefont{Paul et~al.}(2010)\citenamefont{Paul, Thakur, and
  Ghose}}]{Paul2010}
\bibinfo{author}{\bibfnamefont{B.~C.} \bibnamefont{Paul}},
  \bibinfo{author}{\bibfnamefont{P.}~\bibnamefont{Thakur}}, \bibnamefont{and}
  \bibinfo{author}{\bibfnamefont{S.}~\bibnamefont{Ghose}},
  \bibinfo{journal}{Mon. Not. Roy. Astron. Soc.}
  \textbf{\bibinfo{volume}{407}}, \bibinfo{pages}{415} (\bibinfo{year}{2010}),
  \eprint{1004.4256}.

\bibitem[{\citenamefont{Zhang et~al.}(2010)\citenamefont{Zhang, Wu, and
  Yu}}]{Zhang10}
\bibinfo{author}{\bibfnamefont{K.}~\bibnamefont{Zhang}},
  \bibinfo{author}{\bibfnamefont{P.}~\bibnamefont{Wu}}, \bibnamefont{and}
  \bibinfo{author}{\bibfnamefont{H.~W.} \bibnamefont{Yu}},
  \bibinfo{journal}{Phys. Lett. B} \textbf{\bibinfo{volume}{690}},
  \bibinfo{pages}{229} (\bibinfo{year}{2010}), \eprint{1005.4201}.

\bibitem[{\citenamefont{{Paul} et~al.}(2011)\citenamefont{{Paul}, {Ghose}, and
  {Thakur}}}]{Paul2011}
\bibinfo{author}{\bibfnamefont{B.~C.} \bibnamefont{{Paul}}},
  \bibinfo{author}{\bibfnamefont{S.}~\bibnamefont{{Ghose}}}, \bibnamefont{and}
  \bibinfo{author}{\bibfnamefont{P.}~\bibnamefont{{Thakur}}},
  \bibinfo{journal}{Mon. Not. R. Astron. Soc.} \textbf{\bibinfo{volume}{413}},
  \bibinfo{pages}{686} (\bibinfo{year}{2011}), \eprint{1101.1360}.

\bibitem[{\citenamefont{Chattopadhyay and Debnath}(2011)}]{Chatto11}
\bibinfo{author}{\bibfnamefont{S.}~\bibnamefont{Chattopadhyay}}
  \bibnamefont{and} \bibinfo{author}{\bibfnamefont{U.}~\bibnamefont{Debnath}},
  \bibinfo{journal}{Int. J. Mod. Phys. D} \textbf{\bibinfo{volume}{20}},
  \bibinfo{pages}{1135} (\bibinfo{year}{2011}), \eprint{1105.1091}.

\bibitem[{\citenamefont{del Campo et~al.}(2011)\citenamefont{del Campo,
  Guendelman, Kaganovich, Herrera, and Labrana}}]{Campo3}
\bibinfo{author}{\bibfnamefont{S.}~\bibnamefont{del Campo}},
  \bibinfo{author}{\bibfnamefont{E.~I.} \bibnamefont{Guendelman}},
  \bibinfo{author}{\bibfnamefont{A.~B.} \bibnamefont{Kaganovich}},
  \bibinfo{author}{\bibfnamefont{R.}~\bibnamefont{Herrera}}, \bibnamefont{and}
  \bibinfo{author}{\bibfnamefont{P.}~\bibnamefont{Labrana}},
  \bibinfo{journal}{Phys. Lett. B} \textbf{\bibinfo{volume}{699}},
  \bibinfo{pages}{211} (\bibinfo{year}{2011}), \eprint{1105.0651}.

\bibitem[{\citenamefont{del Campo et~al.}(2010)\citenamefont{del Campo,
  Guendelman, Herrera, and Labrana}}]{Campo4}
\bibinfo{author}{\bibfnamefont{S.}~\bibnamefont{del Campo}},
  \bibinfo{author}{\bibfnamefont{E.~I.} \bibnamefont{Guendelman}},
  \bibinfo{author}{\bibfnamefont{R.}~\bibnamefont{Herrera}}, \bibnamefont{and}
  \bibinfo{author}{\bibfnamefont{P.}~\bibnamefont{Labrana}},
  \bibinfo{journal}{JCAP} \textbf{\bibinfo{volume}{06}}, \bibinfo{pages}{026}
  (\bibinfo{year}{2010}), \eprint{1006.5734}.

\bibitem[{\citenamefont{Labrana}(2012)}]{Labrana2012}
\bibinfo{author}{\bibfnamefont{P.}~\bibnamefont{Labrana}},
  \bibinfo{journal}{Phys. Rev. D} \textbf{\bibinfo{volume}{86}},
  \bibinfo{pages}{083524} (\bibinfo{year}{2012}), \eprint{1111.5360}.

\bibitem[{\citenamefont{Cai et~al.}(2012)\citenamefont{Cai, Li, and
  Zhang}}]{Cai2012}
\bibinfo{author}{\bibfnamefont{Y.-F.} \bibnamefont{Cai}},
  \bibinfo{author}{\bibfnamefont{M.}~\bibnamefont{Li}}, \bibnamefont{and}
  \bibinfo{author}{\bibfnamefont{X.}~\bibnamefont{Zhang}},
  \bibinfo{journal}{Phys. Lett. B} \textbf{\bibinfo{volume}{718}},
  \bibinfo{pages}{248} (\bibinfo{year}{2012}), \eprint{1209.3437}.

\bibitem[{\citenamefont{Rudra}(2012)}]{Rudra12}
\bibinfo{author}{\bibfnamefont{P.}~\bibnamefont{Rudra}}, \bibinfo{journal}{Mod.
  Phys. Lett. A} \textbf{\bibinfo{volume}{27}}, \bibinfo{pages}{1250189}
  (\bibinfo{year}{2012}), \eprint{1211.2047}.

\bibitem[{\citenamefont{{Ghose} et~al.}(2012)\citenamefont{{Ghose}, {Thakur},
  and {Paul}}}]{Ghose12}
\bibinfo{author}{\bibfnamefont{S.}~\bibnamefont{{Ghose}}},
  \bibinfo{author}{\bibfnamefont{P.}~\bibnamefont{{Thakur}}}, \bibnamefont{and}
  \bibinfo{author}{\bibfnamefont{B.~C.} \bibnamefont{{Paul}}},
  \bibinfo{journal}{Mon. Not. R. Astron. Soc.} \textbf{\bibinfo{volume}{421}},
  \bibinfo{pages}{20} (\bibinfo{year}{2012}), \eprint{1105.3303}.

\bibitem[{\citenamefont{Liu and Piao}(2013)}]{Liu13}
\bibinfo{author}{\bibfnamefont{Z.-G.} \bibnamefont{Liu}} \bibnamefont{and}
  \bibinfo{author}{\bibfnamefont{Y.-S.} \bibnamefont{Piao}},
  \bibinfo{journal}{Phys. Lett. B} \textbf{\bibinfo{volume}{718}},
  \bibinfo{pages}{734} (\bibinfo{year}{2013}), \eprint{1207.2568}.

\bibitem[{\citenamefont{Aguirre and Kehayias}(2013)}]{Aguirre13}
\bibinfo{author}{\bibfnamefont{A.}~\bibnamefont{Aguirre}} \bibnamefont{and}
  \bibinfo{author}{\bibfnamefont{J.}~\bibnamefont{Kehayias}},
  \bibinfo{journal}{Phys. Rev. D} \textbf{\bibinfo{volume}{88}},
  \bibinfo{pages}{103504} (\bibinfo{year}{2013}), \eprint{1306.3232}.

\bibitem[{\citenamefont{Cai et~al.}(2014)\citenamefont{Cai, Wan, and
  Zhang}}]{Cai2014}
\bibinfo{author}{\bibfnamefont{Y.-F.} \bibnamefont{Cai}},
  \bibinfo{author}{\bibfnamefont{Y.}~\bibnamefont{Wan}}, \bibnamefont{and}
  \bibinfo{author}{\bibfnamefont{X.}~\bibnamefont{Zhang}},
  \bibinfo{journal}{Phys. Lett. B} \textbf{\bibinfo{volume}{731}},
  \bibinfo{pages}{217} (\bibinfo{year}{2014}), \eprint{1312.0740}.

\bibitem[{\citenamefont{Atazadeh et~al.}(2014)\citenamefont{Atazadeh,
  Heydarzade, and Darabi}}]{Ataz14}
\bibinfo{author}{\bibfnamefont{K.}~\bibnamefont{Atazadeh}},
  \bibinfo{author}{\bibfnamefont{Y.}~\bibnamefont{Heydarzade}},
  \bibnamefont{and} \bibinfo{author}{\bibfnamefont{F.}~\bibnamefont{Darabi}},
  \bibinfo{journal}{Phys. Lett. B} \textbf{\bibinfo{volume}{732}},
  \bibinfo{pages}{223} (\bibinfo{year}{2014}), \eprint{1401.7638}.

\bibitem[{\citenamefont{Zhang et~al.}(2014)\citenamefont{Zhang, Wu, and
  Yu}}]{Zhang14}
\bibinfo{author}{\bibfnamefont{K.}~\bibnamefont{Zhang}},
  \bibinfo{author}{\bibfnamefont{P.}~\bibnamefont{Wu}}, \bibnamefont{and}
  \bibinfo{author}{\bibfnamefont{H.}~\bibnamefont{Yu}}, \bibinfo{journal}{JCAP}
  \textbf{\bibinfo{volume}{01}}, \bibinfo{pages}{048} (\bibinfo{year}{2014}),
  \eprint{1311.4051}.

\bibitem[{\citenamefont{Bag et~al.}(2014)\citenamefont{Bag, Sahni, Shtanov, and
  Unnikrishnan}}]{Bag14}
\bibinfo{author}{\bibfnamefont{S.}~\bibnamefont{Bag}},
  \bibinfo{author}{\bibfnamefont{V.}~\bibnamefont{Sahni}},
  \bibinfo{author}{\bibfnamefont{Y.}~\bibnamefont{Shtanov}}, \bibnamefont{and}
  \bibinfo{author}{\bibfnamefont{S.}~\bibnamefont{Unnikrishnan}},
  \bibinfo{journal}{JCAP} \textbf{\bibinfo{volume}{07}}, \bibinfo{pages}{034}
  (\bibinfo{year}{2014}), \eprint{1403.4243}.

\bibitem[{\citenamefont{Huang et~al.}(2015)\citenamefont{Huang, Wu, and
  Yu}}]{Huang15}
\bibinfo{author}{\bibfnamefont{Q.}~\bibnamefont{Huang}},
  \bibinfo{author}{\bibfnamefont{P.}~\bibnamefont{Wu}}, \bibnamefont{and}
  \bibinfo{author}{\bibfnamefont{H.}~\bibnamefont{Yu}}, \bibinfo{journal}{Phys.
  Rev. D} \textbf{\bibinfo{volume}{91}}, \bibinfo{pages}{103502}
  (\bibinfo{year}{2015}), \eprint{1504.05284}.

\bibitem[{\citenamefont{B\"ohmer et~al.}(2015)\citenamefont{B\"ohmer, Tamanini,
  and Wright}}]{Boehmer15}
\bibinfo{author}{\bibfnamefont{C.~G.} \bibnamefont{B\"ohmer}},
  \bibinfo{author}{\bibfnamefont{N.}~\bibnamefont{Tamanini}}, \bibnamefont{and}
  \bibinfo{author}{\bibfnamefont{M.}~\bibnamefont{Wright}},
  \bibinfo{journal}{Phys. Rev. D} \textbf{\bibinfo{volume}{92}},
  \bibinfo{pages}{124067} (\bibinfo{year}{2015}), \eprint{1510.01477}.

\bibitem[{\citenamefont{Labra\~na}(2015)}]{Labrana15}
\bibinfo{author}{\bibfnamefont{P.}~\bibnamefont{Labra\~na}},
  \bibinfo{journal}{Phys. Rev. D} \textbf{\bibinfo{volume}{91}},
  \bibinfo{pages}{083534} (\bibinfo{year}{2015}), \eprint{1312.6877}.

\bibitem[{\citenamefont{Khodadi et~al.}(2016)\citenamefont{Khodadi, Heydarzade,
  Darabi, and Saridakis}}]{Kho16}
\bibinfo{author}{\bibfnamefont{M.}~\bibnamefont{Khodadi}},
  \bibinfo{author}{\bibfnamefont{Y.}~\bibnamefont{Heydarzade}},
  \bibinfo{author}{\bibfnamefont{F.}~\bibnamefont{Darabi}}, \bibnamefont{and}
  \bibinfo{author}{\bibfnamefont{E.~N.} \bibnamefont{Saridakis}},
  \bibinfo{journal}{Phys. Rev. D} \textbf{\bibinfo{volume}{93}},
  \bibinfo{pages}{124019} (\bibinfo{year}{2016}), \eprint{1512.08674}.

\bibitem[{\citenamefont{{Zhang} et~al.}(2016)\citenamefont{{Zhang}, {Wu}, {Yu},
  and {Luo}}}]{Zhang16}
\bibinfo{author}{\bibfnamefont{K.}~\bibnamefont{{Zhang}}},
  \bibinfo{author}{\bibfnamefont{P.}~\bibnamefont{{Wu}}},
  \bibinfo{author}{\bibfnamefont{H.}~\bibnamefont{{Yu}}}, \bibnamefont{and}
  \bibinfo{author}{\bibfnamefont{L.-W.} \bibnamefont{{Luo}}},
  \bibinfo{journal}{Physics Letters B} \textbf{\bibinfo{volume}{758}},
  \bibinfo{pages}{37} (\bibinfo{year}{2016}).

\bibitem[{\citenamefont{R{\'{\i}}os et~al.}(2016)\citenamefont{R{\'{\i}}os,
  Labra{\~{n}}a, and Cid}}]{Rios16}
\bibinfo{author}{\bibfnamefont{C.}~\bibnamefont{R{\'{\i}}os}},
  \bibinfo{author}{\bibfnamefont{P.}~\bibnamefont{Labra{\~{n}}a}},
  \bibnamefont{and} \bibinfo{author}{\bibfnamefont{A.}~\bibnamefont{Cid}},
  \bibinfo{journal}{Journal of Physics: Conference Series}
  \textbf{\bibinfo{volume}{720}}, \bibinfo{pages}{012008}
  (\bibinfo{year}{2016}).

\bibitem[{\citenamefont{Khodadi et~al.}(2018)\citenamefont{Khodadi, Nozari, and
  Saridakis}}]{Khodadi16}
\bibinfo{author}{\bibfnamefont{M.}~\bibnamefont{Khodadi}},
  \bibinfo{author}{\bibfnamefont{K.}~\bibnamefont{Nozari}}, \bibnamefont{and}
  \bibinfo{author}{\bibfnamefont{E.~N.} \bibnamefont{Saridakis}},
  \bibinfo{journal}{Class. Quant. Grav.} \textbf{\bibinfo{volume}{35}},
  \bibinfo{pages}{015010} (\bibinfo{year}{2018}), \eprint{1612.09254}.

\bibitem[{\citenamefont{Martineau and Barrau}(2018)}]{Barrau18}
\bibinfo{author}{\bibfnamefont{K.}~\bibnamefont{Martineau}} \bibnamefont{and}
  \bibinfo{author}{\bibfnamefont{A.}~\bibnamefont{Barrau}},
  \bibinfo{journal}{Universe} \textbf{\bibinfo{volume}{4}},
  \bibinfo{pages}{149} (\bibinfo{year}{2018}), \eprint{1812.05522}.

\bibitem[{\citenamefont{Labrana and Cossio}(2019)}]{Labrana19}
\bibinfo{author}{\bibfnamefont{P.}~\bibnamefont{Labrana}} \bibnamefont{and}
  \bibinfo{author}{\bibfnamefont{H.}~\bibnamefont{Cossio}},
  \bibinfo{journal}{Eur. Phys. J. C} \textbf{\bibinfo{volume}{79}},
  \bibinfo{pages}{303} (\bibinfo{year}{2019}), \eprint{1808.09291}.

\bibitem[{\citenamefont{Brandenberger}(2010)}]{BrandenRev11}
\bibinfo{author}{\bibfnamefont{R.~H.} \bibnamefont{Brandenberger}},
  \bibinfo{journal}{PoS} \textbf{\bibinfo{volume}{ICFI2010}},
  \bibinfo{pages}{001} (\bibinfo{year}{2010}), \eprint{1103.2271}.

\bibitem[{\citenamefont{Brandenberger}(2020)}]{Branden18}
\bibinfo{author}{\bibfnamefont{R.~H.} \bibnamefont{Brandenberger}},
  \emph{\bibinfo{title}{{Beyond Standard Inflationary Cosmology}}}
  (\bibinfo{year}{2020}), \eprint{1809.04926}.

\bibitem[{\citenamefont{{Brandenberger} and {Vafa}}(1989)}]{Branden89}
\bibinfo{author}{\bibfnamefont{R.}~\bibnamefont{{Brandenberger}}}
  \bibnamefont{and} \bibinfo{author}{\bibfnamefont{C.}~\bibnamefont{{Vafa}}},
  \bibinfo{journal}{Nuclear Physics B} \textbf{\bibinfo{volume}{316}},
  \bibinfo{pages}{391} (\bibinfo{year}{1989}).

\bibitem[{\citenamefont{Nayeri et~al.}(2006)\citenamefont{Nayeri,
  Brandenberger, and Vafa}}]{Branden06}
\bibinfo{author}{\bibfnamefont{A.}~\bibnamefont{Nayeri}},
  \bibinfo{author}{\bibfnamefont{R.~H.} \bibnamefont{Brandenberger}},
  \bibnamefont{and} \bibinfo{author}{\bibfnamefont{C.}~\bibnamefont{Vafa}},
  \bibinfo{journal}{Phys. Rev. Lett.} \textbf{\bibinfo{volume}{97}},
  \bibinfo{pages}{021302} (\bibinfo{year}{2006}), \eprint{hep-th/0511140}.

\bibitem[{\citenamefont{Brandenberger}(2008)}]{Branden08}
\bibinfo{author}{\bibfnamefont{R.~H.} \bibnamefont{Brandenberger}}
  (\bibinfo{year}{2008}), \eprint{0808.0746}.

\bibitem[{\citenamefont{Brandenberger}(2011)}]{Branden11}
\bibinfo{author}{\bibfnamefont{R.~H.} \bibnamefont{Brandenberger}},
  \bibinfo{journal}{Class. Quant. Grav.} \textbf{\bibinfo{volume}{28}},
  \bibinfo{pages}{204005} (\bibinfo{year}{2011}), \eprint{1105.3247}.

\bibitem[{\citenamefont{Obied et~al.}(2018)\citenamefont{Obied, Ooguri,
  Spodyneiko, and Vafa}}]{Swamp1}
\bibinfo{author}{\bibfnamefont{G.}~\bibnamefont{Obied}},
  \bibinfo{author}{\bibfnamefont{H.}~\bibnamefont{Ooguri}},
  \bibinfo{author}{\bibfnamefont{L.}~\bibnamefont{Spodyneiko}},
  \bibnamefont{and} \bibinfo{author}{\bibfnamefont{C.}~\bibnamefont{Vafa}}
  (\bibinfo{year}{2018}), \eprint{1806.08362}.

\bibitem[{\citenamefont{Bedroya and Vafa}(2020)}]{TCCVafa}
\bibinfo{author}{\bibfnamefont{A.}~\bibnamefont{Bedroya}} \bibnamefont{and}
  \bibinfo{author}{\bibfnamefont{C.}~\bibnamefont{Vafa}},
  \bibinfo{journal}{JHEP} \textbf{\bibinfo{volume}{09}}, \bibinfo{pages}{123}
  (\bibinfo{year}{2020}), \eprint{1909.11063}.

\bibitem[{\citenamefont{Garg and Krishnan}(2019)}]{Swamp2}
\bibinfo{author}{\bibfnamefont{S.~K.} \bibnamefont{Garg}} \bibnamefont{and}
  \bibinfo{author}{\bibfnamefont{C.}~\bibnamefont{Krishnan}},
  \bibinfo{journal}{JHEP} \textbf{\bibinfo{volume}{11}}, \bibinfo{pages}{075}
  (\bibinfo{year}{2019}), \eprint{1807.05193}.

\bibitem[{\citenamefont{Agrawal et~al.}(2018)\citenamefont{Agrawal, Obied,
  Steinhardt, and Vafa}}]{Swamp3}
\bibinfo{author}{\bibfnamefont{P.}~\bibnamefont{Agrawal}},
  \bibinfo{author}{\bibfnamefont{G.}~\bibnamefont{Obied}},
  \bibinfo{author}{\bibfnamefont{P.~J.} \bibnamefont{Steinhardt}},
  \bibnamefont{and} \bibinfo{author}{\bibfnamefont{C.}~\bibnamefont{Vafa}},
  \bibinfo{journal}{Phys. Lett. B} \textbf{\bibinfo{volume}{784}},
  \bibinfo{pages}{271} (\bibinfo{year}{2018}), \eprint{1806.09718}.

\bibitem[{\citenamefont{Brandenberger and Martin}(2013)}]{Brandenb2012}
\bibinfo{author}{\bibfnamefont{R.~H.} \bibnamefont{Brandenberger}}
  \bibnamefont{and} \bibinfo{author}{\bibfnamefont{J.}~\bibnamefont{Martin}},
  \bibinfo{journal}{Class. Quant. Grav.} \textbf{\bibinfo{volume}{30}},
  \bibinfo{pages}{113001} (\bibinfo{year}{2013}), \eprint{1211.6753}.

\bibitem[{\citenamefont{Bedroya et~al.}(2020)\citenamefont{Bedroya,
  Brandenberger, Loverde, and Vafa}}]{Bedroya20}
\bibinfo{author}{\bibfnamefont{A.}~\bibnamefont{Bedroya}},
  \bibinfo{author}{\bibfnamefont{R.}~\bibnamefont{Brandenberger}},
  \bibinfo{author}{\bibfnamefont{M.}~\bibnamefont{Loverde}}, \bibnamefont{and}
  \bibinfo{author}{\bibfnamefont{C.}~\bibnamefont{Vafa}},
  \bibinfo{journal}{Phys. Rev. D} \textbf{\bibinfo{volume}{101}},
  \bibinfo{pages}{103502} (\bibinfo{year}{2020}), \eprint{1909.11106}.

\bibitem[{\citenamefont{Brandenberger}(2021)}]{Branden21}
\bibinfo{author}{\bibfnamefont{R.}~\bibnamefont{Brandenberger}}
  (\bibinfo{year}{2021}), \eprint{2102.09641}.

\bibitem[{\citenamefont{{Wigner}}(1963)}]{Wigner63}
\bibinfo{author}{\bibfnamefont{E.~P.} \bibnamefont{{Wigner}}},
  \bibinfo{journal}{American Journal of Physics} \textbf{\bibinfo{volume}{31}},
  \bibinfo{pages}{6} (\bibinfo{year}{1963}).

\bibitem[{\citenamefont{Omnes}(1994)}]{Omnes}
\bibinfo{author}{\bibfnamefont{R.}~\bibnamefont{Omnes}},
  \emph{\bibinfo{title}{The Interpretation of Quantum Mechanics}}
  (\bibinfo{publisher}{Princeton University Press}, \bibinfo{year}{1994}).

\bibitem[{\citenamefont{Maudlin}(1995)}]{Maudlin95}
\bibinfo{author}{\bibfnamefont{T.}~\bibnamefont{Maudlin}},
  \bibinfo{journal}{Topoi} \textbf{\bibinfo{volume}{14}}, \bibinfo{pages}{7}
  (\bibinfo{year}{1995}).

\bibitem[{\citenamefont{Becker}(2018)}]{Becker}
\bibinfo{author}{\bibfnamefont{A.}~\bibnamefont{Becker}},
  \emph{\bibinfo{title}{What is Real? The Unfinished Quest for the Meaning of
  Quantum Physics}} (\bibinfo{publisher}{Basic Books, New York},
  \bibinfo{year}{2018}).

\bibitem[{\citenamefont{Norsen}(2017)}]{Norsen}
\bibinfo{author}{\bibfnamefont{T.}~\bibnamefont{Norsen}},
  \emph{\bibinfo{title}{{Foundations of Quantum Mechanics}}}
  (\bibinfo{publisher}{Springer International Publishing AG},
  \bibinfo{year}{2017}).

\bibitem[{\citenamefont{Durr and Lazarovici}(2020)}]{Durr}
\bibinfo{author}{\bibfnamefont{D.}~\bibnamefont{Durr}} \bibnamefont{and}
  \bibinfo{author}{\bibfnamefont{D.}~\bibnamefont{Lazarovici}},
  \emph{\bibinfo{title}{{Understanding Quantum Mechanics}}}
  (\bibinfo{publisher}{Springer International Publishing AG},
  \bibinfo{year}{2020}).

\bibitem[{\citenamefont{Albert}(1994)}]{Albert}
\bibinfo{author}{\bibfnamefont{D.~Z.} \bibnamefont{Albert}},
  \bibinfo{journal}{\emph{Quantum Mechanics and Experience}, (Harvard
  University Press)}  (\bibinfo{year}{1994}).

\bibitem[{\citenamefont{Okon}(2014)}]{okon14}
\bibinfo{author}{\bibfnamefont{E.}~\bibnamefont{Okon}}, \bibinfo{journal}{Rev.
  Mex. Fis. E} \textbf{\bibinfo{volume}{60}}, \bibinfo{pages}{130}
  (\bibinfo{year}{2014}).

\bibitem[{\citenamefont{Bell}(1981)}]{Bell81}
\bibinfo{author}{\bibfnamefont{J.~S.} \bibnamefont{Bell}},
  \emph{\bibinfo{title}{Quantum Mechanics for cosmologists}}, Quantum Gravity 2
  (\bibinfo{publisher}{eds. Isham, C., Penrose, R. and Sciama, D., Oxford
  University Press}, \bibinfo{year}{1981}).

\bibitem[{\citenamefont{Perez et~al.}(2006)\citenamefont{Perez, Sahlmann, and
  Sudarsky}}]{PSS06}
\bibinfo{author}{\bibfnamefont{A.}~\bibnamefont{Perez}},
  \bibinfo{author}{\bibfnamefont{H.}~\bibnamefont{Sahlmann}}, \bibnamefont{and}
  \bibinfo{author}{\bibfnamefont{D.}~\bibnamefont{Sudarsky}},
  \bibinfo{journal}{Class. Quant. Grav.} \textbf{\bibinfo{volume}{23}},
  \bibinfo{pages}{2317} (\bibinfo{year}{2006}), \eprint{gr-qc/0508100}.

\bibitem[{\citenamefont{Sudarsky}(2011)}]{Sudarsky11}
\bibinfo{author}{\bibfnamefont{D.}~\bibnamefont{Sudarsky}},
  \bibinfo{journal}{Int. J. Mod. Phys. D} \textbf{\bibinfo{volume}{20}},
  \bibinfo{pages}{509} (\bibinfo{year}{2011}), \eprint{0906.0315}.

\bibitem[{\citenamefont{Landau et~al.}(2013)\citenamefont{Landau, Le\'on, and
  Sudarsky}}]{Susana13}
\bibinfo{author}{\bibfnamefont{S.}~\bibnamefont{Landau}},
  \bibinfo{author}{\bibfnamefont{G.}~\bibnamefont{Le\'on}}, \bibnamefont{and}
  \bibinfo{author}{\bibfnamefont{D.}~\bibnamefont{Sudarsky}},
  \bibinfo{journal}{Phys. Rev. D} \textbf{\bibinfo{volume}{88}},
  \bibinfo{pages}{023526} (\bibinfo{year}{2013}), \eprint{1107.3054}.

\bibitem[{\citenamefont{{Hartle}}(1993)}]{Hartle93}
\bibinfo{author}{\bibfnamefont{J.~B.} \bibnamefont{{Hartle}}},
  \bibinfo{journal}{arXiv e-prints} \bibinfo{eid}{gr-qc/9304006}
  (\bibinfo{year}{1993}), \eprint{gr-qc/9304006}.

\bibitem[{\citenamefont{Berjon et~al.}(2021)\citenamefont{Berjon, Okon, and
  Sudarsky}}]{Berjon21}
\bibinfo{author}{\bibfnamefont{J.}~\bibnamefont{Berjon}},
  \bibinfo{author}{\bibfnamefont{E.}~\bibnamefont{Okon}}, \bibnamefont{and}
  \bibinfo{author}{\bibfnamefont{D.}~\bibnamefont{Sudarsky}},
  \bibinfo{journal}{Phys. Rev. D} \textbf{\bibinfo{volume}{103}},
  \bibinfo{pages}{043521} (\bibinfo{year}{2021}), \eprint{2009.09999}.

\bibitem[{\citenamefont{Bengochea}(2020)}]{Bengochea20}
\bibinfo{author}{\bibfnamefont{G.~R.} \bibnamefont{Bengochea}},
  \bibinfo{journal}{Rev. Mex. Fis. E} \textbf{\bibinfo{volume}{17}},
  \bibinfo{pages}{263} (\bibinfo{year}{2020}), \eprint{2007.03428}.

\bibitem[{\citenamefont{Bohm}(1952)}]{bohm}
\bibinfo{author}{\bibfnamefont{D.}~\bibnamefont{Bohm}}, \bibinfo{journal}{Phys.
  Rev.} \textbf{\bibinfo{volume}{85}}, \bibinfo{pages}{166}
  (\bibinfo{year}{1952}).

\bibitem[{\citenamefont{Valentini}(2010)}]{Valentini08b}
\bibinfo{author}{\bibfnamefont{A.}~\bibnamefont{Valentini}},
  \bibinfo{journal}{Phys. Rev. D} \textbf{\bibinfo{volume}{82}},
  \bibinfo{pages}{063513} (\bibinfo{year}{2010}), \eprint{0805.0163}.

\bibitem[{\citenamefont{{Pinto-Neto} et~al.}(2012)\citenamefont{{Pinto-Neto},
  {Santos}, and {Struyve}}}]{Neto12}
\bibinfo{author}{\bibfnamefont{N.}~\bibnamefont{{Pinto-Neto}}},
  \bibinfo{author}{\bibfnamefont{G.}~\bibnamefont{{Santos}}}, \bibnamefont{and}
  \bibinfo{author}{\bibfnamefont{W.}~\bibnamefont{{Struyve}}},
  \bibinfo{journal}{\prd} \textbf{\bibinfo{volume}{85}}, \bibinfo{eid}{083506}
  (\bibinfo{year}{2012}), \eprint{1110.1339}.

\bibitem[{\citenamefont{{Goldstein} et~al.}(2015)\citenamefont{{Goldstein},
  {Struyve}, and {Tumulka}}}]{goldstein15}
\bibinfo{author}{\bibfnamefont{S.}~\bibnamefont{{Goldstein}}},
  \bibinfo{author}{\bibfnamefont{W.}~\bibnamefont{{Struyve}}},
  \bibnamefont{and}
  \bibinfo{author}{\bibfnamefont{R.}~\bibnamefont{{Tumulka}}},
  \bibinfo{journal}{arXiv e-prints} \bibinfo{eid}{arXiv:1508.01017}
  (\bibinfo{year}{2015}), \eprint{1508.01017}.

\bibitem[{\citenamefont{{Pinto-Neto} and {Struyve}}(2018)}]{Neto18}
\bibinfo{author}{\bibfnamefont{N.}~\bibnamefont{{Pinto-Neto}}}
  \bibnamefont{and}
  \bibinfo{author}{\bibfnamefont{W.}~\bibnamefont{{Struyve}}},
  \bibinfo{journal}{arXiv e-prints} \bibinfo{eid}{arXiv:1801.03353}
  (\bibinfo{year}{2018}), \eprint{1801.03353}.

\bibitem[{\citenamefont{Vitenti et~al.}(2019)\citenamefont{Vitenti, Peter, and
  Valentini}}]{Valentini19}
\bibinfo{author}{\bibfnamefont{S.~D.~P.} \bibnamefont{Vitenti}},
  \bibinfo{author}{\bibfnamefont{P.}~\bibnamefont{Peter}}, \bibnamefont{and}
  \bibinfo{author}{\bibfnamefont{A.}~\bibnamefont{Valentini}},
  \bibinfo{journal}{Phys. Rev. D} \textbf{\bibinfo{volume}{100}},
  \bibinfo{pages}{043506} (\bibinfo{year}{2019}), \eprint{1901.08885}.

\bibitem[{\citenamefont{Kiefer and Polarski}(2009)}]{kiefer09}
\bibinfo{author}{\bibfnamefont{C.}~\bibnamefont{Kiefer}} \bibnamefont{and}
  \bibinfo{author}{\bibfnamefont{D.}~\bibnamefont{Polarski}},
  \bibinfo{journal}{Adv. Sci. Lett.} \textbf{\bibinfo{volume}{2}},
  \bibinfo{pages}{164} (\bibinfo{year}{2009}), \eprint{0810.0087}.

\bibitem[{\citenamefont{Halliwell}(1989)}]{halliwell}
\bibinfo{author}{\bibfnamefont{J.~J.} \bibnamefont{Halliwell}},
  \bibinfo{journal}{Phys. Rev.} \textbf{\bibinfo{volume}{D39}},
  \bibinfo{pages}{2912} (\bibinfo{year}{1989}).

\bibitem[{\citenamefont{Kiefer}(2000)}]{kiefer2}
\bibinfo{author}{\bibfnamefont{C.}~\bibnamefont{Kiefer}},
  \bibinfo{journal}{Nucl. Phys. Proc. Suppl.} \textbf{\bibinfo{volume}{88}},
  \bibinfo{pages}{255} (\bibinfo{year}{2000}), \eprint{astro-ph/0006252}.

\bibitem[{\citenamefont{Polarski and Starobinsky}(1996)}]{polarski}
\bibinfo{author}{\bibfnamefont{D.}~\bibnamefont{Polarski}} \bibnamefont{and}
  \bibinfo{author}{\bibfnamefont{A.~A.} \bibnamefont{Starobinsky}},
  \bibinfo{journal}{Class. Quant. Grav.} \textbf{\bibinfo{volume}{13}},
  \bibinfo{pages}{377} (\bibinfo{year}{1996}), \eprint{gr-qc/9504030}.

\bibitem[{\citenamefont{Okon and Sudarsky}(2016)}]{okon16}
\bibinfo{author}{\bibfnamefont{E.}~\bibnamefont{Okon}} \bibnamefont{and}
  \bibinfo{author}{\bibfnamefont{D.}~\bibnamefont{Sudarsky}},
  \bibinfo{journal}{Found. Phys.} \textbf{\bibinfo{volume}{46}},
  \bibinfo{pages}{852} (\bibinfo{year}{2016}), \eprint{1512.05298}.

\bibitem[{\citenamefont{{Adler}}(2003)}]{Adler01}
\bibinfo{author}{\bibfnamefont{S.~L.} \bibnamefont{{Adler}}},
  \bibinfo{journal}{Studies in History and Philosophy of Modern Physics}
  \textbf{\bibinfo{volume}{34}}, \bibinfo{pages}{135} (\bibinfo{year}{2003}),
  \eprint{quant-ph/0112095}.

\bibitem[{\citenamefont{Schlosshauer}(2004)}]{schlosshauer}
\bibinfo{author}{\bibfnamefont{M.}~\bibnamefont{Schlosshauer}},
  \bibinfo{journal}{Rev. Mod. Phys.} \textbf{\bibinfo{volume}{76}},
  \bibinfo{pages}{1267} (\bibinfo{year}{2004}), \eprint{quant-ph/0312059}.

\bibitem[{\citenamefont{Everett}(1957)}]{Everett}
\bibinfo{author}{\bibfnamefont{H.}~\bibnamefont{Everett}},
  \bibinfo{journal}{Rev. Mod. Phys.} \textbf{\bibinfo{volume}{29}},
  \bibinfo{pages}{454} (\bibinfo{year}{1957}).

\bibitem[{\citenamefont{Mukhanov}(2005)}]{mukhanov2005}
\bibinfo{author}{\bibfnamefont{V.}~\bibnamefont{Mukhanov}},
  \emph{\bibinfo{title}{Physical Foundations of Cosmology}}
  (\bibinfo{publisher}{New York: Cambridge University Press},
  \bibinfo{year}{2005}).

\bibitem[{\citenamefont{{Kent}}(1990)}]{kent}
\bibinfo{author}{\bibfnamefont{A.}~\bibnamefont{{Kent}}},
  \bibinfo{journal}{International Journal of Modern Physics A}
  \textbf{\bibinfo{volume}{5}}, \bibinfo{pages}{1745} (\bibinfo{year}{1990}),
  \eprint{gr-qc/9703089}.

\bibitem[{\citenamefont{{Stapp}}(2002)}]{stapp}
\bibinfo{author}{\bibfnamefont{H.~P.} \bibnamefont{{Stapp}}},
  \bibinfo{journal}{Canadian Journal of Physics} \textbf{\bibinfo{volume}{80}},
  \bibinfo{pages}{1043} (\bibinfo{year}{2002}), \eprint{quant-ph/0110148}.

\bibitem[{\citenamefont{Pearle}(1976)}]{Pearle76}
\bibinfo{author}{\bibfnamefont{P.}~\bibnamefont{Pearle}},
  \bibinfo{journal}{Phys. Rev. D} \textbf{\bibinfo{volume}{13}},
  \bibinfo{pages}{857} (\bibinfo{year}{1976}).

\bibitem[{\citenamefont{Ghirardi et~al.}(1986)\citenamefont{Ghirardi, Rimini,
  and Weber}}]{Ghirardi86}
\bibinfo{author}{\bibfnamefont{G.}~\bibnamefont{Ghirardi}},
  \bibinfo{author}{\bibfnamefont{A.}~\bibnamefont{Rimini}}, \bibnamefont{and}
  \bibinfo{author}{\bibfnamefont{T.}~\bibnamefont{Weber}},
  \bibinfo{journal}{Phys.Rev.} \textbf{\bibinfo{volume}{D34}},
  \bibinfo{pages}{470} (\bibinfo{year}{1986}).

\bibitem[{\citenamefont{Pearle}(1989)}]{Pearle89}
\bibinfo{author}{\bibfnamefont{P.~M.} \bibnamefont{Pearle}},
  \bibinfo{journal}{Phys.Rev.} \textbf{\bibinfo{volume}{A39}},
  \bibinfo{pages}{2277} (\bibinfo{year}{1989}).

\bibitem[{\citenamefont{Diosi}(1987)}]{Diosi87}
\bibinfo{author}{\bibfnamefont{L.}~\bibnamefont{Diosi}},
  \bibinfo{journal}{Phys.Lett.} \textbf{\bibinfo{volume}{A120}},
  \bibinfo{pages}{377} (\bibinfo{year}{1987}).

\bibitem[{\citenamefont{Diosi}(1989)}]{Diosi89}
\bibinfo{author}{\bibfnamefont{L.}~\bibnamefont{Diosi}},
  \bibinfo{journal}{Phys.Rev.} \textbf{\bibinfo{volume}{A40}},
  \bibinfo{pages}{1165} (\bibinfo{year}{1989}).

\bibitem[{\citenamefont{Penrose}(1996)}]{Penrose96}
\bibinfo{author}{\bibfnamefont{R.}~\bibnamefont{Penrose}},
  \bibinfo{journal}{Gen.Rel.Grav.} \textbf{\bibinfo{volume}{28}},
  \bibinfo{pages}{581} (\bibinfo{year}{1996}).

\bibitem[{\citenamefont{Okon and Sudarsky}(2014)}]{Beneficios}
\bibinfo{author}{\bibfnamefont{E.}~\bibnamefont{Okon}} \bibnamefont{and}
  \bibinfo{author}{\bibfnamefont{D.}~\bibnamefont{Sudarsky}},
  \bibinfo{journal}{Found. Phys.} \textbf{\bibinfo{volume}{44}},
  \bibinfo{pages}{114} (\bibinfo{year}{2014}), \eprint{1309.1730}.

\bibitem[{\citenamefont{{Bassi} and {Ghirardi}}(2003)}]{Bassi1}
\bibinfo{author}{\bibfnamefont{A.}~\bibnamefont{{Bassi}}} \bibnamefont{and}
  \bibinfo{author}{\bibfnamefont{G.}~\bibnamefont{{Ghirardi}}},
  \bibinfo{journal}{Phys. Rept.} \textbf{\bibinfo{volume}{379}},
  \bibinfo{pages}{257} (\bibinfo{year}{2003}), \eprint{quant-ph/0302164}.

\bibitem[{\citenamefont{{Bassi} et~al.}(2013)\citenamefont{{Bassi}, {Lochan},
  {Satin}, {Singh}, and {Ulbricht}}}]{Bassi2}
\bibinfo{author}{\bibfnamefont{A.}~\bibnamefont{{Bassi}}},
  \bibinfo{author}{\bibfnamefont{K.}~\bibnamefont{{Lochan}}},
  \bibinfo{author}{\bibfnamefont{S.}~\bibnamefont{{Satin}}},
  \bibinfo{author}{\bibfnamefont{T.~P.} \bibnamefont{{Singh}}},
  \bibnamefont{and}
  \bibinfo{author}{\bibfnamefont{H.}~\bibnamefont{{Ulbricht}}},
  \bibinfo{journal}{Reviews of Modern Physics} \textbf{\bibinfo{volume}{85}},
  \bibinfo{pages}{471} (\bibinfo{year}{2013}), \eprint{1204.4325}.

\bibitem[{\citenamefont{Leon and Sudarsky}(2010)}]{Daniel10}
\bibinfo{author}{\bibfnamefont{G.}~\bibnamefont{Leon}} \bibnamefont{and}
  \bibinfo{author}{\bibfnamefont{D.}~\bibnamefont{Sudarsky}},
  \bibinfo{journal}{Class. Quant. Grav.} \textbf{\bibinfo{volume}{27}},
  \bibinfo{pages}{225017} (\bibinfo{year}{2010}), \eprint{1003.5950}.

\bibitem[{\citenamefont{Diez-Tejedor and Sudarsky}(2012)}]{Tejedor12}
\bibinfo{author}{\bibfnamefont{A.}~\bibnamefont{Diez-Tejedor}}
  \bibnamefont{and} \bibinfo{author}{\bibfnamefont{D.}~\bibnamefont{Sudarsky}},
  \bibinfo{journal}{JCAP} \textbf{\bibinfo{volume}{07}}, \bibinfo{pages}{045}
  (\bibinfo{year}{2012}), \eprint{1108.4928}.

\bibitem[{\citenamefont{{Diez-Tejedor}
  et~al.}(2012)\citenamefont{{Diez-Tejedor}, {Le{\'o}n}, and
  {Sudarsky}}}]{Tejedor12B}
\bibinfo{author}{\bibfnamefont{A.}~\bibnamefont{{Diez-Tejedor}}},
  \bibinfo{author}{\bibfnamefont{G.}~\bibnamefont{{Le{\'o}n}}},
  \bibnamefont{and}
  \bibinfo{author}{\bibfnamefont{D.}~\bibnamefont{{Sudarsky}}},
  \bibinfo{journal}{General Relativity and Gravitation}
  \textbf{\bibinfo{volume}{44}}, \bibinfo{pages}{2965} (\bibinfo{year}{2012}),
  \eprint{1106.1176}.

\bibitem[{\citenamefont{{Ca{\~n}ate} et~al.}(2013)\citenamefont{{Ca{\~n}ate},
  {Pearle}, and {Sudarsky}}}]{Pedro13}
\bibinfo{author}{\bibfnamefont{P.}~\bibnamefont{{Ca{\~n}ate}}},
  \bibinfo{author}{\bibfnamefont{P.}~\bibnamefont{{Pearle}}}, \bibnamefont{and}
  \bibinfo{author}{\bibfnamefont{D.}~\bibnamefont{{Sudarsky}}},
  \bibinfo{journal}{Phys.Rev. D} \textbf{\bibinfo{volume}{87}},
  \bibinfo{eid}{104024} (\bibinfo{year}{2013}), \eprint{1211.3463}.

\bibitem[{\citenamefont{Bengochea et~al.}(2015)\citenamefont{Bengochea,
  {Ca{\~n}ate}, and Sudarsky}}]{Bengochea15}
\bibinfo{author}{\bibfnamefont{G.~R.} \bibnamefont{Bengochea}},
  \bibinfo{author}{\bibfnamefont{P.}~\bibnamefont{{Ca{\~n}ate}}},
  \bibnamefont{and} \bibinfo{author}{\bibfnamefont{D.}~\bibnamefont{Sudarsky}},
  \bibinfo{journal}{Phys. Lett.} \textbf{\bibinfo{volume}{B743}},
  \bibinfo{pages}{484} (\bibinfo{year}{2015}), \eprint{1410.4212}.

\bibitem[{\citenamefont{Le\'on and Sudarsky}(2015)}]{Leon15}
\bibinfo{author}{\bibfnamefont{G.}~\bibnamefont{Le\'on}} \bibnamefont{and}
  \bibinfo{author}{\bibfnamefont{D.}~\bibnamefont{Sudarsky}},
  \bibinfo{journal}{JCAP} \textbf{\bibinfo{volume}{06}}, \bibinfo{pages}{020}
  (\bibinfo{year}{2015}), \eprint{1503.01417}.

\bibitem[{\citenamefont{Leon and Bengochea}(2016)}]{Leon16}
\bibinfo{author}{\bibfnamefont{G.}~\bibnamefont{Leon}} \bibnamefont{and}
  \bibinfo{author}{\bibfnamefont{G.~R.} \bibnamefont{Bengochea}},
  \bibinfo{journal}{Eur. Phys. J.} \textbf{\bibinfo{volume}{C76}},
  \bibinfo{pages}{29} (\bibinfo{year}{2016}), \eprint{1502.04907}.

\bibitem[{\citenamefont{{Le{\'o}n}}(2017)}]{Leon17}
\bibinfo{author}{\bibfnamefont{G.}~\bibnamefont{{Le{\'o}n}}},
  \bibinfo{journal}{European Physical Journal C} \textbf{\bibinfo{volume}{77}},
  \bibinfo{eid}{705} (\bibinfo{year}{2017}), \eprint{1705.03958}.

\bibitem[{\citenamefont{{Landau} et~al.}(2012)\citenamefont{{Landau},
  {Sc{\'o}ccola}, and {Sudarsky}}}]{Landau12}
\bibinfo{author}{\bibfnamefont{S.~J.} \bibnamefont{{Landau}}},
  \bibinfo{author}{\bibfnamefont{C.~G.} \bibnamefont{{Sc{\'o}ccola}}},
  \bibnamefont{and}
  \bibinfo{author}{\bibfnamefont{D.}~\bibnamefont{{Sudarsky}}},
  \bibinfo{journal}{Physical Review D} \textbf{\bibinfo{volume}{85}},
  \bibinfo{eid}{123001} (\bibinfo{year}{2012}), \eprint{1112.1830}.

\bibitem[{\citenamefont{Benetti et~al.}(2016)\citenamefont{Benetti, Landau, and
  Alcaniz}}]{Benetti16}
\bibinfo{author}{\bibfnamefont{M.}~\bibnamefont{Benetti}},
  \bibinfo{author}{\bibfnamefont{S.~J.} \bibnamefont{Landau}},
  \bibnamefont{and} \bibinfo{author}{\bibfnamefont{J.~S.}
  \bibnamefont{Alcaniz}}, \bibinfo{journal}{JCAP}
  \textbf{\bibinfo{volume}{12}}, \bibinfo{pages}{035} (\bibinfo{year}{2016}),
  \eprint{1610.03091}.

\bibitem[{\citenamefont{Bengochea and Le\'on}(2017)}]{Bengo17}
\bibinfo{author}{\bibfnamefont{G.~R.} \bibnamefont{Bengochea}}
  \bibnamefont{and} \bibinfo{author}{\bibfnamefont{G.}~\bibnamefont{Le\'on}},
  \bibinfo{journal}{Phys. Lett. B} \textbf{\bibinfo{volume}{774}},
  \bibinfo{pages}{338} (\bibinfo{year}{2017}), \eprint{1708.07527}.

\bibitem[{\citenamefont{Ca\~nate et~al.}(2018)\citenamefont{Ca\~nate, Ramirez,
  and Sudarsky}}]{Pedro18}
\bibinfo{author}{\bibfnamefont{P.}~\bibnamefont{Ca\~nate}},
  \bibinfo{author}{\bibfnamefont{E.}~\bibnamefont{Ramirez}}, \bibnamefont{and}
  \bibinfo{author}{\bibfnamefont{D.}~\bibnamefont{Sudarsky}},
  \bibinfo{journal}{JCAP} \textbf{\bibinfo{volume}{08}}, \bibinfo{pages}{043}
  (\bibinfo{year}{2018}), \eprint{1802.02238}.

\bibitem[{\citenamefont{Ju\'arez-Aubry
  et~al.}(2018)\citenamefont{Ju\'arez-Aubry, Kay, and Sudarsky}}]{Benito18}
\bibinfo{author}{\bibfnamefont{B.~A.} \bibnamefont{Ju\'arez-Aubry}},
  \bibinfo{author}{\bibfnamefont{B.~S.} \bibnamefont{Kay}}, \bibnamefont{and}
  \bibinfo{author}{\bibfnamefont{D.}~\bibnamefont{Sudarsky}},
  \bibinfo{journal}{Phys. Rev. D} \textbf{\bibinfo{volume}{97}},
  \bibinfo{pages}{025010} (\bibinfo{year}{2018}), \eprint{1708.09371}.

\bibitem[{\citenamefont{{Piccirilli} et~al.}(2019)\citenamefont{{Piccirilli},
  {Le{\'o}n}, {Landau}, {Benetti}, and {Sudarsky}}}]{Picci19}
\bibinfo{author}{\bibfnamefont{M.~P.} \bibnamefont{{Piccirilli}}},
  \bibinfo{author}{\bibfnamefont{G.}~\bibnamefont{{Le{\'o}n}}},
  \bibinfo{author}{\bibfnamefont{S.~J.} \bibnamefont{{Landau}}},
  \bibinfo{author}{\bibfnamefont{M.}~\bibnamefont{{Benetti}}},
  \bibnamefont{and}
  \bibinfo{author}{\bibfnamefont{D.}~\bibnamefont{{Sudarsky}}},
  \bibinfo{journal}{International Journal of Modern Physics D}
  \textbf{\bibinfo{volume}{28}}, \bibinfo{pages}{1950041}
  (\bibinfo{year}{2019}), \eprint{1709.06237}.

\bibitem[{\citenamefont{Le\'on et~al.}(2015)\citenamefont{Le\'on, Kraiselburd,
  and Landau}}]{Lucila15}
\bibinfo{author}{\bibfnamefont{G.}~\bibnamefont{Le\'on}},
  \bibinfo{author}{\bibfnamefont{L.}~\bibnamefont{Kraiselburd}},
  \bibnamefont{and} \bibinfo{author}{\bibfnamefont{S.~J.}
  \bibnamefont{Landau}}, \bibinfo{journal}{Phys. Rev. D}
  \textbf{\bibinfo{volume}{92}}, \bibinfo{pages}{083516}
  (\bibinfo{year}{2015}), \eprint{1509.08399}.

\bibitem[{\citenamefont{{Mariani} et~al.}(2016)\citenamefont{{Mariani},
  {Bengochea}, and {Le{\'o}n}}}]{Mariani16}
\bibinfo{author}{\bibfnamefont{M.}~\bibnamefont{{Mariani}}},
  \bibinfo{author}{\bibfnamefont{G.~R.} \bibnamefont{{Bengochea}}},
  \bibnamefont{and}
  \bibinfo{author}{\bibfnamefont{G.}~\bibnamefont{{Le{\'o}n}}},
  \bibinfo{journal}{Physics Letters B} \textbf{\bibinfo{volume}{752}},
  \bibinfo{pages}{344} (\bibinfo{year}{2016}), \eprint{1412.6471}.

\bibitem[{\citenamefont{Le\'on et~al.}(2017)\citenamefont{Le\'on, Majhi, Okon,
  and Sudarsky}}]{Maj17}
\bibinfo{author}{\bibfnamefont{G.}~\bibnamefont{Le\'on}},
  \bibinfo{author}{\bibfnamefont{A.}~\bibnamefont{Majhi}},
  \bibinfo{author}{\bibfnamefont{E.}~\bibnamefont{Okon}}, \bibnamefont{and}
  \bibinfo{author}{\bibfnamefont{D.}~\bibnamefont{Sudarsky}},
  \bibinfo{journal}{Phys. Rev. D} \textbf{\bibinfo{volume}{96}},
  \bibinfo{pages}{101301} (\bibinfo{year}{2017}), \eprint{1607.03523}.

\bibitem[{\citenamefont{{Le{\'o}n} et~al.}(2018)\citenamefont{{Le{\'o}n},
  {Majhi}, {Okon}, and {Sudarsky}}}]{ModosB}
\bibinfo{author}{\bibfnamefont{G.}~\bibnamefont{{Le{\'o}n}}},
  \bibinfo{author}{\bibfnamefont{A.}~\bibnamefont{{Majhi}}},
  \bibinfo{author}{\bibfnamefont{E.}~\bibnamefont{{Okon}}}, \bibnamefont{and}
  \bibinfo{author}{\bibfnamefont{D.}~\bibnamefont{{Sudarsky}}},
  \bibinfo{journal}{\prd} \textbf{\bibinfo{volume}{98}}, \bibinfo{eid}{023512}
  (\bibinfo{year}{2018}), \eprint{1712.02435}.

\bibitem[{\citenamefont{{Le{\'o}n} et~al.}(2016)\citenamefont{{Le{\'o}n},
  {Bengochea}, and {Landau}}}]{Bouncing16}
\bibinfo{author}{\bibfnamefont{G.}~\bibnamefont{{Le{\'o}n}}},
  \bibinfo{author}{\bibfnamefont{G.~R.} \bibnamefont{{Bengochea}}},
  \bibnamefont{and} \bibinfo{author}{\bibfnamefont{S.~J.}
  \bibnamefont{{Landau}}}, \bibinfo{journal}{European Physical Journal C}
  \textbf{\bibinfo{volume}{76}}, \bibinfo{eid}{407} (\bibinfo{year}{2016}),
  \eprint{1605.03632}.

\bibitem[{\citenamefont{{Josset} et~al.}(2017)\citenamefont{{Josset}, {Perez},
  and {Sudarsky}}}]{Josset17}
\bibinfo{author}{\bibfnamefont{T.}~\bibnamefont{{Josset}}},
  \bibinfo{author}{\bibfnamefont{A.}~\bibnamefont{{Perez}}}, \bibnamefont{and}
  \bibinfo{author}{\bibfnamefont{D.}~\bibnamefont{{Sudarsky}}},
  \bibinfo{journal}{\prl} \textbf{\bibinfo{volume}{118}}, \bibinfo{eid}{021102}
  (\bibinfo{year}{2017}), \eprint{1604.04183}.

\bibitem[{\citenamefont{Leon and Piccirilli}(2020)}]{Leon2020}
\bibinfo{author}{\bibfnamefont{G.}~\bibnamefont{Leon}} \bibnamefont{and}
  \bibinfo{author}{\bibfnamefont{M.~P.} \bibnamefont{Piccirilli}},
  \bibinfo{journal}{Phys. Rev. D} \textbf{\bibinfo{volume}{102}},
  \bibinfo{pages}{043515} (\bibinfo{year}{2020}), \eprint{2006.03092}.

\bibitem[{\citenamefont{{Martin} et~al.}(2012)\citenamefont{{Martin}, {Vennin},
  and {Peter}}}]{Martin12}
\bibinfo{author}{\bibfnamefont{J.}~\bibnamefont{{Martin}}},
  \bibinfo{author}{\bibfnamefont{V.}~\bibnamefont{{Vennin}}}, \bibnamefont{and}
  \bibinfo{author}{\bibfnamefont{P.}~\bibnamefont{{Peter}}},
  \bibinfo{journal}{\prd} \textbf{\bibinfo{volume}{86}}, \bibinfo{eid}{103524}
  (\bibinfo{year}{2012}), \eprint{1207.2086}.

\bibitem[{\citenamefont{{Das} et~al.}(2013)\citenamefont{{Das}, {Lochan},
  {Sahu}, and {Singh}}}]{Das13}
\bibinfo{author}{\bibfnamefont{S.}~\bibnamefont{{Das}}},
  \bibinfo{author}{\bibfnamefont{K.}~\bibnamefont{{Lochan}}},
  \bibinfo{author}{\bibfnamefont{S.}~\bibnamefont{{Sahu}}}, \bibnamefont{and}
  \bibinfo{author}{\bibfnamefont{T.~P.} \bibnamefont{{Singh}}},
  \bibinfo{journal}{\prd} \textbf{\bibinfo{volume}{88}}, \bibinfo{eid}{085020}
  (\bibinfo{year}{2013}), \eprint{1304.5094}.

\bibitem[{\citenamefont{Markkanen et~al.}(2015)\citenamefont{Markkanen,
  Rasanen, and Wahlman}}]{Syksy15}
\bibinfo{author}{\bibfnamefont{T.}~\bibnamefont{Markkanen}},
  \bibinfo{author}{\bibfnamefont{S.}~\bibnamefont{Rasanen}}, \bibnamefont{and}
  \bibinfo{author}{\bibfnamefont{P.}~\bibnamefont{Wahlman}},
  \bibinfo{journal}{Phys. Rev. D} \textbf{\bibinfo{volume}{91}},
  \bibinfo{pages}{084064} (\bibinfo{year}{2015}), \eprint{1407.4691}.

\bibitem[{\citenamefont{Alexander et~al.}(2016)\citenamefont{Alexander, Jyoti,
  and Magueijo}}]{Stephon16}
\bibinfo{author}{\bibfnamefont{S.}~\bibnamefont{Alexander}},
  \bibinfo{author}{\bibfnamefont{D.}~\bibnamefont{Jyoti}}, \bibnamefont{and}
  \bibinfo{author}{\bibfnamefont{J.}~\bibnamefont{Magueijo}},
  \bibinfo{journal}{Phys. Rev. D} \textbf{\bibinfo{volume}{94}},
  \bibinfo{pages}{043502} (\bibinfo{year}{2016}), \eprint{1602.01216}.

\bibitem[{\citenamefont{{Drossel} and {Ellis}}(2018)}]{Ellis18}
\bibinfo{author}{\bibfnamefont{B.}~\bibnamefont{{Drossel}}} \bibnamefont{and}
  \bibinfo{author}{\bibfnamefont{G.}~\bibnamefont{{Ellis}}},
  \bibinfo{journal}{New Journal of Physics} \textbf{\bibinfo{volume}{20}},
  \bibinfo{eid}{113025} (\bibinfo{year}{2018}), \eprint{1807.08171}.

\bibitem[{\citenamefont{Martin and Vennin}(2020)}]{MartinShadow}
\bibinfo{author}{\bibfnamefont{J.}~\bibnamefont{Martin}} \bibnamefont{and}
  \bibinfo{author}{\bibfnamefont{V.}~\bibnamefont{Vennin}},
  \bibinfo{journal}{Phys. Rev. Lett.} \textbf{\bibinfo{volume}{124}},
  \bibinfo{pages}{080402} (\bibinfo{year}{2020}), \eprint{1906.04405}.

\bibitem[{\citenamefont{Bengochea
  et~al.}(2020{\natexlab{a}})\citenamefont{Bengochea, Leon, Pearle, and
  Sudarsky}}]{Bengo20Letter}
\bibinfo{author}{\bibfnamefont{G.~R.} \bibnamefont{Bengochea}},
  \bibinfo{author}{\bibfnamefont{G.}~\bibnamefont{Leon}},
  \bibinfo{author}{\bibfnamefont{P.}~\bibnamefont{Pearle}}, \bibnamefont{and}
  \bibinfo{author}{\bibfnamefont{D.}~\bibnamefont{Sudarsky}}
  (\bibinfo{year}{2020}{\natexlab{a}}), \eprint{2006.05313}.

\bibitem[{\citenamefont{Bengochea
  et~al.}(2020{\natexlab{b}})\citenamefont{Bengochea, Le\'on, Pearle, and
  Sudarsky}}]{Bengo20Long}
\bibinfo{author}{\bibfnamefont{G.~R.} \bibnamefont{Bengochea}},
  \bibinfo{author}{\bibfnamefont{G.}~\bibnamefont{Le\'on}},
  \bibinfo{author}{\bibfnamefont{P.}~\bibnamefont{Pearle}}, \bibnamefont{and}
  \bibinfo{author}{\bibfnamefont{D.}~\bibnamefont{Sudarsky}},
  \bibinfo{journal}{Eur. Phys. J. C} \textbf{\bibinfo{volume}{80}},
  \bibinfo{pages}{1021} (\bibinfo{year}{2020}{\natexlab{b}}),
  \eprint{2008.05285}.

\bibitem[{\citenamefont{Martin and Vennin}(2021{\natexlab{a}})}]{Martin20R}
\bibinfo{author}{\bibfnamefont{J.}~\bibnamefont{Martin}} \bibnamefont{and}
  \bibinfo{author}{\bibfnamefont{V.}~\bibnamefont{Vennin}},
  \bibinfo{journal}{Eur. Phys. J. C} \textbf{\bibinfo{volume}{81}},
  \bibinfo{pages}{64} (\bibinfo{year}{2021}{\natexlab{a}}),
  \eprint{2010.04067}.

\bibitem[{\citenamefont{Gundhi et~al.}(2021)\citenamefont{Gundhi, Gaona-Reyes,
  Carlesso, and Bassi}}]{Bassi21}
\bibinfo{author}{\bibfnamefont{A.}~\bibnamefont{Gundhi}},
  \bibinfo{author}{\bibfnamefont{J.~L.} \bibnamefont{Gaona-Reyes}},
  \bibinfo{author}{\bibfnamefont{M.}~\bibnamefont{Carlesso}}, \bibnamefont{and}
  \bibinfo{author}{\bibfnamefont{A.}~\bibnamefont{Bassi}},
  \bibinfo{journal}{Phys. Rev. Lett.} \textbf{\bibinfo{volume}{127}},
  \bibinfo{pages}{091302} (\bibinfo{year}{2021}), \eprint{2102.07688}.

\bibitem[{\citenamefont{Martin and Vennin}(2021{\natexlab{b}})}]{Martin21}
\bibinfo{author}{\bibfnamefont{J.}~\bibnamefont{Martin}} \bibnamefont{and}
  \bibinfo{author}{\bibfnamefont{V.}~\bibnamefont{Vennin}},
  \bibinfo{journal}{Eur. Phys. J. C} \textbf{\bibinfo{volume}{81}},
  \bibinfo{pages}{516} (\bibinfo{year}{2021}{\natexlab{b}}),
  \eprint{2103.01697}.

\bibitem[{\citenamefont{Le\'on and Bengochea}(2021)}]{GLGB21}
\bibinfo{author}{\bibfnamefont{G.}~\bibnamefont{Le\'on}} \bibnamefont{and}
  \bibinfo{author}{\bibfnamefont{G.~R.} \bibnamefont{Bengochea}},
  \bibinfo{journal}{Eur. Phys. J. C} \textbf{\bibinfo{volume}{81}},
  \bibinfo{pages}{1055} (\bibinfo{year}{2021}), \eprint{2107.05470}.

\bibitem[{\citenamefont{Ju\'arez-Aubry
  et~al.}(2020)\citenamefont{Ju\'arez-Aubry, Miramontes, and
  Sudarsky}}]{Benito20}
\bibinfo{author}{\bibfnamefont{B.~A.} \bibnamefont{Ju\'arez-Aubry}},
  \bibinfo{author}{\bibfnamefont{T.}~\bibnamefont{Miramontes}},
  \bibnamefont{and} \bibinfo{author}{\bibfnamefont{D.}~\bibnamefont{Sudarsky}},
  \bibinfo{journal}{J. Math. Phys.} \textbf{\bibinfo{volume}{61}},
  \bibinfo{pages}{032301} (\bibinfo{year}{2020}), \eprint{1907.09960}.

\bibitem[{\citenamefont{Eppley and Hannah}(1977)}]{Eppley77}
\bibinfo{author}{\bibfnamefont{K.}~\bibnamefont{Eppley}} \bibnamefont{and}
  \bibinfo{author}{\bibfnamefont{E.}~\bibnamefont{Hannah}},
  \bibinfo{journal}{Foundations of Physics} \textbf{\bibinfo{volume}{7}},
  \bibinfo{pages}{51} (\bibinfo{year}{1977}).

\bibitem[{\citenamefont{Page and Geilker}(1981)}]{Page81}
\bibinfo{author}{\bibfnamefont{D.~N.} \bibnamefont{Page}} \bibnamefont{and}
  \bibinfo{author}{\bibfnamefont{C.~D.} \bibnamefont{Geilker}},
  \bibinfo{journal}{Phys. Rev. Lett.} \textbf{\bibinfo{volume}{47}},
  \bibinfo{pages}{979} (\bibinfo{year}{1981}).

\bibitem[{\citenamefont{{Mattingly}}(2005)}]{Mattingly05}
\bibinfo{author}{\bibfnamefont{J.}~\bibnamefont{{Mattingly}}},
  \emph{\bibinfo{title}{{Is Quantum Gravity Necessary?}}}
  (\bibinfo{year}{2005}), vol.~\bibinfo{volume}{11}, pp.
  \bibinfo{pages}{327--338}.

\bibitem[{\citenamefont{Mattingly}(2006)}]{Mattingly06}
\bibinfo{author}{\bibfnamefont{J.}~\bibnamefont{Mattingly}},
  \bibinfo{journal}{Phys. Rev.} \textbf{\bibinfo{volume}{D73}},
  \bibinfo{pages}{064025} (\bibinfo{year}{2006}), \eprint{gr-qc/0601127}.

\bibitem[{\citenamefont{Kent}(2018)}]{Kent18}
\bibinfo{author}{\bibfnamefont{A.}~\bibnamefont{Kent}},
  \bibinfo{journal}{Class. Quant. Grav.} \textbf{\bibinfo{volume}{35}},
  \bibinfo{pages}{245008} (\bibinfo{year}{2018}), \eprint{1807.08708}.

\bibitem[{\citenamefont{Tilloy and Di\'{o}si}(2016)}]{Tilloy16}
\bibinfo{author}{\bibfnamefont{A.}~\bibnamefont{Tilloy}} \bibnamefont{and}
  \bibinfo{author}{\bibfnamefont{L.}~\bibnamefont{Di\'{o}si}},
  \bibinfo{journal}{Phys. Rev.} \textbf{\bibinfo{volume}{D93}},
  \bibinfo{pages}{024026} (\bibinfo{year}{2016}), \eprint{1509.08705}.

\bibitem[{\citenamefont{Carlip}(2008)}]{Carlip08}
\bibinfo{author}{\bibfnamefont{S.}~\bibnamefont{Carlip}},
  \bibinfo{journal}{Class. Quant. Grav.} \textbf{\bibinfo{volume}{25}},
  \bibinfo{pages}{154010} (\bibinfo{year}{2008}), \eprint{0803.3456}.

\bibitem[{\citenamefont{Albers et~al.}(2008)\citenamefont{Albers, Kiefer, and
  Reginatto}}]{Albers08}
\bibinfo{author}{\bibfnamefont{M.}~\bibnamefont{Albers}},
  \bibinfo{author}{\bibfnamefont{C.}~\bibnamefont{Kiefer}}, \bibnamefont{and}
  \bibinfo{author}{\bibfnamefont{M.}~\bibnamefont{Reginatto}},
  \bibinfo{journal}{Phys. Rev.} \textbf{\bibinfo{volume}{D78}},
  \bibinfo{pages}{064051} (\bibinfo{year}{2008}), \eprint{0802.1978}.

\bibitem[{\citenamefont{Ford}(2005)}]{Ford05Review}
\bibinfo{author}{\bibfnamefont{L.~H.} \bibnamefont{Ford}},
  \bibinfo{journal}{100 Years Of Relativity : space-time structure: Einstein
  and beyond} pp. \bibinfo{pages}{293--310} (\bibinfo{year}{2005}),
  \eprint{gr-qc/0504096}.

\bibitem[{\citenamefont{{Ford}}(2005)}]{Ford05ReviewBook}
\bibinfo{author}{\bibfnamefont{L.~H.} \bibnamefont{{Ford}}},
  \emph{\bibinfo{title}{{Spacetime in Semiclassical Gravity}}}
  (\bibinfo{year}{2005}), pp. \bibinfo{pages}{293--310}.

\bibitem[{\citenamefont{{Piscicchia} et~al.}(2017)\citenamefont{{Piscicchia},
  {Bassi}, {Curceanu}, {Grande}, {Donadi}, {Hiesmayr}, and
  {Pichler}}}]{sandro2017}
\bibinfo{author}{\bibfnamefont{K.}~\bibnamefont{{Piscicchia}}},
  \bibinfo{author}{\bibfnamefont{A.}~\bibnamefont{{Bassi}}},
  \bibinfo{author}{\bibfnamefont{C.}~\bibnamefont{{Curceanu}}},
  \bibinfo{author}{\bibfnamefont{R.}~\bibnamefont{{Grande}}},
  \bibinfo{author}{\bibfnamefont{S.}~\bibnamefont{{Donadi}}},
  \bibinfo{author}{\bibfnamefont{B.}~\bibnamefont{{Hiesmayr}}},
  \bibnamefont{and}
  \bibinfo{author}{\bibfnamefont{A.}~\bibnamefont{{Pichler}}},
  \bibinfo{journal}{Entropy} \textbf{\bibinfo{volume}{19}},
  \bibinfo{pages}{319} (\bibinfo{year}{2017}), \eprint{1710.01973}.

\bibitem[{\citenamefont{Lewis et~al.}(2000)\citenamefont{Lewis, Challinor, and
  Lasenby}}]{Lewis:1999bs}
\bibinfo{author}{\bibfnamefont{A.}~\bibnamefont{Lewis}},
  \bibinfo{author}{\bibfnamefont{A.}~\bibnamefont{Challinor}},
  \bibnamefont{and} \bibinfo{author}{\bibfnamefont{A.}~\bibnamefont{Lasenby}},
  \bibinfo{journal}{\apj} \textbf{\bibinfo{volume}{538}}, \bibinfo{pages}{473}
  (\bibinfo{year}{2000}), \eprint{astro-ph/9911177}.

\end{thebibliography}
\bibliographystyle{apsrev}
\end{document}